\documentclass{aa}  

\usepackage{graphicx}
\usepackage[varg]{txfonts}
\usepackage{amsmath}
\usepackage{amssymb}
\usepackage[none]{hyphenat}
\usepackage{enumitem}
\usepackage{hyperref}
\usepackage{bm}

\usepackage{caption}
\usepackage{subcaption}

\hypersetup{
    colorlinks=true,    
    linkcolor=blue,
    citecolor=blue,
    urlcolor=blue,
}

\begin{document} 

   \title{Informed total-error-minimizing priors: \\
   Interpretable cosmological parameter constraints \\
   despite complex nuisance effects}
   
   \titlerunning{Informed total-error-minimizing priors}

   \author{Bernardita Ried Guachalla\inst{1,2,3,4}\fnmsep\thanks{bried@stanford.edu}
          \and
          Dylan Britt\inst{1,2,3,4}
          \and
          Daniel Gruen\inst{1,5}
          \and
          Oliver Friedrich\inst{1,5}
    }

    \institute{
            University Observatory, Faculty of Physics, Ludwig-Maximilians-Universit\"{a}t, Scheinerstra{\ss}e 1, 81679 Munich, Germany
        \and
            Department of Physics, Stanford University, 382 Via Pueblo Mall, Stanford, CA 94305, USA
         \and
             Kavli Institute for Particle Astrophysics \& Cosmology, 452 Lomita Mall, Stanford, CA 94305, USA
        \and
            SLAC National Accelerator Laboratory, 2575 Sand Hill Road, Menlo Park, CA 94025, USA
        \and
            Excellence Cluster ORIGINS, Boltzmannstra{\ss}e 2, 85748 Garching, Germany
    }

   \date{Received XXX; accepted YYY}

    \abstract
    {While Bayesian inference techniques are standard in cosmological analyses, it is common to interpret resulting parameter constraints with a frequentist intuition. 
    This intuition can fail, for example, when marginalizing high-dimensional parameter spaces onto subsets of parameters, because of what has come to be known as projection effects or prior volume effects. 
    We present the method of informed total-error-minimizing (ITEM) priors to address this problem. 
    An ITEM prior is a prior distribution on a set of nuisance parameters, such as those describing astrophysical or calibration systematics, intended to enforce the validity of a frequentist interpretation of the posterior constraints derived for a set of target parameters (e.g., cosmological parameters). 
    Our method works as follows. 
    For a set of plausible nuisance realizations, we generate target parameter posteriors using several different candidate priors for the nuisance parameters. 
    We reject candidate priors that do not accomplish the minimum requirements of bias (of point estimates) and coverage (of confidence regions among a set of noisy realizations of the data) for the target parameters on one or more of the plausible nuisance realizations. 
    Of the priors that survive this cut, we select the ITEM prior as the one that minimizes the total error of the marginalized posteriors of the target parameters. 
    As a proof of concept, we applied our method to the density split statistics measured in Dark Energy Survey Year 1 data. 
    We demonstrate that the ITEM priors substantially reduce prior volume effects that otherwise arise and that they allow for sharpened yet robust constraints on the parameters of interest.}


    \maketitle


\section{Introduction}
\label{sec:Intro}

Present and near-future wide-area surveys of galaxies will provide an unprecedented volume of data over most of the extragalactic sky. 
Examples of these surveys are the Dark Energy Survey (\href{https://www.darkenergysurvey.org/}{DES}; \citealt{DES2016}, \citealt{Sevilla_Noarbe_2021}), the \textit{Vera C. Rubin} Observatory's Legacy Survey of Space and Time (\href{http://www.lsst.org/lsst}{LSST}; \citealt{2009arXiv0912.0201L}, \citealt{Ivezi2019}), the Dark Energy Spectroscopic Instrument (\href{https://www.desi.lbl.gov/}{DESI}) survey (\citealt{desicollaboration2016desi}), the space mission \href{www.cosmos.esa.int/web/euclid, www.euclid-ec.org}{Euclid} (\citealt{laureijs2011euclid}), the 4m Multi-Object Spectroscopic Telescope (\href{https://www.4most.eu/cms/}{4MOST}) (\citealt{de_Jong_2012}), the Kilo-Degree Survey (\href{http://kids.strw.leidenuniv.nl/}{KiDS}; \citealt{de_Jong_2012}), the Hyper Suprime-Cam (\href{https://hsc.mtk.nao.ac.jp/ssp/}{HSC}; \citealt{Aihara_2017}), and the \href{https://roman.gsfc.nasa.gov/}{Nancy Grace Roman} Space Telescope (\citealt{2021MNRAS.507.1746E}). 
The promise of precision cosmology with the resulting data can only be realized by accurately accounting for the nuisance effects -- both astrophysical and in the calibration uncertainty of data -- that become relevant in increasingly complex models. 
Examples of astrophysical effects include intrinsic alignments of galaxies as a nuisance effect for weak lensing analyses (see \citealt{Troxel_2015} and \citealt{lamman2023ia} for recent reviews and \citealt{secco2021dark}, \citealt{Samuroff_2023}, \citealt{Asgari_2021} and \citealt{dalal2023hyper} for the latest analyses); baryonic physics that impact the statistics of the cosmic matter density field, such as star formation, radiative cooling, and feedback (e.g., \citealt{Cui_2014}; \citealt{Velliscig_2014}; \citealt{Mummery_2017}: \citealt{2022A&A...660A..27T}); galaxy bias and other parameters describing the galaxy-matter connection (see \citealt{Schmidt_2021} for a recent review, \citealt{Scherrer_1998}, \citealt{Dekel_1999}, \citealt{Baldauf_2011}, \citealt{Heymans_2021}, \citealt{ishikawa2021halomodel} and \citealt{Pandey_2022}); and systematic effects on the baryon acoustic oscillation signal (\citealt{Seo_2007}, \citealt{Ding_2018}, \citealt{rosell2021dark} and \citealt{Lamman_2023}). 
Calibration-related nuisance effects include measurement biases on galaxy shapes (see \citealt{mandelbaum2019widefield} for a review, \citealt{maccrann2020des}, \citealt{Georgiou_2021} and \citealt{li2021threeyear}), the estimation of redshift distributions of photometric galaxy samples (see \citealt{Newman_2022} for a review and \citealt{Tanaka_2017}, \citealt{Huang_2017}, \citealt{Hildebrandt_2021}, \citealt{Myles_2021}, \citealt{Cordero_2022}), and systematic clustering of galaxies induced by observational effects (e.g. \citealt{Cawthon_2022}, \citealt{Baleato_Lizancos_2023}). 

Cosmological inference from survey data is performed through likelihood analyses.
The first step is to predict a model data vector as a function of parameters (both cosmological and nuisance) using a theoretical modeling pipeline. 
The second step is to infer the posterior probability distribution of the parameters using the observed data vector as a noisy measurement.
Finally, one obtains a marginalized posterior for a subset of parameters by projecting out the remaining parameters.
The results are summarized in the form of point estimates (best estimate of a parameter value) with their confidence errors.
It might be desirable for these marginalized posteriors to have certain properties:

\begin{figure*}
  \centering
  \begin{subfigure}[b]{0.48\textwidth}
  \centering
    \includegraphics[width=0.85\textwidth]{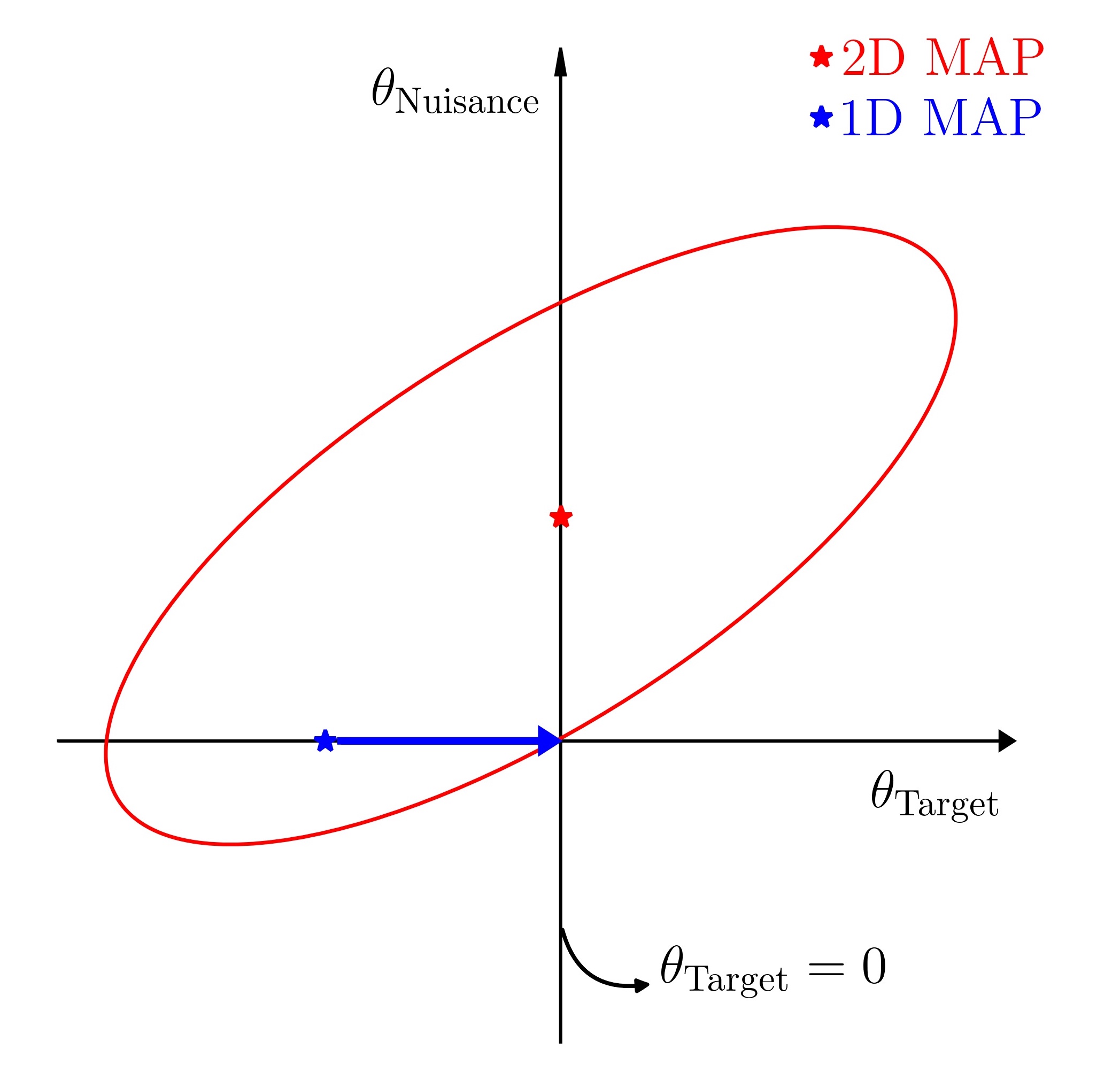}
  \end{subfigure}
  \hfill
  \begin{subfigure}[b]{0.48\textwidth}
    \includegraphics[width=0.85\textwidth]{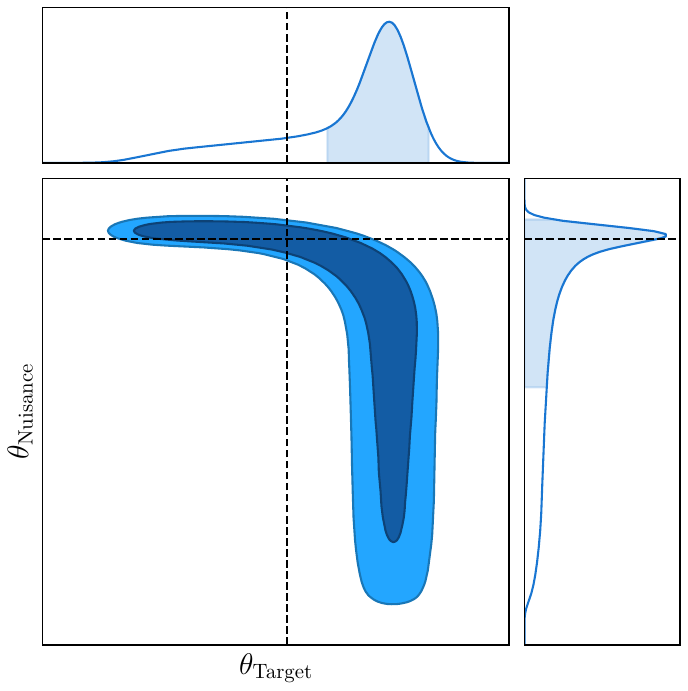}
  \end{subfigure}
  \caption{Sketches on how the treatment of nuisance parameters can cause systematic errors on the target parameter posteriors.
  For both panels, we assume a 2D space where the parameter $\pmb{\theta} = (\theta_{\rm Target}, \theta_{\rm Nuisance})$.
  \textit{Left:} \textbf{Bias due to underfitting a nuisance effect}. 
  In a 2D model, let 0 be the real value of the parameter $\theta_{\rm Target}$, despite the true value of $\theta_{\rm Nuisance}$.
  When $\theta_{\rm Nuisance}$ is a free parameter of the 2D Gaussian model presented in the panel, the true value of the target parameter is recovered with the 2D Maximum a Posteriori (2D MAP) corresponding to the red mark. 
  With a simpler nuisance model that fixes $\theta_{\rm Nuisance}=0$ and thus underfits the nuisance, the 1D MAP (blue mark) of $\theta_{\rm Target}$ is biased towards negative values as a result. 
  \textit{Right:} \textbf{Prior volume effect}. 
  In this toy illustration, we show the 2D and 1D marginalized posteriors of a distribution with the 1$\sigma$ and 2$\sigma$ confidence contours with blue and light blue respectively.
  Such posteriors can, for example, result from a model that is highly non-Gaussian in the full posterior space. 
  The black dashed line represents the fiducial value, and the intersection of them is inside the 1$\sigma$ contour.
  If we only have access to the 1D marginalized posterior of $\theta_{\rm Target}$, we would hardly estimate the fiducial value with our point estimators.
  In other words, even if the posterior in the entire parameter space is centered around the correct parameters, the posterior marginalized over the nuisance parameters can be off of the correct target parameters. 
  This bias error is called the prior volume effect, and in this paper, we propose a pipeline to overcome it by constraining systematically the priors on the nuisance parameters.
  }
  \label{fi:sketchs}
\end{figure*}

\begin{figure*}
  \includegraphics[width=0.7\textwidth]{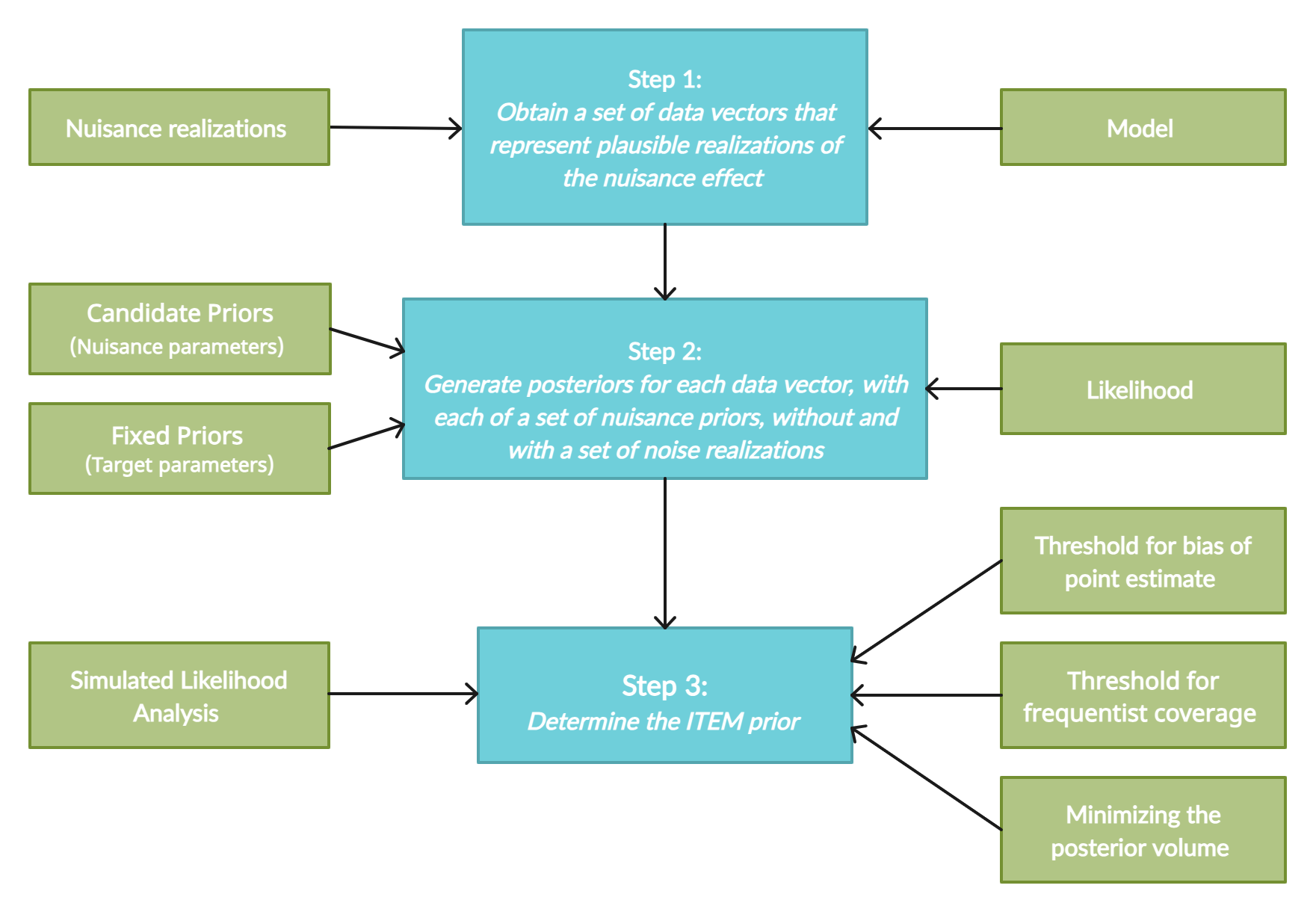}
  \centering
    \caption{Basic steps and concepts present in our ITEM prior methodology. In light blue, we summarize the steps of our pipeline, and in green, we highlight the ingredients of the methodology.
    }
    \label{fi:flowchart}
\end{figure*}

\begin{itemize}
    \item Unbiasedness: We require the point estimate of each cosmological parameter to have a bias, with respect to the fiducial parameters, smaller than some fraction of the width of the confidence interval (e.g., to be inside the 1$\sigma$ confidence interval or region if the space is two or more dimensions).
    In Sect. \ref{sec:crit_1} we review the point estimators that are currently used in cosmology.
    
    \item Reliability: In repeated trials (with the same realization of the nuisance, independent noise, and the same fiducial values for the cosmological parameters), we impose that the "true" values of a set of cosmological parameters be within a derived confidence region of those parameters for at least some desired fraction of time. 
    In other words, the frequentist coverage of the independent realizations should be consistent with the confidence level (the idea of matching Bayesian and frequentist coverage probabilities has been discussed extensively in \citealt{percival2021matching}).
    
    \item Robustness: We demand that the previous two requirements hold across a list of plausible nuisance and physical realizations of the fiducial configuration (e.g., under two different fiducial nuisance parameters, we still can recover unbiased and reliable marginal posteriors on the cosmology).
    
    \item Precision: Among all the possible priors whose resulting confidence regions fulfill the above requirements, we prefer the one that, on average, results in the smallest total (i.e., statistical and systematic) uncertainty on the cosmological parameters.
\end{itemize}

There are two ways in which the treatment of nuisance parameters can cause systematic errors and therefore violate these statements. 
The first one is to assume a nuisance parameter model that is overly simplistic, for instance, by fixing a nuisance parameter that indeed needs to be free to accurately describe the observed data. 
In the left panel of Fig. \ref{fi:sketchs}, we present a simple 2D sketch of this situation. There, the assumption is that a nuisance parameter, $\theta_{\rm Nuisance}$, equivalent to zero leads to an unavoidable bias on the cosmological parameter $\theta_{\rm Target}$.\footnote{Given that our work may be relevant to areas beyond cosmology, we use the term "target parameter" and "cosmological parameter" interchangeably for the rest of this paper.} 
We refer to this effect as "underfitting" of the nuisance, following \citet{Bernstein_2010}.
With the coming cosmological surveys, the complexity of the data is expected to increase, and therefore, unknown or not modeled nuisance effects could become underfitted. 

The second treatment involves assuming a prior distribution such that the estimated parameters on the marginalized posterior are biased with respect to the "true" values. 
This systematic error is known as prior volume effect or projection effect, and it is shown in the right panel of Fig. \ref{fi:sketchs}.
We emphasize that the prior volume effect is a continuous one (i.e., there are cases in which this error is larger than others), and it is not limited to cases where distributions are non-symmetric or exclusively non-Gaussian.
To avoid confusion, we limit ourselves to the definition of prior volume effect stated above, in which the point estimates calculated from the marginalized posteriors will be substantially offset from the true value being predicted. 
(For real-life examples of the effect, see, e.g.,~\citet{secco2021dark,Amon_2022} on intrinsic alignments, \citet{Sugiyama_2020,Pandey_2022} on galaxy bias, or \citet{Sartoris2020} on galaxy cluster mass profiles.
For studies on overcoming this effect in weak lensing analyses, see \citet{Joachimi_2021}, \citet{Chintalapati_2022}, and \citet{Campos_2023}.)

Another consequence of the prior volume effect is that the marginalized coverage fraction (the fraction of marginalized confidence intervals in repeated trials that encompass the true target parameters) may not be consistent with the associated credible level. 
The Bayesian and frequentist interpretations imply that the full posterior's confidence region does not necessarily have the same credible region. 
If this is the case, these two effects combine when marginalizing, and the marginalized coverage fraction is less intuitively characterized.
This can be rephrased as follows: The marginalized 1$\sigma$ confidence contours obtained with repeated Bayesian analyses in an ensemble of realizations do not necessarily encloses the fiducial parameter configuration with a frequency of $\sim68 \%$. 
The discrepancy may not constitute a problem from a purely Bayesian standpoint since Bayesian statistics do not claim to satisfy frequentist expectations. 
However, this does not change the fact that one may want to make probabilistic statements about the true value of a physical parameter and that marginalized parameter constraints quoted in cosmological publications are interpreted in a frequentist way by a large fraction (if not the majority) of the cosmology audience. 
Projection effects in high-dimensional nuisance parameter spaces may hence cause a buildup of wrong intuitions in the inferred parameters.
Therefore, in the absence of projections, we would only find the general mismatch between the confidence and credible regions of the marginalized posteriors, and this mismatch can be overcome by using solutions such as the one presented in \citet{percival2021matching} and \citet{Chintalapati_2022}.

The situation is further complicated by the fact that a precise model for nuisance effects with a finite set of parameters and well-motivated Bayesian priors is often not available. 
Commonly, at best, some plausible configurations of the nuisance effect are known. 
These could, for example, be a set of summary statistics measured from a range of plausible hydrodynamical simulations (\citealt{van_Daalen_2019}, \citealt{Walther_2021}) or a compilation of different models and parameters that have been found to approximately describe the nuisance, as is the case for intrinsic alignment \citep{Hirata2004, Kiessling_2015, Blazek_2019}, the galaxy-matter connection  (\citealt{szewciw2021accurate}, \citealt{Voivodic_2021}, \citealt{friedrich2021pdf}, \citealt{britt2024bounds}), and in the calibration of redshift distributions \citep{Cordero_2022}. 
Additionally, simulation-based inference techniques, such as simulation-based calibration (SBC; \citealt{2018Talts}), have offered an alternative way to discriminate among priors, and the cosmology community has recently applied SBC in a number of studies (\citealt{novaes2024cosmologyhscy1weak}, \citealt{Ho_2024}, \citealt{nguyen2024informationextractedgalaxyclustering}, \citealt{Ivanov2024}, \citealt{yao2024simulationbasedstacking}).
SBC does not make any assumptions about the analytical likelihood functions (\citealt{Cranmer_2020}), yet it evaluates whether a probabilistic model accurately infers parameters from data by comparing the inferred posterior distributions with the priors.
However, it has limitations, such as the fact that the SBC method quantifies biases in a one-dimensional posterior relative to a baseline prior distribution using a rank histogram (\citealt{yao2023discriminative}).
Current high-dimensional cosmological models, however, include multiple nuisance parameters, which requires us to diagnose biases in the joint distribution.

In this paper, our objective is to make statements about the marginalized posteriors of the target parameters that fulfill the above criteria of unbiasedness, reliability, robustness, and precision. 
These are pragmatic objectives and not necessarily consistent with common Bayesian approaches. 
We chose to do this by defining the prior on nuisance effects in such a way that our objectives are met instead of by directly using our prior knowledge of their parameter values.
Since the nuisance priors we constructed to satisfy these criteria are informed by knowledge of possible configurations of the nuisance effects and they minimize the systematic and statistical errors, we call them informed total-error-minimizing (ITEM) priors.
How we constructed these priors has two important consequences for the interpretation of the resulting parameter constraints:

\begin{itemize}
    \item In the presence of limited information about the nuisance effect, the considered realizations can be used to tune the priors of the nuisance parameters to ensure certain properties of the resulting constraints on the target parameters. 
    Hence, we do not assume the ITEM prior directly represents information about the nuisance effect or our knowledge about it. It is merely a tool for interpretable inference of the target parameters.
    As a consequence, we relinquish the ability to derive meaningful posterior constraints on the nuisance parameters.
    
    \item The coverage fraction and bias of any procedure to derive parameter constraints in a set of repeated experiments will be sensitive to the "true" values of cosmological and nuisance parameters that underlie those experiments. 
    Given a set of plausible realizations of cosmology and nuisances, we can therefore only ensure a minimum coverage and a maximum bias of the constraints derived from ITEM priors.
\end{itemize}

This paper is structured as follows: We start by describing the general methodology for obtaining ITEM priors in Sect. \ref{sec:methodology}. 
In Sect. \ref{sec:testing_in_DSS}, we investigate the performance of the method in a test case: cosmological parameter constraints marginalized over models for non-Poissonian shot-noise in density split statistics (\citealt{Friedrich2018, Gruen2018}). 
We explore the change in simulated and DES Y1 results of these statistics when using ITEM priors defined from a simplified set of plausible nuisance realizations. Finally, a discussion and conclusions are given in Sect. \ref{sec:d_c}.


\section{Methodology}
\label{sec:methodology}

Assume that some observational data can be described in a data vector $\pmb{\hat{\xi}}$ of $N_d$ data points. 
Let $\pmb{\xi}[\pmb{\theta}]$ be a model for this data vector that depends on a parameter vector $\pmb{\theta}$ of $N_p$ parameters, and let $\pmb{\text{C}}$ be the covariance matrix of $\pmb{\hat{\xi}}$. 
In many cases, it is reasonable to assume that the likelihood $\mathcal{L}$ of finding $\pmb{\hat{\xi}}$ given the parameters $\pmb{\theta}$ is a multivariate Gaussian distribution: 
\begin{equation}
    \ln{\mathcal{L}(\pmb{\hat{\xi}}|\pmb{\theta})} = -\dfrac{1}{2} \sum_{i,j} (\hat{\xi}_i - \xi_i[\theta]) \text{C}^{-1}_{ij} (\hat{\xi}_j - \xi_j[\theta]) + \mathcal{C} \; ,
\end{equation}
where $\text{C}^{-1}_{ij}$ are the elements of the inverse of the covariance matrix $\pmb{\text{C}}$, and $\mathcal{C}$ is a constant as long as the covariance does not depend on the model parameters. 
We derived the posterior distribution $\mathcal{P}(\pmb{\hat{\xi}}|\pmb{\theta})$ using Bayes' theorem:
\begin{equation}
    \mathcal{P}(\pmb{\theta}|\pmb{\hat{\xi}}) \propto \mathcal{L}(\pmb{\hat{\xi}}|\pmb{\theta}) \Pi(\pmb{\theta}) \; ,
\end{equation}
with $\Pi(\pmb{\theta})$ being a prior probability distribution incorporating a priori knowledge or assumptions on $\pmb{\theta}$. 

In many situations, parameters $\pmb{\theta}$ can be divided into two types: target parameters $\pmb{\theta}_{T}$ and nuisance parameters $\pmb{\theta}_{N}$, as done in Fig. \ref{fi:sketchs} ($\theta_{\rm Target}$ and $\theta_{\rm Nuisance}$ respectively). 
This distinction is subjective and depends on which parameters one would like to make well-founded statements about (i.e., robust, valid, reliable, and/or precise estimates).
Also, this division is based on which parameters are needed to describe the complexities of the experiment but are not considered intrinsically of interest. 
For example, in a cosmological experiment, the target parameters are usually those describing the studied cosmological model, such as the $\Lambda$-Cold Dark Matter ($\Lambda$CDM) and its variations. 
This could, for example, include the present-day density parameters ($\Omega_x$, with $x$ representing a specific content of the universe), the Hubble constant ($H_0$), and the equation of state parameter of dark energy ($w$). 
On the other hand, the nuisance parameters account for observational and astrophysical effects, or uncertainties in the modeling of the data vector. 

Recent cosmological analyses have chosen an approximate nuisance model with a limited number of free parameters for all these complex effects. 
These either assume a wide flat prior distribution over the nuisance parameters or an informative prior distribution set by simulation and/or calibration measurements \citep{planck_2018, Asgari_2021, Sugiyama_2022, Abbott_2022}.
As we explained in Sect. \ref{sec:Intro}, precarious modeling of nuisance parameters can cause systematic errors, leading to the following consequences:

\begin{itemize}
    \item Limited modeling of the nuisance effects: 
    The assumed model $\pmb{\xi}[\pmb{\theta}]$ may not sufficiently describe the nuisance effects. 
    For example, there could be an oversight of one underlying source of the nuisance, the model of a nuisance effect may be truncated at too low an order, or it can be entirely neglected. 
    This results in an intrinsic bias on the inferred target parameters, as would, for instance, be revealed by performing the analysis on simulated data that include the full nuisance effect. 
    \item Prior-volume effects: 
    The prior distribution $\Pi(\pmb{\theta}_N)$ over one or multiple nuisance parameters could result in biases in the marginalized posterior when projecting to lower dimensions.
\end{itemize}

\begin{figure*}

  \includegraphics[width=1.0\textwidth]{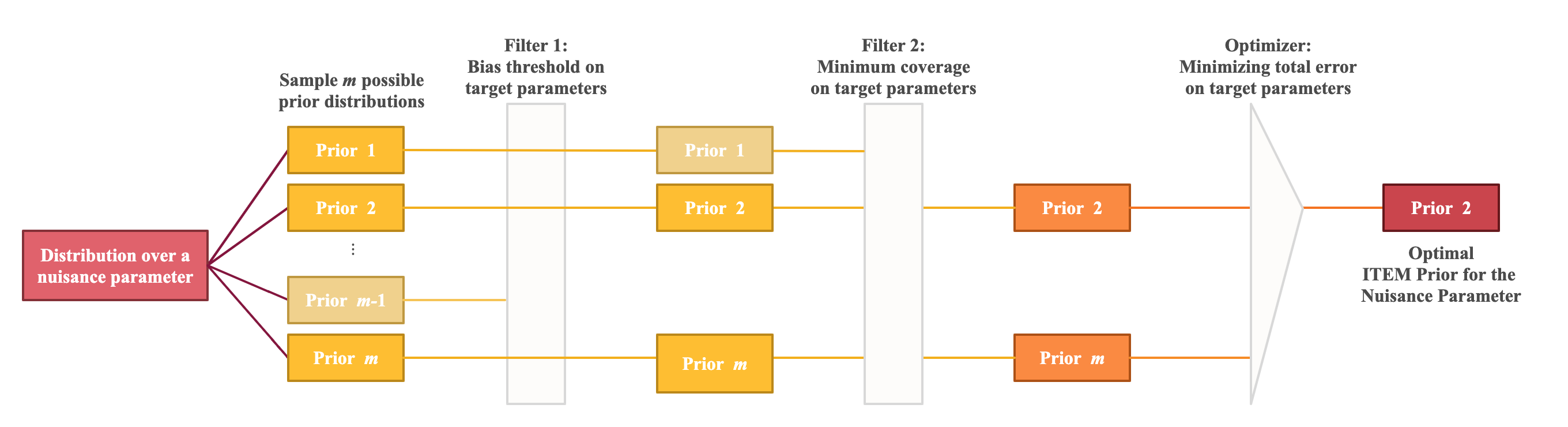}
  \centering
    \caption{Flowchart of the procedure to obtain the ITEM prior. First, a set of $m$ candidate prior distributions is proposed for the nuisance parameters. 
    Then, the first criterion filters some of the priors by requiring that the bias on the target parameters in a simulated likelihood analysis be less than $x^*$ times its posterior's statistical uncertainty (see Eq. \ref{eq:x_thershold}). 
    Second, we require that the remaining priors result in a minimum coverage $y^*$ (see Eq. \ref{eq:y_thershold}) over the target quantities. 
    This leads to $m' \leq m$ candidate priors, which are finally optimized by minimizing the target parameters' uncertainty (see Sect. \ref{sec:crit_3}).} 
    \label{fi:flowchart_2}
\end{figure*}

In this work, we introduce a procedure that determines a prior distribution on the nuisance parameters that ensures the criteria presented in Sect. \ref{sec:Intro} for marginalized posteriors of the target parameters. 
We will call the resulting priors {Informed Total-Error-Minimizing} priors, or ITEM priors. 
In Sect. \ref{sec:step_1}, we describe how to obtain different data vectors given a model and plausible realizations of the nuisance effect.
In Sect. \ref{sec:step_2}, we generate posteriors for each data vector given three ingredients: the likelihood of the model, a set of fixed prior distributions for the target parameters, and a collection of nuisance candidate priors.
In Sect. \ref{sec:step_3} we determine the ITEM prior after requiring the posteriors to account for specific criteria based on maximum biases, a minimum coverage fraction, and the minimization of the posterior uncertainty.
The steps are summarized in Fig. \ref{fi:flowchart}.


\subsection{Step 1: Obtain a set of data vectors that represent plausible realizations of the nuisance effect}
\label{sec:step_1}

We assumed that at fixed target (e.g.~cosmological) parameters $\pmb{\theta}_{T}$, the range of plausible realizations of a nuisance effect is well represented by a set of $n$ different resulting data vectors:
\begin{equation}
\pmb{\hat{\xi}}^i = \pmb{\xi}[\pmb{\theta}_{T}, \mathrm{nuisance}\;\mathrm{realization}\;i] , \text{ for } i = 1, ..., n\ .
\end{equation}
These could, for example,~be based on a model for the nuisance effect with different nuisance parameter values. 
Alternatively, the $\pmb{\hat{\xi}}^i$ could result from a set of simulations with different assumptions that produce realizations of the nuisance effect (e.g., the impact of baryonic physics on the power spectrum from a variety of hydrodynamical simulations, as done in \citealt{Chisari2018} and \citealt{Huang_2019}). 
In some other cases, direct measurements of the nuisance effect may be used, and for example a bootstrapping of those measurements could yield a set of possible realizations. 
The ITEM prior is constructed such (cf. Sects. \ref{sec:crit_1}, \ref{sec:crit_2}, and \ref{sec:crit_3}) that if the most extreme among the reasonable nuisance realizations are included in the process, our desired properties of the posterior derived with the resulting ITEM prior are likely to hold for any intermediate nuisance realizations as well.

For the sake of notation, let us assume that the nuisance realizations are given in terms of particular nuisance parameter combinations $\pmb{\theta}_{N}^i$ of the considered nuisance model: $\pmb{\hat{\xi}}^i = \pmb{\hat{\xi}}[\pmb{\theta}_{T}, \pmb{\theta}_{N}^i]$. 
However, we note that this is not an assumption that a model to be used in the analysis shares those same nuisance parameters.
For the following tests to be meaningful, we require these data vectors to contain no noise, or noise that is negligible compared to the covariance considered in the likelihood analysis. 

In principle, it is possible to consider more than one realization of the target parameters as well (e.g., more than one cosmology for the same nuisance parametrization). 
This extra layer of complexity can be added to the pipeline, but as a first pass, we will limit ourselves to a fixed target parameter vector for simplicity, yet suggest doing this in future works. 


\subsection{Step 2: Generate posteriors for each data vector, with each of a set of nuisance priors, with and without a set of noise realizations}
\label{sec:step_2} 

We chose our ITEM prior among the space of nuisance priors, that is,~the space of the parameters that describe a family of considered prior distributions $\Pi(\pmb{\theta}_{N})$. 
The details of that family may depend on: the preferred functional form of the priors, on consistency relations that are a priori known, and on practical considerations such as limits on the computational power.
As mentioned before, the distinction between target and nuisance is subjective, and these labels may change between iterations of the pipeline (i.e., a prior on a target parameter can be fixed for one analysis, but it can be studied and optimized as a nuisance in another). 

For example, we may search for an ITEM prior among a set of uniform prior distributions for one particular nuisance parameter, $\Pi(\theta_{\rm Nuisance}) \sim \mathcal{U}(a, b)$, where $a$ and $b$ are the lower and upper bounds, respectively. 
In that case, we propose to sample uniformly and independently over both $a$ and $b$, subject to the consistency relation that $a < b$. 
In practice, we can for example sample $m_a$ lower bounds ($a_i \in [a_1, a_2, ..., a_{m_a}]$) and $m_b$ upper bounds ($b_j \in [b_1, b_2, ..., b_{m_b}]$). 
Besides requiring that $a_i < b_j$ for all tested pairs one could demand, if there was some reason to, that other requirements are met, for example a minimum and maximum value for the prior width $b_j - a_i$, or that some nuisance parameter values are contained within the considered priors. 
More generally, we aim to use candidate priors that cover a reasonable range of nuisance realizations, ensuring that these realizations are enclosed within a region of non-negligible probability and can therefore be sampled.
A Kullback-Leibler divergence test (\citealt{Kullback1951}) could also be used to help determine the ranges these priors could span.
When combining all these configurations, we will have up to $m_a \cdot m_b$ possible priors with different widths and midpoints. 

This idea can be extended to many dimensions, and the combination of all possible priors that could be used will lead to a total number of priors * denote by $m$.
Therefore, for any given measurement of the data $\pmb{\hat{\xi}}$, this will result in $n \cdot m$ different posteriors on the full set of parameters $\pmb{\theta}$.
\begin{equation}
    \mathcal{P}_j(\pmb{\theta}|\pmb{\hat{\xi}}^i) \propto \mathcal{L}(\pmb{\hat{\xi}}^i|\pmb{\theta}) \Pi_j(\pmb{\theta}_{N}) \ ,
\end{equation}
for $i = 1, ..., n $ and $j = 1, ..., m$.

In the next steps of our pipeline, we generate such sets of posteriors for different data vector realizations to determine which priors meet the criteria we outlined in Sect. \ref{sec:Intro}. 
To obtain all these posteriors, instead of running a large number of Monte-Carlo-Markov-Chains (MCMCs) for the $n \cdot m$ total posteriors, we propose to do an importance sampling (\citealt{Siegmund_1976}, \citealt{importance_sampling1}) over the posterior derived from the widest prior (i.e., the prior on the nuisance that encloses all the other priors on its prior volume).
We refer to Sect. 3.4.2 from \citet{Campos_2023} for a summary on importance sampling. 
This sampling technique is straightforward when using uniform priors, but becomes highly time-consuming with more complex priors.
In such cases, neural networks and importance-weighted variational autoencoders can help accelerate the process.
Recent work on likelihood emulators in cosmology has been proposed to speed up inference (see, e.g., \citealt{To2023}).
Additionally, we generate a collection of $n_{\mathrm{noise}}$ realizations by adding multi-variate Gaussian noise (according to our fiducial covariance) to each noiseless $\pmb{\hat{\xi}}^i$. 
For this paper, we focus on a family of uniform priors; however, it is straightforward to generalize our pipeline to, for example, Gaussian prior distributions or other functional forms commonly used in cosmological analyses.


\subsection{Step 3: Determine the ITEM prior}
\label{sec:step_3}

Our goal is to find a prior on the nuisance parameters that returns robust, valid, reliable, and precise posteriors for the target parameters. 
To determine the ITEM prior, we apply filters to the considered family of nuisance prior distributions. 
The first and second ones are to set an upper limit for the bias and a lower limit for the coverage fraction of the posterior respectively. 
The third and final one is an optimization that minimizes the uncertainty on the target parameters. 
Figure \ref{fi:flowchart_2} synthesizes these steps in a diagram. 


\subsubsection{Criterion 1: Maximum bias for the point estimate}
\label{sec:crit_1}

Constraints on target parameters are commonly reported as confidence intervals around a point estimate for those parameters. For example, that point estimate may be the maximum a posteriori probability (MAP) estimator, the maximum of a marginalized posterior on the target parameters, or the 1D marginalized mean (that we will refer as mean for simplicity). 
To devise a measure for the bias with respect to the true values of the target parameters let us return to the exercise that was already mentioned in Sect. \ref{sec:Intro}: 
Consider a noise-free realization $\bm{\xi}$ of a data vector, and a data vector model $\bm{\xi}[\pmb{\theta}]$, which we assume to model the data perfectly, and the fiducial parameters $\pmb{\theta}$ are known. 
We use this model to run an MCMC around $\bm{\xi}$ as if the latter was a noisy measurement of the data. 
If the model $\bm{\xi}[\pmb{\theta}]$ is non-linear in the nuisance parameters, then for example the MAP of the marginalized posterior for the target parameters obtained from that MCMC can be biased with respect to the true parameters (even though $\bm{\xi}$ is noiseless and the model $\bm{\xi}[\pmb{\theta}]$ is assumed to be perfect). 

In Sect. \ref{sec:step_1} we have obtained a set of realizations $\pmb{\hat{\xi}}^i$ (with $i = 1,\ \dots\ ,\ n$) of the data, corresponding to $n$ different realizations of the considered nuisance effects. 
For each marginalized posterior from $\pmb{\hat{\xi}}^i$ we can obtain a point estimator.
The difference between these estimates and the true target parameters underlying our data vectors will then act as a measure of how much projection effects bias our inference. 

If $\pmb{\theta}_{T}$ is the true target vector and $\hat{\pmb{\theta}}_{T}^{i}$ is the vector estimate from the nuisance realization $i$, we then use the Mahalanobis distance (\citealt{mahalanobis1936generalized, etherington2019mahalanobis}) to quantify the bias between the two, namely:
\begin{equation}
    x_i = \sqrt{(\pmb{\theta}_{T} - \hat{\pmb{\theta}}_{T}^{ i})^{\top} \textbf{C}_i^{-1} (\pmb{\theta}_{T} - \hat{\pmb{\theta}}_{T}^{ i})}\ ,
    \label{eq:mahalanobis}
\end{equation}
where $\textbf{C}_i^{-1}$ is the inverse covariance matrix of the parameter posterior, which we directly estimate from the chain for the $i$-th nuisance realization. 
We note that Eq. \ref{eq:mahalanobis} is related to the $\chi$-square test. 

The quantity $x_{i}$ measures how much the point estimate deviates from the truth compared to the overall extension and orientation of the posterior constraints. 
The first filter we apply to our family of candidate priors demands all $x_{i}$ to be smaller than a maximum threshold, which we denote as $x^*$. 
This means that we calculate $x_{i}$ for each realization of the nuisance, and we then determine the maximum:
\begin{equation}
    x = \max_{i} \{ x_{i} \}
    \label{eq:x_thershold}
\end{equation}
of those Mahalanobis distances. 
All candidate priors that do not meet the requirement 
\begin{equation}
    x \leq x^*
\end{equation}
will be excluded. 
In other words, we demand that the prior results in less than $x^*$ bias for any realization of the nuisance with respect to the target parameter uncertainties. 
That leaves us with an equal or smaller number of candidate priors, depending on the considered threshold quantity $x^*$. 
For example, in the DES Y3 2-point function analyses, this threshold was chosen as $0.3 \sigma_{\rm 2D}$ (with the subscript "2D" standing for the 68$\%$ credible interval) for the joint, marginalized posterior of $S_8$ and $\Omega_m$ (\citealt{krause2021}).
If none of the considered priors meet the bias requirement, this suggests that, given the setup of the problem, a less stringent bias threshold should be chosen and taken into account when interpreting the posterior.


\subsubsection{Criterion 2: Minimum frequentist coverage}
\label{sec:crit_2}

We now apply a second filter to the candidate priors: a minimum coverage probability for the $1$$\sigma$ confidence region of the marginalized posterior of the target parameters (i.e. the smallest sub-volume of target parameter space that still encloses about $68$\% of the posterior). 
This coverage probability measures how often the true target parameters would be found within that confidence region in independent repeated trials. 
In a Bayesian analysis, it can in general not be expected that this coverage matches the confidence level of the considered region (in our case $68$\%). 
This has been extensively discussed in \citet{percival2021matching}, for example. 
The presence of projection effects does impact this additionally when studying the marginalized posteriors. 

To measure the coverage probability, we proceed as follows for a given prior: 
For each of the $n_{\mathrm{noise}}$ noise realizations per data vector $\pmb{\hat{\xi}}^i$ described in Sect. \ref{sec:step_2}, we determine the $1$$\sigma$ confidence region (e.g., the contour in the target parameter space or the 1D confidence limit of each parameter). 
The fraction of these confidence regions that enclose the true target parameters is an estimate for the coverage probability.
We will denote with $y_i$ the fraction corresponding to the $i$-th nuisance realization.
With an approach similar to our previous filter, we considered the minimum coverage per nuisance realization,
\begin{equation}
    y = \min_{i} \{ y_{i} \}\ ,
    \label{eq:y_thershold}
\end{equation}
and then we required each nuisance realization to have a coverage of at least $y^*$,
\begin{equation}
    y \geq y^*,
\end{equation}
in order for a candidate nuisance prior to pass our coverage criterion. 
In other words, we demanded that a prior result in a coverage of at least $y^*$.

It is worth mentioning that the value of $y^*$ is not necessarily the probability associated with a confidence level of $1$$\sigma$ (i.e., $\sim 68\%$) because the coverage fraction will scatter around that value following a binomial distribution.
As an example, for 100 independent noisy realizations, the probability of having $z$ out of the 100 realizations covering the true parameters will be $P(z) = $ Bin$(z; n=100, p=0.68)$. 
Therefore, ensuring a minimum coverage such that the cumulative binomial distribution at that value is statistically meaningful should be enough.
Using our same previous example, a coverage of 62 or more out of 100 realizations will happen with 90$\%$ probability. 
The larger the number of nuisance realizations used to derive an ITEM prior, the smaller the coverage threshold should be chosen to avoid false positives of low coverage. 
The smaller the number of nuisance realizations used to derive an ITEM prior, the higher the coverage threshold should be chosen to avoid false positives of low coverage. 

In the presence of a bias on the target parameters, however, there will be a mismatch between the coverage probability and the confidence level of the posterior.
If an otherwise accurate multivariate Gaussian posterior is not, on average, centered on the true parameter values, any of its confidence regions will have a lower coverage than nominally expected.
To account for this effect in the coverage test without double-penalizing bias, and later to find the prior that minimizes posterior volume in a total error sense, we propose to appropriately raise the confidence level to be used. 
For our implementation of this, we made the assumption of a Gaussian posterior. 
For an unbiased posterior, this means that among many realizations of the noisy data vector, the log-posterior value of the true parameters is a $\chi^2$ distributed random variable with the number of degrees of freedom equal to the number of target parameters. 
In the case of a biased analysis, the log-posterior of the true parameters instead follows a non-central $\chi^2$ distribution. 

To calculate the required confidence level, we proceeded as follows:
\begin{itemize}
    \item We calculated for which value the cumulative distribution function of a non-central $\chi^2$-distribution representing the biased marginalized posterior takes a value of $68\%$.
    \item At that value, we evaluated the cumulative distribution function of a central $\chi^2$-distribution (with the same number of degrees of freedom). This will always be higher than $68\%$.
\end{itemize}
It is this higher confidence level that we use to calculate the enlarged "total error" $1$$\sigma$ confidence region for the noisy realizations of our data vectors $\bm{\xi}_i$ and to measure the coverage probabilities.  

\begin{figure}
  \includegraphics[width=0.95\columnwidth]{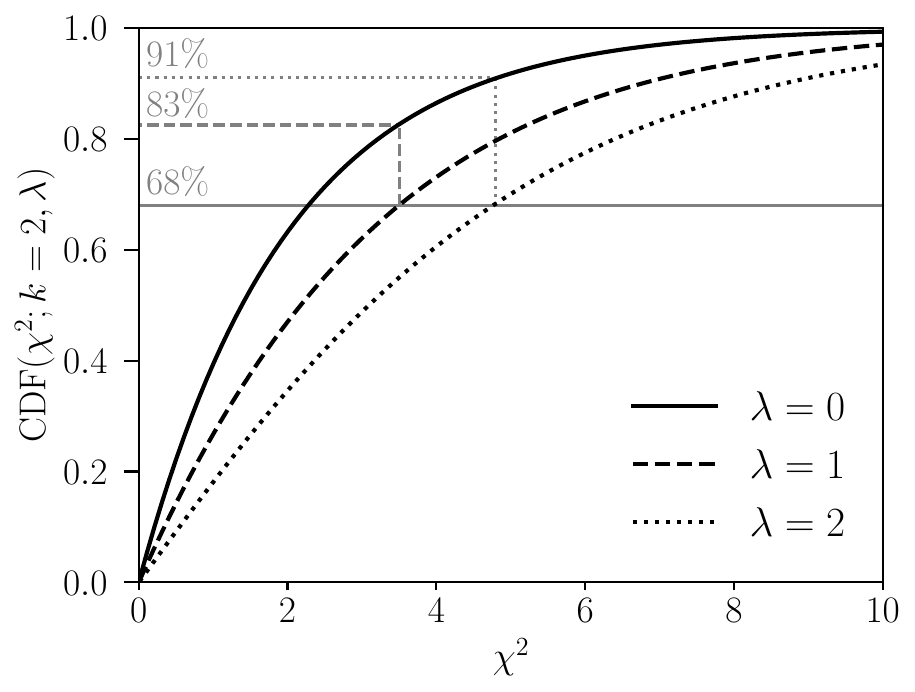}
  \centering
    \caption{
    Cumulative distribution functions of two degrees of freedom central and non-central $\chi^2$-distributions.
    The plotted non-central $\chi^2$-distribution shows that for the same significance threshold, the central $\chi^2$-distribution will always have a larger CDF.
    }
    \label{fi:CDF_chi_squared}
\end{figure}

The $\chi^2$-distribution with $k$ degrees of freedom is the distribution of a sum of the squares of $k$ independent standard normal random variables.
Its cumulative distribution function (CDF) is
\begin{equation}
    {\rm CDF}(\chi^2;\,k)={\frac {\gamma ({\frac {k}{2}},\,{\frac {\chi^2}{2}})}{\Gamma ({\frac {k}{2}})}},
\end{equation}
where $\gamma$ is the lower incomplete gamma function and $\Gamma$ is the gamma function.
In Fig. \ref{fi:CDF_chi_squared}, we plot with a continuous line the CDF of a $\chi^2$-distribution with two degrees of freedom ($k = 2$) for visualization purposes. 

The non-central $\chi^2$-distribution with $k$ degrees of freedom is an extension of the central $\chi^2$-distribution, and it has a non-centrality parameter $\lambda_i$ defined as
\begin{equation}
    \lambda_i = (\pmb{\theta}_{T} - \hat{\pmb{\theta}}_{T}^{ i})^{\top} \textbf{C}_i^{-1} (\pmb{\theta}_{T} - \hat{\pmb{\theta}}_{T}^{ i})\ ,
    \label{eq:noncentrality}
\end{equation}
where $\pmb{\theta}_{T}$ is the true target parameter vector, $\hat{\pmb{\theta}}_{T}^{i}$ is the target parameter estimate for nuisance realization $i$, and $\textbf{C}_i^{-1}$ is the associated inverse covariance matrix. 
We point out that $\lambda_i = x_i^2$ from Eq. \ref{eq:mahalanobis}. 
In Fig. \ref{fi:CDF_chi_squared} we show the CDF of two non-central $\chi^2$-distributions with two degrees of freedom, but different non-centrality parameters.
For CDF($\chi^2; k = 2, \lambda =1, 2$) = 0.68, it is clear that CDF($\chi^2; k = 2$) > 0.68. 
This augmented cumulative probability accounts for the overall bias, including the one induced by projection effects. 

%


\subsubsection{Criterion 3: Minimizing the posterior volume}
\label{sec:crit_3}

The candidate priors that have passed our previous two filters follow our definitions of {unbiasedness}, {reliability}, and {robustness}. 
The ITEM prior corresponds to the one among them that yields the smallest posterior uncertainty on the target parameters. 
This final selection step is necessary because it is in fact easier to meet our previous criteria with wide posteriors that result from unconstrained nuisance priors. 

We quantify the target parameter random error by approximating the volume that the $1$$\sigma$ confidence contours (68$\%$) enclose. 
For a Gaussian posterior of any dimensionality, the volume would be given in terms of the parameter covariance matrix $\textbf{C}_i$ as (see \citealt{Huterer_2001} for a derivation)
\begin{equation}
    V_i = \pi \hspace{0.05 cm} \text{det($\textbf{C}_i$)}^{1/2}\ ,
\end{equation}
and for simplicity, we used that formula also for our (potentially non-Gaussian) posteriors.

We let $m'$ be the number of candidate priors that passed the filters listed in Sects.~\ref{sec:crit_1} and \ref{sec:crit_2}.
We let $V_{i, j}$ be the volume enclosed by the $1$$\sigma$ ellipsoid of the $i$-th posterior distributions at the $j$-th prior, $j = 1, ..., m'$.
Then, we considered the maximum of these volumes across all noiseless data vectors: 

\begin{equation}
    V_j^{\rm max} = 
    \max_{i}(V_{i, j}) .
    \label{eq:volume_random}
\end{equation}

As done in Sect. \ref{sec:crit_2}, we accounted for the bias when optimizing the posterior uncertainty by considering the volume enclosed by the extended central $\chi^2$-distribution. This allows us to minimize the total error, including random and systematic contributions. 
The re-scaled volume includes a factor equivalent to (see Appendix \ref{app:volume}) 
\begin{equation}
    \tilde{V}_{i, j} \propto V_{i, j} \cdot {\rm CDF}^{-1}(0.68; k = N_T, \lambda_i)^{N_T/2}. \label{eqn:volume_rescaling}
\end{equation}
We chose the prior that has the minimum volume $\smash{\displaystyle \tilde{V}_{\rm ITEM} =  \min_{j} (\tilde{V}_j^{\rm max})}$ as the ITEM prior.


\section{ITEM priors for density split statistics}
\label{sec:testing_in_DSS}

\begin{table*}
\caption{\label{t7}Parameters used in the simulated likelihood analyses for the $\alpha$ model.}
        \centering
        \label{tab:table_1}
        \begin{tabular}{lccc} 
                \hline
                 & Parameter vector 1 & Parameter vector 2 & DSS original\\
                 & Non-stochasticity & Buzzard stochasticity & Prior distribution  \\
                \hline
                \hline
                \textbf{Target parameters} &  &  & \\
                Cosmological parameters &  &  & \\
                $\Omega_m$ & 0.286 & 0.286 & $\mathcal{U}$[0.1,0.9] \\
                $\sigma_8$ & 0.820 & 0.820   & $\mathcal{U}$[0.2,1.6] \\
                \hline
                \textbf{Nuisance parameters} &  &  &  \\
                Tracer galaxies &  &  &  \\
                $b$ & 1.54 & 1.54  & $\mathcal{U}$[0.8,2.5] \\
                
                 &  &  & \\
                 
                Stochasticity &  &  &  \\
                $\alpha_0$ & 1.00 & 1.26 & $\mathcal{U}$[0.1,3.0]   \\
                $\alpha_1$ & 0.00 & 0.29 & $\mathcal{U}$[-1.0,4.0]      \vspace{0.3 cm} \\
        \end{tabular}
\tablefoot{
From left to right, we list the fiducial values and prior ranges (using $\mathcal{U}$ to denote a uniform prior) used to simulate the synthetic data for the two cases of stochasticity.
The nuisance realizations and the priors are chosen to be either the same or the derived results from \citealt{Friedrich2018, Gruen2018}. 
In our work, the cosmological parameters would be the target ones, while the stochasticity parameters would be the nuisance ones. 
$\Omega_m, S_8$ and $b$ have fixed priors and unchanged values for the parameter vectors.
}
\end{table*}

In this section we examine the performance of the ITEM prior when analyzing a higher-order statistic of the matter density field called Density Split Statistics. 
We do this as a proof of concept under simplified assumptions, particularly on the nuisance realization deemed plausible.
Readers are referred to \citealt{britt2024bounds} for more careful analysis of potential nuisance effects relevant for those statistics.


\subsection{Density split statistics}
\label{sec:DSS}

We test our ITEM prior methodology in considering as the relevant nuisance effect the galaxy-matter connection models employed for the Density Split Statistics (DSS, \citealt{Friedrich2018, Gruen2018} for the Dark Energy Survey Year 1, DES Y1, analysis). The data vector of DSS is a compressed version of the joint PDF of projected matter density and galaxy count. These studies consider two ways of generalizing a linear, deterministic galaxy-matter connection, which is their most relevant nuisance effect.
The first one accounts for an extension of the Poissonian shot-noise in the distribution of galaxy counts by adding two free parameters: $\alpha_0$ and $\alpha_1$.
For simplicity, we call this the $\alpha$ model.
The second one more simply parametrizes galaxy stochasticity via one correlation coefficient $r$.
We refer to this as the $r$ model.
(For a derivation and explanation of these models, see Appendix \ref{app:APP_2}.)


\subsection{Simulated likelihood analysis with wide priors}
\label{sec:modeling_nuisance_DSS}

In this subsection, we present results of simulated likelihood analyses of DSS with the wide prior distributions used in the original DES Y1 analysis \citep{Friedrich2018, Gruen2018}. 
Our study focuses on the five quantities listed in Table \ref{tab:table_1}. 
The cosmological parameters are $\Omega_m$\footnote{The present-day energy density of matter in units of the critical energy density.} and $\sigma_8$\footnote{The present-day linear root-mean-square amplitude of relative matter density fluctuations in spheres of radius 8 $h^{-1}$ Mpc.}. 
The parameters describing the galaxy-matter connection are the linear bias $b$\footnote{The square root of the ratio of the galaxy and matter auto-correlation functions on large scales.}, and the stochasticity parameters $\alpha_0$ and $\alpha_1$. 
These two last parameters were introduced in the DSS to accurately model the galaxy-matter connection. However, the study did not introduce a physically motivated prior distribution of these parameters. For a more detailed explanation on the $\alpha$ parameters, including their physical meaning and origin, we refer to \cite{friedrich2021pdf}.
As a derived parameter, we also considered $S_8$\footnote{The parameter $S_8$ has been widely used because it removes degeneracies between $\Omega_m$ and $\sigma_8$ present in analysis of cosmic structure. Some analyses have found there to be a tension between the $S_8$ measured in the late universe (\citealt{planck_2018}, \citealt{Heymans_2021}, \citealt{Lemos_2021} and \citealt{dalal2023hyper}) and from the CMB (\citealt{Raveri_2019}, \citealt{Asgari_2021} and \citealt{Anchordoqui_2021}), rendering it of particular interest to have an interpretable posterior for.}: 
\begin{equation}
    S_8 = \sigma_8 \sqrt{\Omega_m/0.3} \ . 
\end{equation}

Both $\Omega_m$ and $S_8$ are the target parameters (as presented in, e.g., \citealt{krause2021}), while $\alpha_0$ and $\alpha_1$ are the nuisance parameters whose priors we are optimizing. 
The prior on $b$ remains unchanged, since we find its joint posterior with cosmological parameters to be symmetric and well constrained, at least in the case without modeling the excess skewness  of the matter density distribution $\Delta S_3$ as an additional free parameter, which is the only case we consider here.

The first step of the ITEM prior method is the use of different realizations of the nuisance. 
We limit our analysis by considering two realizations of the stochasticity listed in the second and third columns of Table \ref{tab:table_1}. 
The first one corresponds to a configuration in which there is no stochasticity in the distribution of galaxies, that is where the shot-noise of galaxies follows a Poisson distribution, which is often assumed to be correct at sufficiently large scales (but which has been shown to fail at exactly those scales, see e.g. \citealt{Friedrich2018, MacCrann_2018}).
That is equivalent to setting $\alpha_0 = 1.0$ and $\alpha_1 = 0.0$, since these two values turn Eq. \ref{eq:Poissonian} into a Poisson distribution. 
We call this the Non-stochasticity case.
The second corresponds to the best-fit shot-noise parameters found by \citet{Friedrich2018} on \textsc{redMaGiC} mock catalogs (\citealt{Rozo_2016}), constructed given realistic DES Y1-like survey simulations called Buzzard-v1.1 (\citealt{derose2019buzzard}). 
These Buzzard mock galaxy catalogs (\citealt{wechsler2021addgals}) have been used extensively in DES analyses (\citealt{2021arXiv210513547D}). 
The values for the stochasticity parameters are $\alpha_0 = 1.26$ and $\alpha_1 = 0.29$. 
We call this the Buzzard stochasticity case. 

These two realizations should be considered as a simple test case for the methodology.
This work, as mentioned previously, emphasizes that the resulting ITEM prior is limited because of the few, and potentially unrealistic, nuisance realizations considered. There is a wide range of plausible stochasticity configurations that could be taken into account (see e.g.\ the findings of \citealt{friedrich2021pdf}). 
In a companion paper (\citealt{britt2024bounds}) we consider diverse realizations of halo occupation distributions (HODs) to derive such a set of stochasticity configurations. 

\begin{figure}
        \includegraphics[width=0.9\columnwidth]{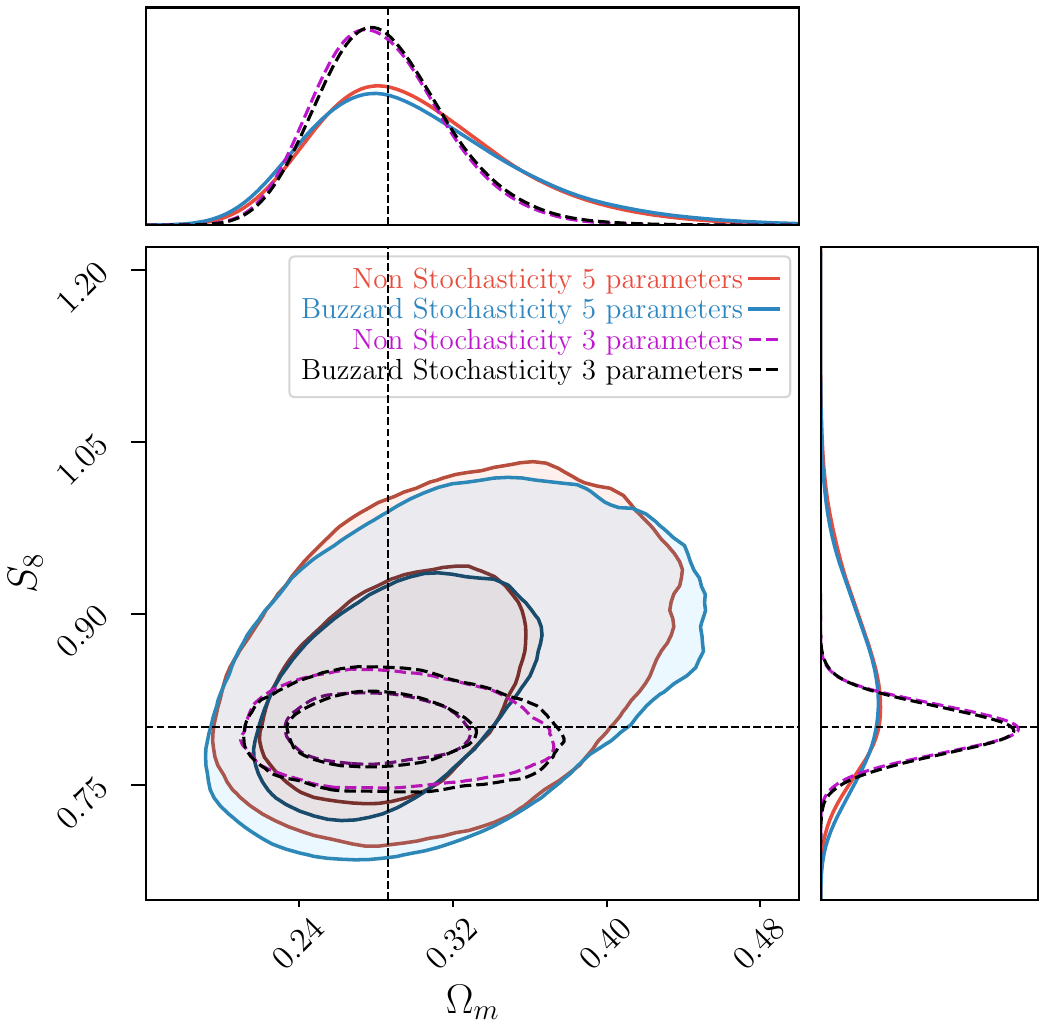}
    \caption{Marginalized ($\Omega_m, S_8$) posteriors with 1$\sigma$ and 2$\sigma$ contours from the simulated DSS $\alpha$ model of synthetic, noiseless data vectors using the fiducial values and priors from Table \ref{tab:table_1}. 
    The red and blue contours show the case with 5 degrees of freedom, while the purple and black contours show the case with 3 degrees of freedom, with $\alpha_0 = [1.0, 1.26]$ and $\alpha_1 = [0.0, 0.29]$ respectively.}
    \label{fi:5_vs_3_parameters}
\end{figure}

\begin{figure}
  \includegraphics[width=0.95\columnwidth]{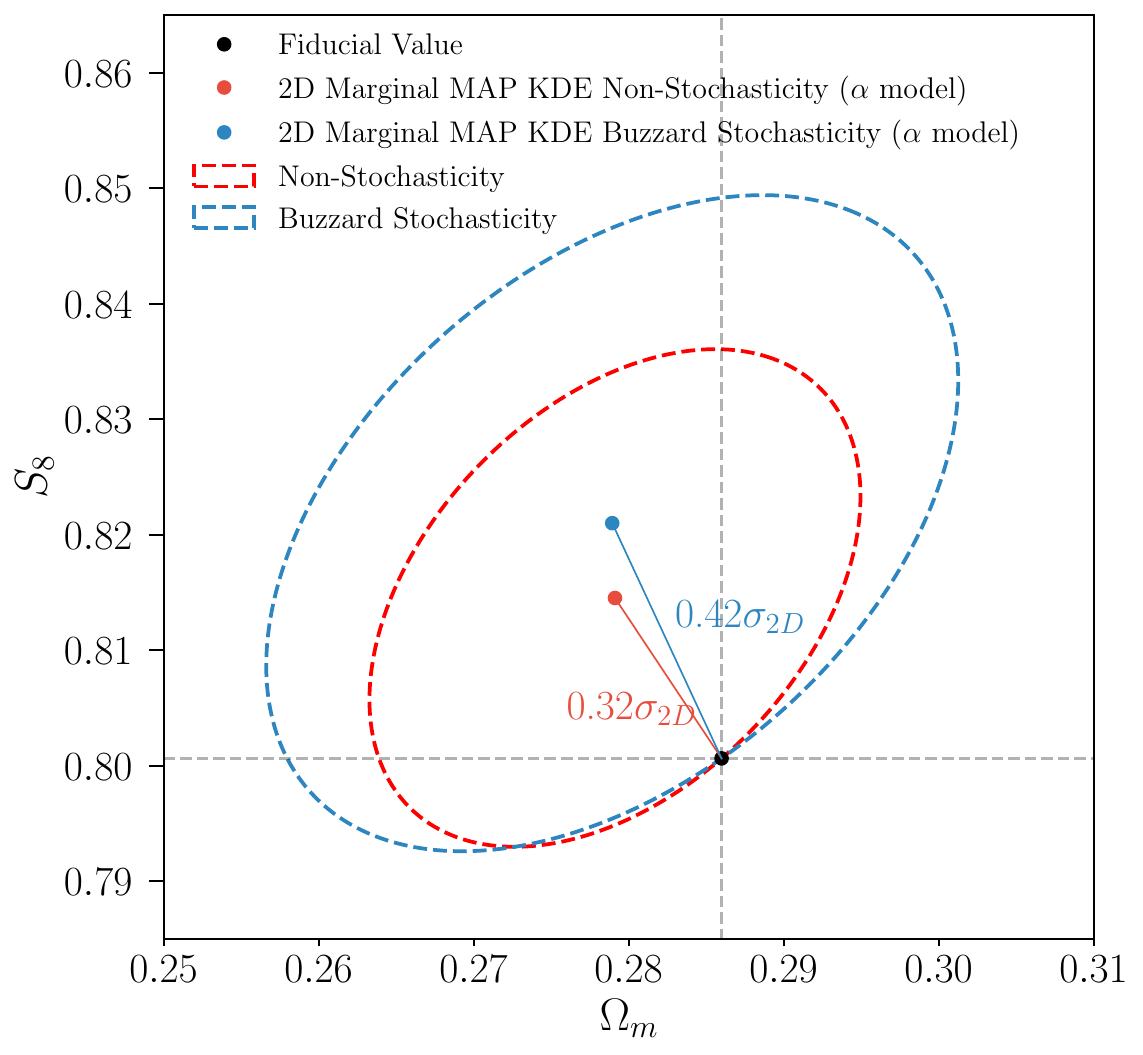}
  
  \includegraphics[width=0.95\columnwidth]{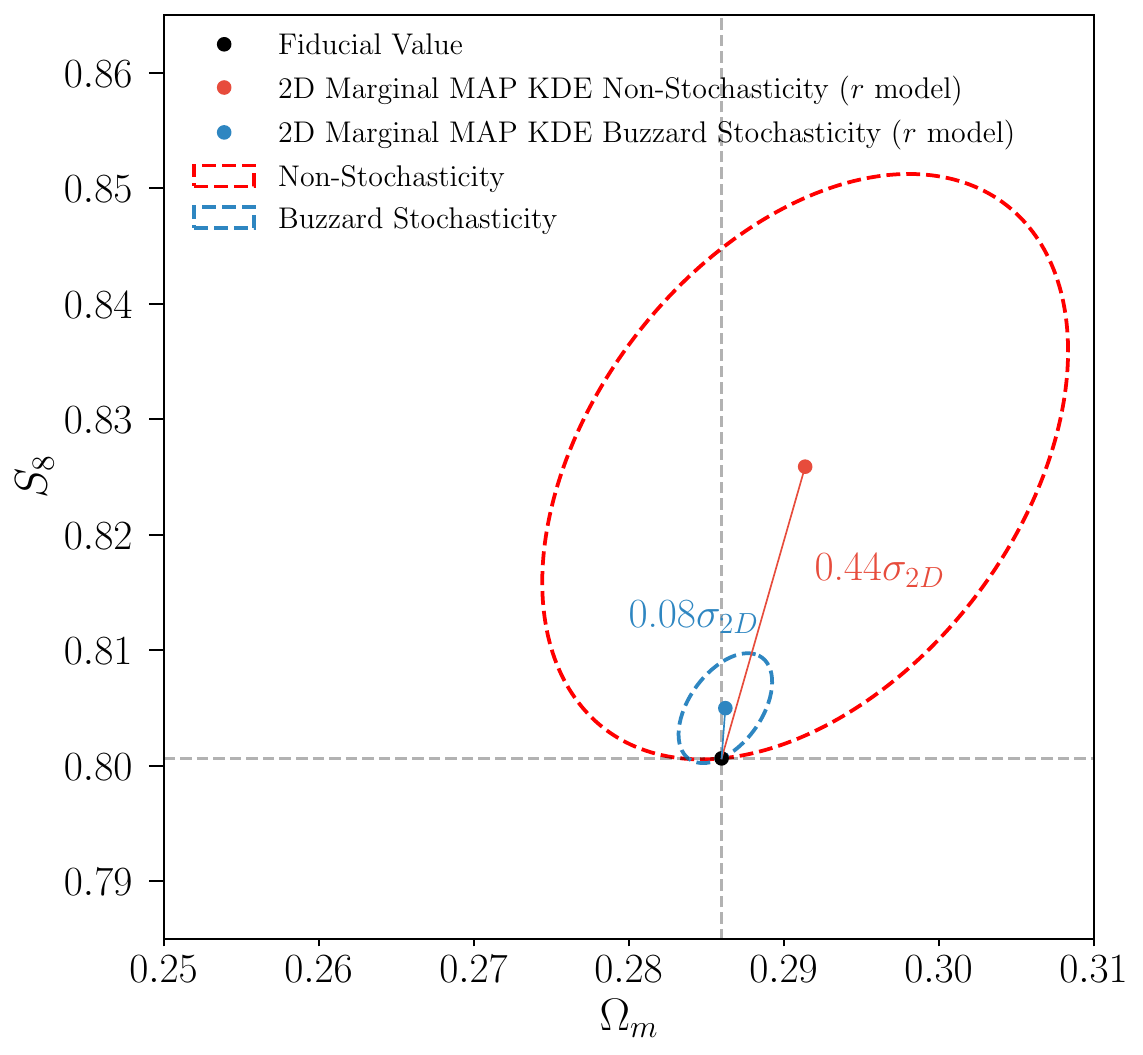}
   \caption{
    Parameter biases in simulated likelihood analyses for the $\alpha$ model (top) and the $r$ model (bottom). 
    In red and blue, we show the ellipses for the 2D marginalized constraints of the non-stochasticity and the Buzzard stochasticity configurations, respectively. 
    These figures follow Fig. 4 from \citet{krause2021}.
    For the $\alpha$ model, these are centered on their corresponding 2D marginal MAP. 
    Because of prior volume effects, the marginalized constraints are not centered on the input cosmology.
    For the $r$ model, there is a systematic bias in the non-stochasticity case that is explained by the sharp upper bound from the prior on $r$.}
  \label{fi:original_bias}
\end{figure}

\begin{figure}
  \includegraphics[width=1.0\columnwidth]{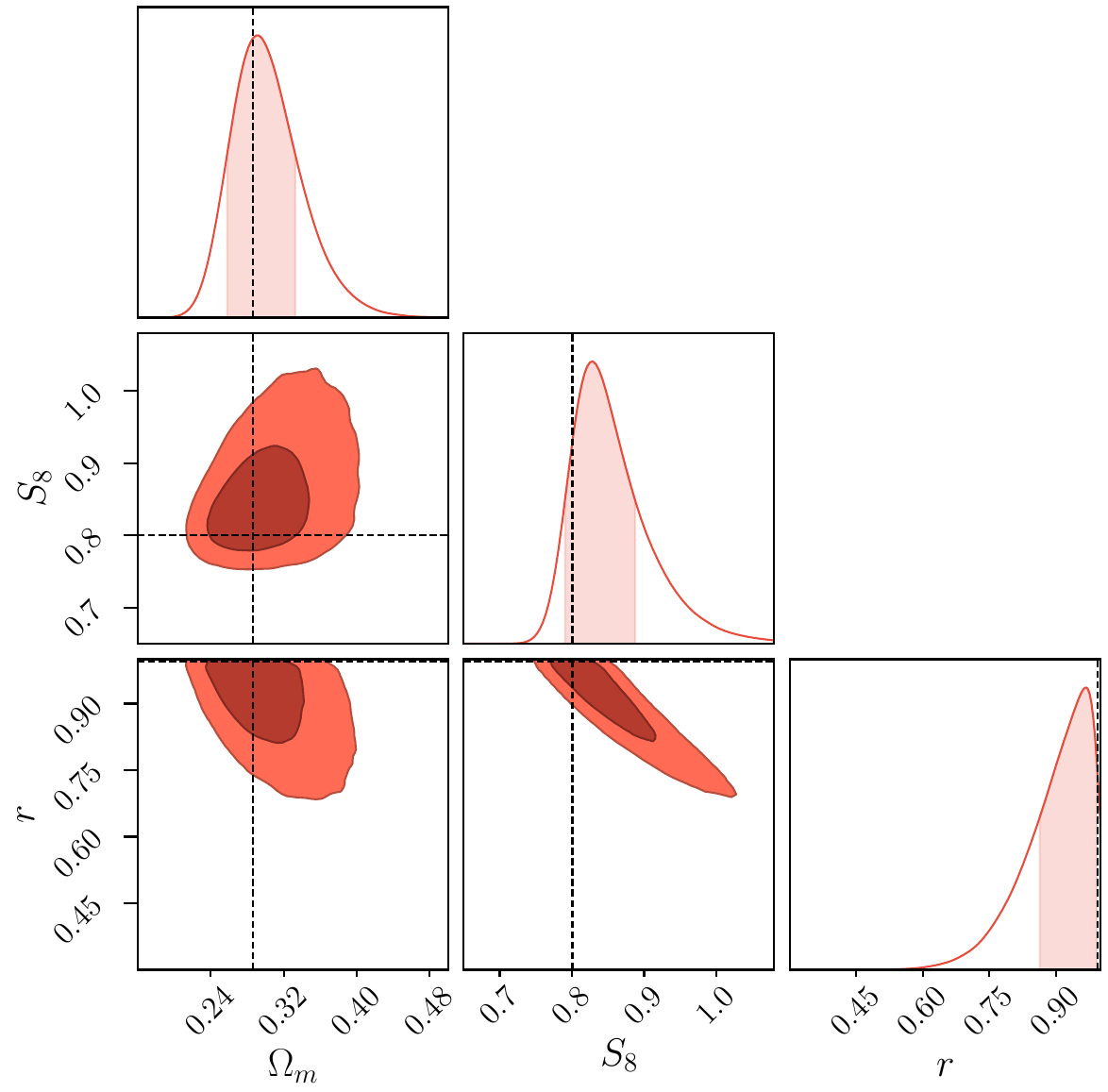}
  \centering
    \caption{
    Marginalized posteriors ($\Omega_m, S_8, r$) from the simulated DSS $r$ model using a non-stochastic parameter vector from the $\alpha$ model. 
    The fiducial values are represented by the dashed lines. Hence for $r_{\rm fiducial} = \max[r] = 1.0$, it is in the right edge of the plot.
    Because $r>1$ is unphysical, projection effects cause a systematic offset between the mean or peak of the 1D marginalized posteriors, particularly in $r$ and $S_8$, and their input values.
    }
    \label{fi:prior_vol_in_dss_2}
\end{figure}

\begin{figure}
\centering
    \includegraphics[width=0.85\columnwidth]{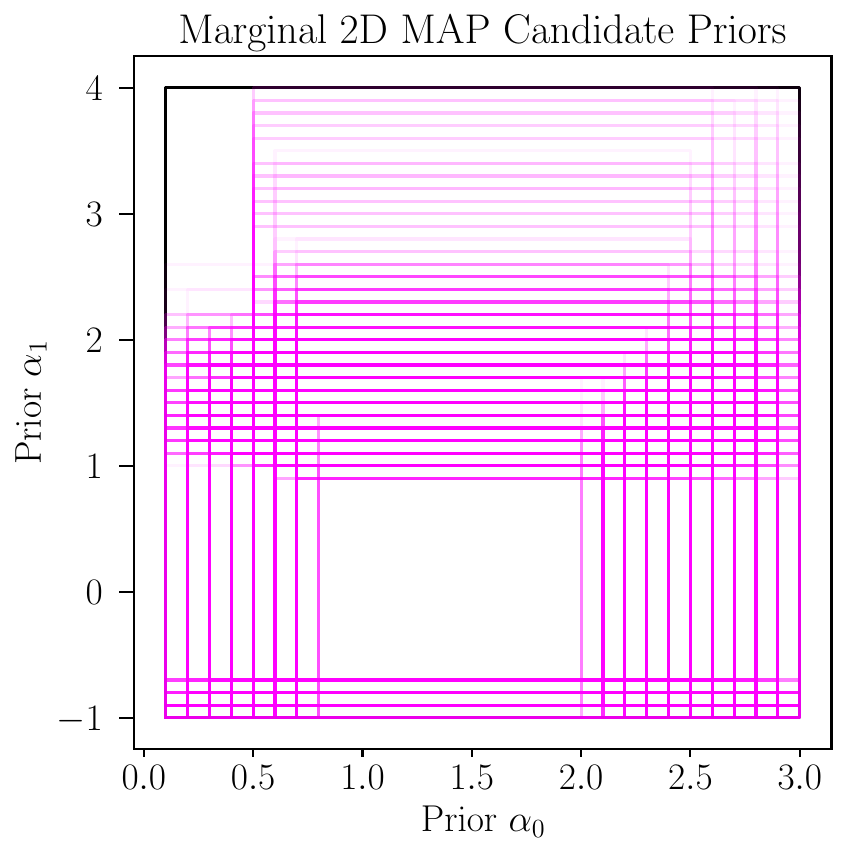}
    \caption{\label{t7}
    Distribution of the candidate priors for the $\alpha$ model that pass the first filter. 
    Each rectangle represents a combination of lower and upper bounds of $\alpha_0$ and $\alpha_1$.
    }
    \label{fi:A_2}
\end{figure}

For our simulated likelihood analysis, we obtain noiseless data vectors for each nuisance realization from the model prediction \citep{Friedrich2018} for the Buzzard-measured values of the stochasticity parameters.
Additionally, we used the priors presented in the last column of Table \ref{tab:table_1}, which are the ones used originally in the DES Y1 analysis of DSS (\citealt{Gruen2018}). 
To obtain the corresponding posterior, we run MCMCs using the {\scshape emcee} algorithm (\citealt{Foreman_Mackey_2013}), leaving the stochasticity parameters free with the original broad priors from the last column in Table \ref{tab:table_1}.
We point to \citet{survey2023des} for a discussion of the use of posterior samplers. 
We also run two additional MCMCs fixing the stochasticity parameters to the values listed in Table \ref{tab:table_1} for comparison purposes. 
In Fig. \ref{fi:5_vs_3_parameters}, we show the four estimated marginalized posteriors of $\Omega_m$ and $S_8$.
We find that in the cases in which we considered free stochasticity parameters, the posterior returns much larger uncertainties over both cosmological parameters, with a somewhat shifted mean. 

Subsequently, we fit a 2D Gaussian distribution centered at the 2D marginalized MAP and with the chains' parameter covariance matrix, and evaluate the off-centering of the chains from the true cosmological parameters the model was evaluated for when generating the data vector. 
Figure \ref{fi:original_bias} illustrates the biases for each chain obtained with the original wide priors on stochasticity parameters.
Both cases are biased: 
the bias of the Buzzard stochasticity case (in blue) reaches a value of $0.42 \sigma_{2D}$, more than the defined DES threshold of 0.3$\sigma_{2D}$.
The bias, in both cases, is systematically pointing towards lower and higher values on $\Omega_m$ and $S_8$, respectively. 

In an additional case study, we assume an alternative parametrization of galaxy stochasticity (the $r$ model, introduced in Appendix \ref{app:APP_2}) as a basis, yet still use the data vectors obtained previously with the $\alpha$ model.
The $r$ model is simpler than the $\alpha$ model because it only considers one correlation parameter $r$ for the effect of stochasticity. 
This test checks whether the conditions demanded by the ITEM formalism can be met even when a model does not produce the data, which happens often with real data, and how much information the ITEM prior can recover in such a case.

The original DSS prior used on $r$ follows a uniform distribution: $\Pi(r) \sim \mathcal{U}[0.0, 1.0]$, where $r = 1$ recovers the non-stochasticity case (see Appendix \ref{app:APP_2}).
The sharp upper bound results in a projection effect as seen in Fig. \ref{fi:prior_vol_in_dss_2}, with posteriors biased towards large values of $S_8$, particularly in the non-stochastic case.
We also illustrate the bias for both chains in the bottom panel of Fig. \ref{fi:original_bias} with the original wide prior on $r$.


\subsection{ITEM priors on the density split statistics}

In this subsection we detail the procedure to obtain ITEM priors for two nuisance models: 
first in Sect. \ref{sec:ITEM_alpha}, a simple approach considering the $\alpha$ model as a baseline, and second in Sect. \ref{sec:case_2}, a case when the data vectors are simulated with the $\alpha$ model, but analyzed with the $r$ model, leading to a systematic error.


\subsubsection{ITEM prior for the $\alpha$ model}
\label{sec:ITEM_alpha}

In the original DSS work, the nuisance parameters $\alpha_0$ and $\alpha_1$ had uniform priors. We made the same choice here and only varied the limits of these priors.\footnote{We note that one could also vary, e.g., the mean and width of a Gaussian prior.} 

We started by sampling the candidate priors as done in the example in Sect. \ref{sec:step_2}. 
We shrank the bounds of the priors (min$[\alpha_0]$, max$[\alpha_0]$, min$[\alpha_1]$ and max$[\alpha_1]$) with a step of 0.1 units per cut until they reached the closest rounded fiducial value of the parameter vector:
\begin{align*}
         & \min[\alpha_0] \in \{0.1, 0.2, ...,  0.9\} \\
         & \max[\alpha_0] \in \{1.4, 1.5, ..., 3.0\} \\
         &  \\
         & \min[\alpha_1] \in \{-1.0, -0.9, ..., -0.1\} \\
     & \max[\alpha_1] \in \{0.3, 0.4, ..., 4.0\} \ .
\end{align*}
The final number of candidate priors corresponds to 58,140 when combining each different bound.
Next, we apply an importance sampling to the original 5D wide posteriors described in Sect. \ref{sec:modeling_nuisance_DSS} and follow the three steps described in Sect. \ref{sec:methodology} to obtain the ITEM priors:

\begin{figure}
  \includegraphics[width=0.95\columnwidth]{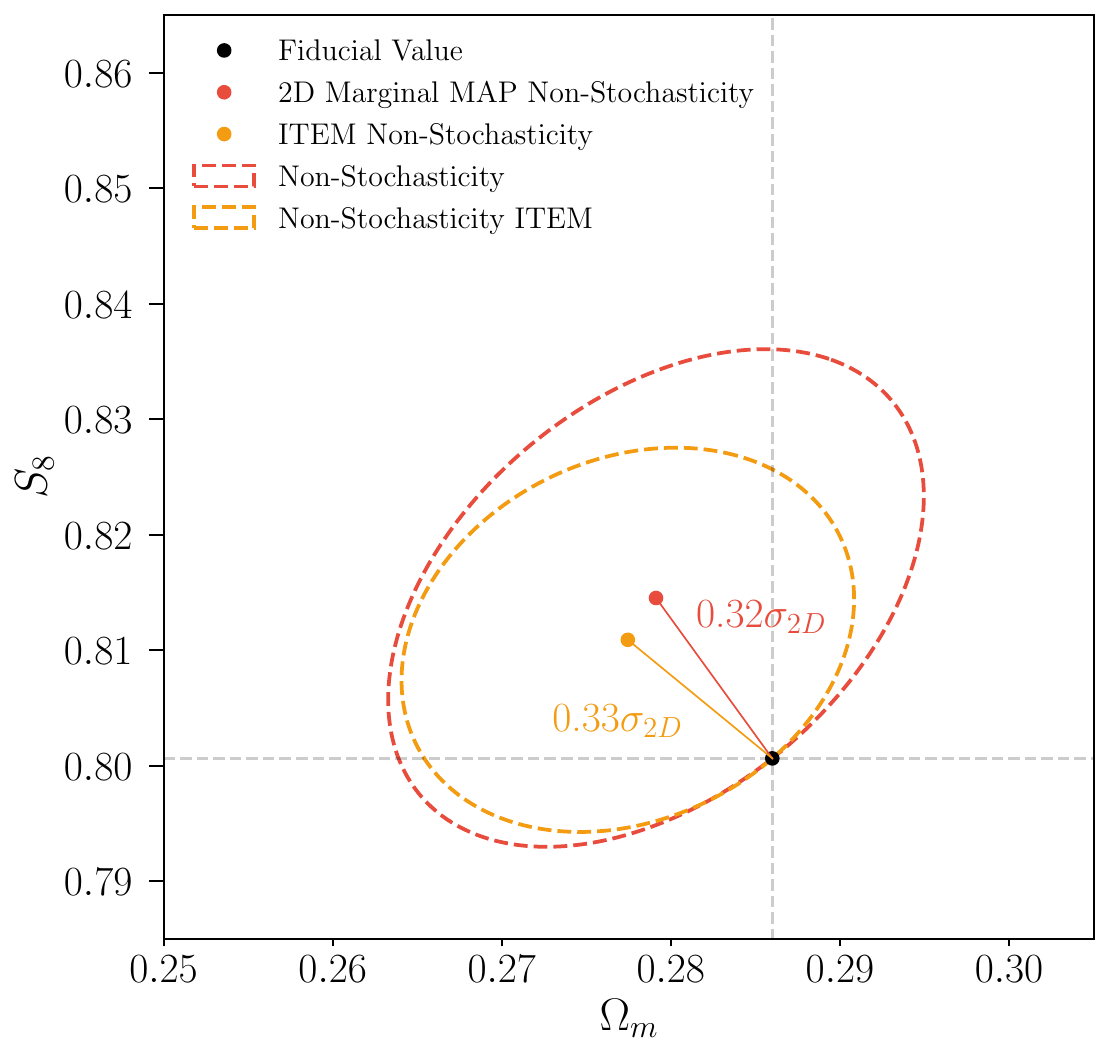}
  
  \includegraphics[width=0.95\columnwidth]{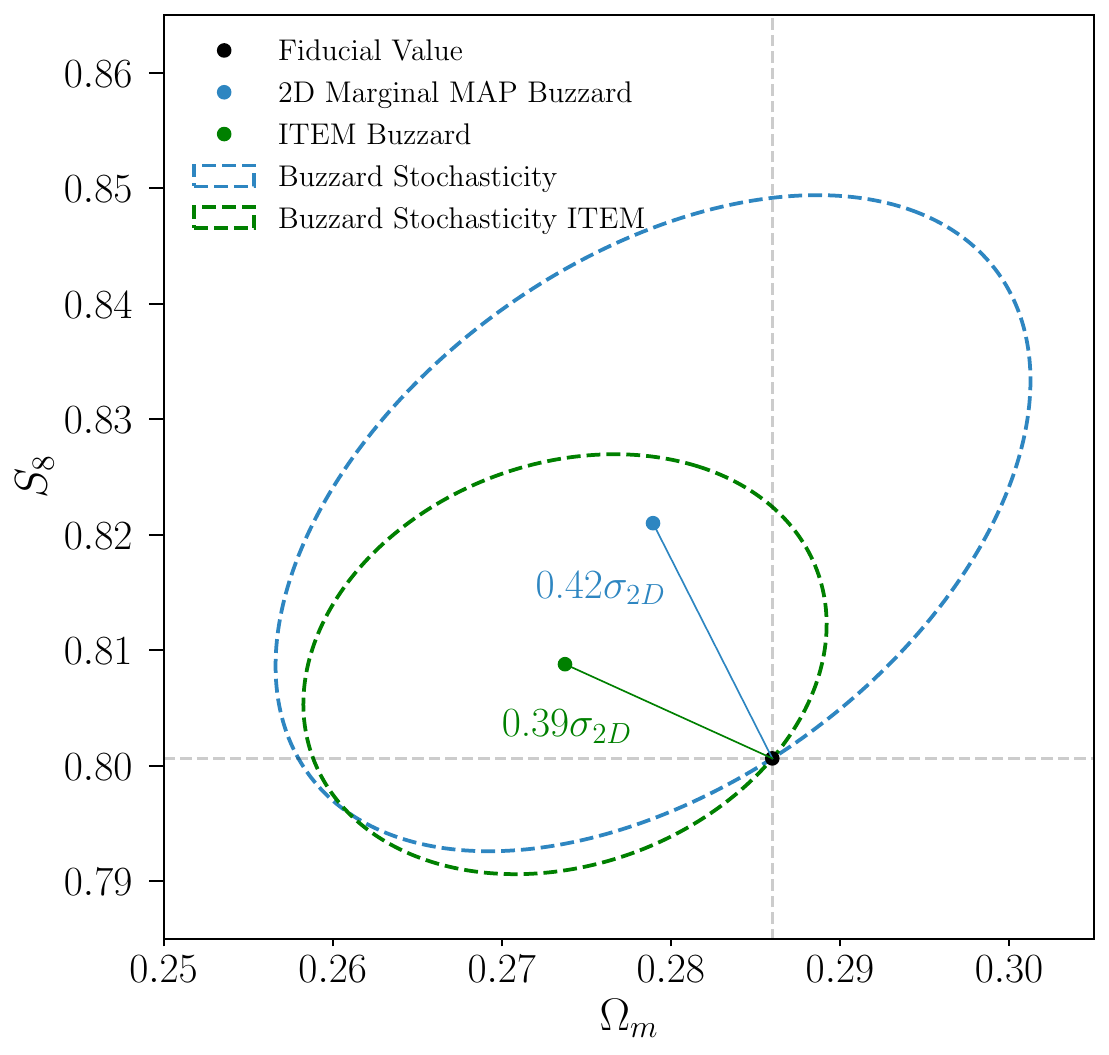}
   \caption{
   Parameter biases of the data vectors with the original and ITEM priors. 
   The red and blue ellipses show wide contours (same from Fig. \ref{fi:original_bias}), while the yellow and green ones were obtained after enforcing the $0.4 \sigma_{2D}$ bias threshold and 1D coverage threshold for the 2D marginalized constraints. 
   Both are centered in their respective 2D marginalized MAP. 
   Due to parameter volume effects, the marginalized constraints from the baseline prior analysis are not centered on the input cosmology. 
   \textit{Top:} Simulated likelihood analysis for the data vector without shot-noise ($\alpha_0 = 1.0$, $\alpha_1 = 0.0$). 
   \textit{Bottom:} Simulated likelihood analysis for the data vector with Buzzard stochasticity ($\alpha_0 = 1.26$, $\alpha_1 = 0.29$). 
   The dashed horizontal and vertical lines indicate the fiducial parameter values.}
  \label{fi:simulated_likelihood_analyses}
\end{figure}

\begin{table*}
    \caption{\label{t7}Ranges of the original and ITEM priors for the $\alpha$ model with a maximum bias of 0.4$\sigma_{2D}$ and a minimum coverage of 56\% for both $\Omega_m$ and $S_8$. 
    }
    \centering
        \label{tab:table_2}
    \begin{tabular}{lcccccccc}
        \hline
        Parameter & Prior & Prior width & Bias [$\sigma_{2D}$] & $\Omega_m$ coverage & $S_8$ coverage & $\overline{\sigma}_{\Omega_m}$  & $\overline{\sigma}_{S_8}$ & $\tilde{V}/\tilde{V}_{\rm Original}$ \\ 
        \hline
        \vspace{-0.2 cm} & & & & & & & \\
        \textbf{Original priors} & & & & $ $& & & \\
        $\alpha_0$ & $\mathcal{U}$[0.1, 3.0] & 2.9 & \Large{$\underset{0.32/0.42}{}$} & \Large{$\underset{76 \% / 76 \%}{}$} &  \Large{$\underset{72 \% / 52 \%}{}$} &  \Large{$\underset{0.052}{}$} & \Large{$\underset{0.068}{}$} & \Large{$\underset{1.00}{}$} \\
        $\alpha_1$ & $\mathcal{U}$[-1.0, 4.0] & 5.0 & & & & & \\
        & & & & & & & \\
        \textbf{ITEM prior} & & & & & & & & \\
        $\alpha_0$ & $\mathcal{U}$[0.6, 2.0] & 1.4 & 
        \Large{$\underset{0.33/0.39}{}$}
        & 
        \Large{$\underset{80 \% / 64 \%}{}$}
        &  
        \Large{$\underset{68 \% / 56 \%}{}$}
        & 
        \Large{$\underset{0.040}{}$}
        & 
        \Large{$\underset{0.049}{}$}
        & 
        \Large{$\underset{0.58}{}$} \\
        \vspace{-0.2 cm}
        $\alpha_1$ & $\mathcal{U}$[-1.0, 1.1] & 2.1 & & & & \\  
        & & & & & & & \\ 
        \hline
    \end{tabular}
    \tablefoot{The bias and error coverage quantities from both nuisance realizations are illustrated for each prior as non-stochasticity/Buzzard stochasticity cases respectively. 
    In the final three columns, we include the average error of the final estimate for $\Omega_m$ and $S_8$ from both nuisance realizations and the 2D approximated, normalized and re-scaled volume of the 2D marginalized posterior $\tilde{V}$ defined in Sect. \ref{sec:crit_3}.}
\end{table*}

\begin{table*}
        \centering
    \caption{Ranges of the original and ITEM priors for the $r$ model with a maximum bias of 0.5$\sigma_{2D}$ and a minimum coverage of 56\% for both $\Omega_m$ and $S_8$.
    }
        \label{tab:table_3}
        \begin{tabular}{lcccccccc}
                \hline
            Parameter & Prior & Prior width & Bias [$\sigma_{2D}$] & $\Omega_m$ coverage & $S_8$ coverage & $\overline{\sigma}_{\Omega_m}$  & $\overline{\sigma}_{S_8}$  & $\tilde{V}/\tilde{V}_{\rm Original}$ \\ 
                \hline
            \vspace{-0.2 cm}& & & & & & & \\
                \textbf{Original prior}  & & & & & & & \\
                $r$ & $\mathcal{U}$[0.00, 1.00] & 1.00 & 0.44/0.08 & $84\% / 76\%$ & $56\% / 56\%$ & 0.039 & 0.059 & 1.00 \\
            & & & & & & & \\

                \textbf{ITEM prior} & & & & & & & \\
                $r$ & $\mathcal{U}$[0.68, 1.00] & 0.32 & 0.50/0.09 & $84\% / 76\%$ & $56\% / 56\%$ & 0.037 & 0.053 & 0.91 \\
            \vspace{-0.2 cm} & & & & & & & \\
                \hline
        \end{tabular}
    \tablefoot{As in Table \ref{tab:table_2}, the bias and coverage quantities from both nuisance realizations are separated with a backslash.}
\end{table*}

\begin{figure}
  \includegraphics[width=0.95\columnwidth]{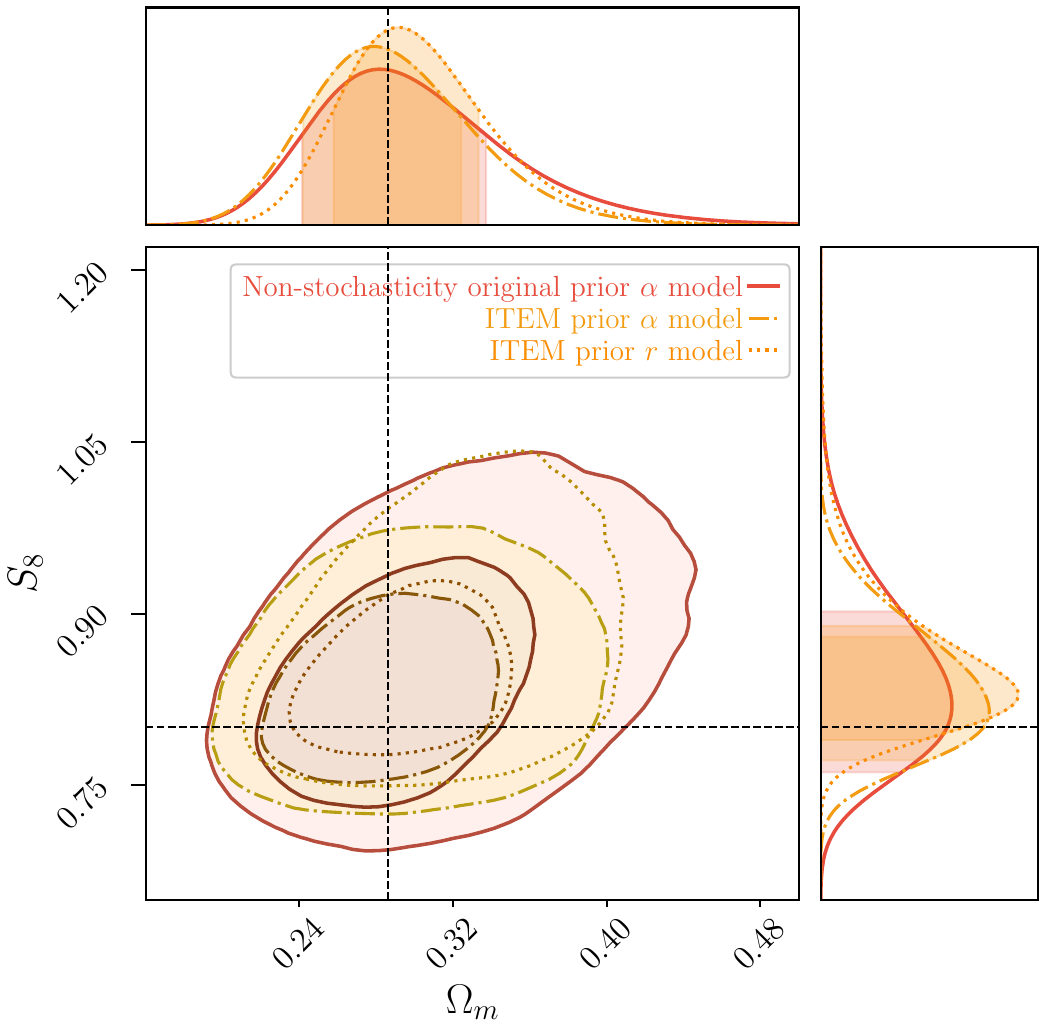}
  
  \includegraphics[width=0.95\columnwidth]{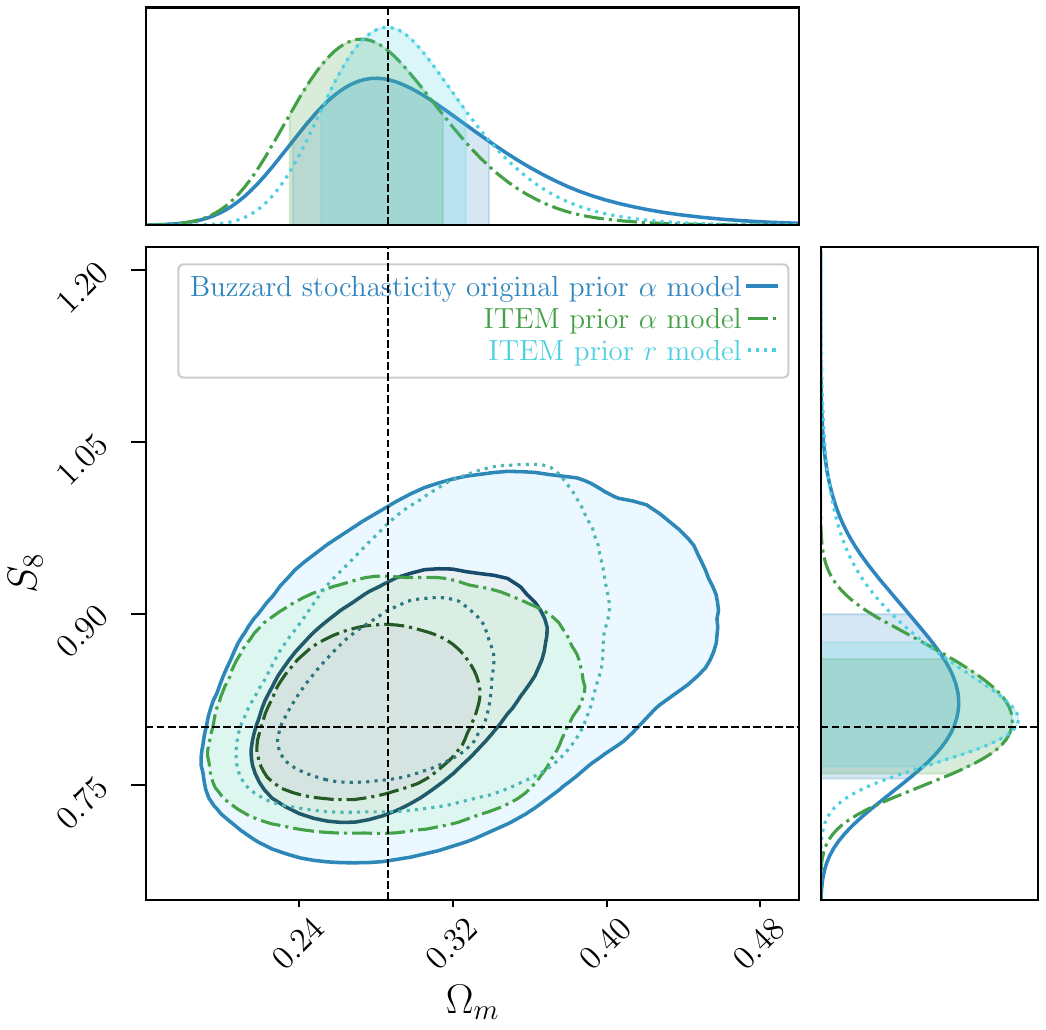}
  \caption{
  Marginalized posterior distributions for the chains from the $\alpha$ and $r$ models from DSS.
   These have been plotted with the widened $1$$\sigma$ and $2$$\sigma$ confidence levels to account for the bias as presented in Sect. \ref{sec:crit_2}. 
   In the top and bottom panels, we illustrate the non-stochasticity and the Buzzard stochasticity respectively. 
   The larger contours correspond to the posterior simulated using the original DSS $\alpha$ model priors. 
   The narrow continuous contours illustrate the one with the ITEM prior on the $\alpha$ model, and the discontinuous contours for the $r$ model.
  }
  \label{fi:ITEM_marginalized_posteriors}
\end{figure}

First, we estimate, for each candidate prior, the maximum Mahalanobis distance $x$ from Eq. \ref{eq:x_thershold} for the different posteriors. 
We set the upper threshold to 0.4$\sigma_{2D}$ for the first filter of the ITEM priors. 
A total of 1,980 priors had a bias of less than $0.4 \sigma_{2D}$ ($\sim$3$\%$ of the original set, indicating that prior volume related biases are prevalent in this inference problem). 
In Fig. \ref{fi:A_2}, we show their distribution on the prior space. 
These mainly preferred lower values for the upper prior bound on $\alpha_1$, indicating that the original asymmetric prior caused a prior volume related bias.

Second, we run 50 noisy simulated realizations, 25 for each nuisance realization, following the criterion from Sect. \ref{sec:crit_2}.
These noisy realizations were simulated by adding multivariate Gaussian noise with the covariance matrix from \citet{Gruen2018} to the noiseless data vectors. 
We choose the threshold $z = 14$ such that the cumulative binomial distribution for independent 25 realizations is at least 90$\%$ (this corresponds to a 14/25 = 56$\%$ coverage). 
Only a small fraction of priors should thus be excluded due to chance fluctuations in the number of contours that contain the input parameter values. 
In this step, we considered the systematic bias as explained in Sect. \ref{sec:crit_2}.
A total of 1,868 out of the 1,980 remaining candidate priors qualified under the 1D coverage requirement for both nuisance realizations.
In Fig. \ref{fi:coverages_alpha} we show the distribution of the coverages among the tested priors. 

The final step corresponds to the minimization of the error from Sect. \ref{sec:crit_3}.
The resulting ITEM priors for $\alpha_0$ and $\alpha_1$ are listed in the second row of Table \ref{tab:table_2}.
We first note that the prior widths were reduced to $\sim$48$\%$ and $42\%$ their original ones respectively.
In particular, the prior in $\alpha_0$ remained centered around 1.3, while all upper bounds such that $\max[\alpha_1]>1.1$ were rejected.
The biases obtained with the original and ITEM priors are displayed in Fig. \ref{fi:simulated_likelihood_analyses}.
While the bias in $\Omega_m$ slightly increased, the one on $S_8$ decreased as expected from the imposed threshold.
The coverages for both nuisance realizations did not change more than $\pm 12\%$ with respect to the coverages from the original priors. 
In the last three columns of Table \ref{tab:table_2}, we present the average standard deviations and the statistical uncertainty of the target parameters.
These were reduced, corresponding to an improvement that is formally equivalent to an experiment surveying more than three times the volume. 
Finally, the marginalized distributions are shown in Fig. \ref{fi:ITEM_marginalized_posteriors}. 

We note that in a proper implementation of the ITEM prior program one should consider an exhaustive list of possible nuisance realizations,
instead of just the two realizations considered in our example.
In the companion study \citealt{britt2024bounds}, we include shot-noise realizations generated from a range of plausible HOD descriptions to obtain fiducial stochasticity parameters and derive a more realistic result than the one presented here. 
The nuisance realizations derived there will be used in a more thorough application of the ITEM prior method in the future.

Due to recent discussions on the impact of using different point estimators in cosmological analyses (e.g. \citealt{Amon_2022}), in Appendix \ref{app:APP_3} we explore the ITEM priors obtained when, instead of using the marginalized MAP, we use the MAP and means as point estimators. 


\subsubsection{ITEM prior for $r$ model with a different basis model}
\label{sec:case_2}

In this subsection we investigate how the ITEM priors can help in cases where the assumed nuisance models are not the generators of the data. 
We use the same two-parameter vectors from Sect. \ref{sec:ITEM_alpha} (the non-stochasticity and the Buzzard stochasticity) and follow the same procedure. 
However, we use the $r$ model presented in Appendix \ref{app:APP_r_model} to obtain posteriors. 
The parameters considered are $\Omega_m, \sigma_8, b$ and $r$. 
We use the same prior distribution over $\Omega_m, \sigma_8$ and $b$, and change the prior on $r$.
We then obtain two posteriors using the non-stochasticity and the Buzzard realization of the nuisance in the simulated data.
Next, we sample different priors for $r$: we changed the lower bound min$[r]$ and kept max$[r]$ fixed to 1.0:
\begin{equation*}
    \text{min}[r] \in \{0.00, 0.01, ..., 0.99\}\ .
\end{equation*}
The total number of priors is 99. 
Then, we repeat the three steps of the ITEM prior pipeline.

\begin{figure}
  \includegraphics[width=\columnwidth]{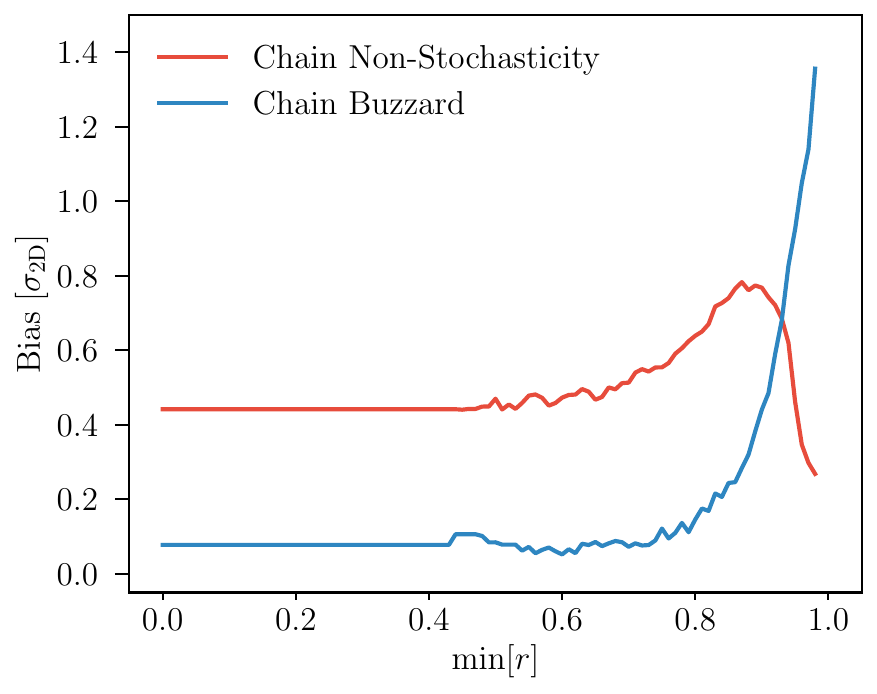}
  \centering
    \caption{Biases of the non-stochasticity and Buzzard chains for different lower bounds of $\Pi(r)$.
    After a min[$r$] $\sim$ 0.9, we see a trade-off between the two stochasticity cases.
    }
    \label{fi:biases_on_r}
\end{figure}

In this case there is a larger systematic bias due to the sharp upper bound on $r$ and the nuisance realization from a different model.
In Fig. \ref{fi:biases_on_r} we show the resulting bias for different lower bounds on min[$r$].
Because of the systematic error present in the data vector, there is a clear trade-off.
On one side, for the non-stochasticity case (red line in Fig. \ref{fi:biases_on_r}), priors with a higher lower bound on $r$ (from $\min[r]$ = 0.9 and above) will be substantially less biased.
This is expected as priors with higher $\min[r]$ will exclusively select values close to $r=1$, near the fiducial values of the parameter vector (we recall that $\alpha_0=1.0$, $\alpha_1=0.0$ is algebraically equivalent to $r=1$ as shown in Appendix \ref{app:APP_2}).
On the other side, the Buzzard stochasticity case (blue line in Fig. \ref{fi:biases_on_r}) becomes highly biased for $\min[r]>0.9$.
Those priors on $r$ restrict the parameter space substantially to a ``more'' Poissonian configuration of the galaxy matter connection (see Appendix \ref{app:APP_2}), increasing their bias drastically.
To account for the systematic bias, we choose a flexible bias threshold of $0.5 \sigma_{2D}$, which reduces the number of candidate priors to 68. 

Subsequently, and as we did in Sect. \ref{sec:ITEM_alpha}, we run 50 noisy simulated realizations, 25 for each nuisance realization using the same data vectors.
We require the noisy realizations to have at least a 56$\%$ 1D coverage for both $\Omega_m$ and $S_8$ out of the 25 simulated realizations.
This second filter did not reduce the number of candidate priors because all of them fulfilled the requirement. 

Finally, we determine the ITEM prior by optimizing the total error from Sect. \ref{sec:crit_3}. 
In Table \ref{tab:table_3}, we summarize the resulting prior, with the corresponding biases, coverage, and uncertainties. 
We find similar statistics, but we used a prior that is substantially less wide (reduction of 68$\%$).
For comparison purposes, in Fig. \ref{fi:ITEM_marginalized_posteriors} we show the marginalized posterior distributions.


\section{Discussion and conclusion}
\label{sec:d_c}

In the era of precision cosmology, observations require increasingly complex models that account for systematic astrophysical and observational effects.
Prior distributions are assigned to the nuisance parameters modeling these phenomena based on limits in which they are plausible, ranges of simulation results, or actual calibration uncertainty. However, the analyses using such priors often yield biased or overly uncertain marginalized constraints on cosmological parameters due to so-called projection or prior volume effects.
In this paper, we have presented a procedure to overcome projection effects through a pragmatic choice of nuisance parameter priors based on a series of simulated likelihood analyses. 
We named this the ITEM priors method.

The ITEM approach performs these simulated likelihood tests in order to find a prior that, when applied, leads to a posterior with some desired statistical properties and interpretability. 
We proposed to divide the parameters of a given model into two types: the target parameters (the ones of interest) and the nuisance parameters (the ones to optimize the prior for).
Our approach then finds an equilibrium between the underfitting and overly free treatment of nuisance effects.
This is done by enforcing three requirements on the target parameter posteriors found when applying the nuisance parameter prior in question in the analysis:
\begin{itemize}
    \item The biases on the target parameters need to be lower than a certain fraction of their statistical uncertainties.
    \item In the presence of noise, we still need to cover the original fiducial parameter values a certain fraction of times in some desired confidence region.
    \item With these two criteria being met, we prefer nuisance parameter priors that lead to small target parameter posterior volumes.
\end{itemize}

We tested the ITEM prior pipeline using two different nuisance realizations from the density split statistics analysis framework (\citealt{Friedrich2018, Gruen2018}).
We simulated two realistic data vectors, where one 
does not include stochasticity between galaxy counts and the underlying matter density contrast, and
the other considers a specific stochasticity found in an N-body simulation based on a mock catalog. 
For both cases, we applied the above criteria to a list of candidate priors in order to limit bias, ensure coverage, and minimize uncertainties. 
The result from applying the ITEM prior to the constraining power on target parameters corresponds to an improvement equivalent to an experiment surveying more than three times the cosmic volume.
Additionally, we studied the case in which the model used to treat the nuisance effect is not the same as the one used to generate the data, a common situation in practice.
The range of the resulting ITEM prior is less than half of the original prior's size, resulting in a smaller total uncertainty of the target parameters with a similar level of bias and coverage as obtained with the original prior.

One of the main limitations of this method is that its results may depend on the fiducial values that are assigned to the parameter of interest and that it formally requires a complete set of plausible realizations of the nuisance effect.
In this first study, we fixed the target parameters and chose just two nuisance realizations; however, one can extend our method to explore a variety of target and nuisance realizations in future works that apply it to actual inference.

Some potential extensions to our work include incorporating Bayesian evidence to compare and/or select priors in the different ITEM steps. 
Modern sampling techniques, such as the nested sampling algorithm (\citealt{Skilling2004}) and its extensions (\citealt{Chopin_2010}, \citealt{Brewer2010}, \citealt{Habeck2015}), provide the Bayes factor as part of their analysis pipelines. 
Well-known samplers, such as MultiNest (\citealt{Feroz_2009}) and more recently dynesty (\citealt{Speagle_2020}) and Nautilus (\citealt{Lange2023}), have become popular tools within the cosmology community. 
We anticipate their usage will continue to grow, and therefore implementing the ITEM procedure should not incur more computational cost than what is usually invested.

To summarize, in this work, we aimed at exploring an alternative approach to assigning prior distributions to nuisance parameters.
The method we have presented is flexible, as it allows us to define requirements and propose priors, given some set of plausible nuisance realizations, for a reliable and information-maximizing statistical inference.

\begin{acknowledgements}

We would like to thank Elisabeth Krause, Steffen Hagstotz, Raffaella Capasso, Elisa Legnani, and members of the Astrophysics, Cosmology, and Artificial Intelligence (ACAI) group at LMU Munich for helpful discussions and feedback. \\

BRG was funded by the Chilean National Agency for Research and Development (ANID) - Subdirección de Capital Humano / Magíster Nacional / 2021 - ID 22210491, the German Academic Exchange Service (DAAD, Short-Term Research Grant 2021 No. 57552337) and by the U.S. Department of Energy through grant DE-SC0013718 and under DE-AC02-76SF00515 to SLAC National Accelerator Laboratory, and by the Kavli Institute for Particle Astrophysics and Cosmology (KIPAC).
BRG also gratefully acknowledges support from the Program for Astrophysics Visitor Exchange at Stanford (PAVES). 
DB acknowledges support provided by the KIPAC Giddings Fellowship, the Graduate Research \& Internship Program in Germany (operated by The Europe Center at Stanford), the German Academic Exchange Service (DAAD, Short-Term Research Grant 2023 No. 57681230), and the Bavaria California Technology Center (BaCaTec).
DG and OF were supported by the Excellence Cluster ORIGINS, which is funded by the Deutsche Forschungsgemeinschaft (DFG, German Research Foundation) under Germany's Excellence Strategy (EXC 2094-390783311). 
OF also gratefully acknowledges support through a Fraunhofer-Schwarzschild-Fellowship at Universitätssternwarte München (LMU observatory). \\

This work used the following packages: {\scshape matplotlib} (\citealt{matpltlib}), {\scshape ChainConsumer} (\citealt{chainconsumer}), {\scshape numpy} (\citealt{Harris_2020}), {\scshape reback2020pandas} (\citealt{reback2020pandas}), {\scshape scipy} (\citealt{2020SciPy}), {\scshape incredible} and {\scshape emcee} (\citealt{Foreman_Mackey_2013}) software packages.

\end{acknowledgements}

\section*{Data availability}

The Dark Energy Survey Year 1 data (Density Split Statistics) used in this article is publicly available at \url{https://des.ncsa.illinois.edu/releases/y1a1/density}.


\bibliographystyle{aa}
\bibliography{refs}

\begin{thebibliography}{104}
\expandafter\ifx\csname natexlab\endcsname\relax\def\natexlab#1{#1}\fi

\bibitem[{Abbott {et~al.}(2022)Abbott, Aguena, Alarcon, Allam, Alves, Amon,
  Andrade-Oliveira, Annis, Avila, Bacon, Baxter, Bechtol, Becker, Bernstein,
  Bhargava, Birrer, Blazek, Brandao-Souza, Bridle, Brooks, Buckley-Geer, Burke,
  Camacho, Campos, Rosell, Kind, Carretero, Castander, Cawthon, Chang, Chen,
  Chen, Choi, Conselice, Cordero, Costanzi, Crocce, da~Costa, da~Silva~Pereira,
  Davis, Davis, Vicente, DeRose, Desai, Valentino, Diehl, Dietrich, Dodelson,
  Doel, Doux, Drlica-Wagner, Eckert, Eifler, Elsner, Elvin-Poole, Everett,
  Evrard, Fang, Farahi, Fernandez, Ferrero, Fert{\'{e}}, Fosalba, Friedrich,
  Frieman, Garc{\'{\i}}a-Bellido, Gatti, Gaztanaga, Gerdes, Giannantonio,
  Giannini, Gruen, Gruendl, Gschwend, Gutierrez, Harrison, Hartley, Herner,
  Hinton, Hollowood, Honscheid, Hoyle, Huff, Huterer, Jain, James, Jarvis,
  Jeffrey, Jeltema, Kovacs, Krause, Kron, Kuehn, Kuropatkin, Lahav, Leget,
  Lemos, Liddle, Lidman, Lima, Lin, MacCrann, Maia, Marshall, Martini,
  McCullough, Melchior, Mena-Fern{\'{a}}ndez, Menanteau, Miquel, Mohr, Morgan,
  Muir, Myles, Nadathur, Navarro-Alsina, Nichol, Ogando, Omori, Palmese,
  Pandey, Park, Paz-Chinch{\'{o}}n, Petravick, Pieres, Malag{\'{o}}n, Porredon,
  Prat, Raveri, Rodriguez-Monroy, Rollins, Romer, Roodman, Rosenfeld, Ross,
  Rykoff, Samuroff, S{\'{a}}nchez, Sanchez, Sanchez, Cid, Scarpine, Schubnell,
  Scolnic, Secco, Serrano, Sevilla-Noarbe, Sheldon, Shin, Smith, Soares-Santos,
  Suchyta, Swanson, Tabbutt, Tarle, Thomas, To, Troja, Troxel, Tucker,
  Tutusaus, Varga, Walker, Weaverdyck, Wechsler, Weller, Yanny, Yin, Zhang, \&
  and}]{Abbott_2022}
Abbott, T., Aguena, M., Alarcon, A., {et~al.} 2022, PRD, 105

\bibitem[{Abbott {et~al.}(2023)Abbott, Aguena, Alarcon, Alves, Amon,
  Andrade-Oliveira, Asgari, Avila, Bacon, Bechtol, Becker, Bernstein, Bertin,
  Bilicki, Blazek, Bocquet, Brooks, Burger, Burke, Camacho, Campos, Rosell,
  Kind, Carretero, Castander, Cawthon, Chang, Chen, Choi, Conselice, Cordero,
  da~Costa, Pereira, Dalal, Davis, deJong, DeRose, Desai, Diehl, Dodelson,
  Doel, Doux, Drlica-Wagner, Dvornik, Eckert, Eifler, Elvin-Poole, Everett,
  Fang, Ferrero, Ferté, Flaugher, Friedrich, Frieman, García-Bellido, Gatti,
  Giannini, Giblin, Gruen, Gruendl, Gutierrez, Harrison, Hartley, Herner,
  Heymans, Hildebrandt, Hinton, Hoekstra, Hollowood, Honscheid, Huang, Huff,
  Huterer, James, Jarvis, Jeffrey, Jeltema, Joachimi, Joudaki, Kannawadi,
  Krause, Kuehn, Kuijken, Kuropatkin, Leget, Lemos, Li, Li, Liddle, Lima, Lin,
  Lin, MacCrann, Mahony, Marshall, McCullough, Mena-Fernández, Menanteau,
  Miquel, Mohr, Muir, Myles, Napolitano, Navarro-Alsina, Ogando, Palmese,
  Pandey, Park, Paterno, Peacock, Petravick, Pieres, Malagón, Porredon, Prat,
  Radovich, Raveri, Reischke, Rollins, Romer, Roodman, Rykoff, Samuroff,
  Sánchez, Sanchez, Sanchez, Schneider, Secco, Sevilla-Noarbe, Shan, Sheldon,
  Shin, Sifón, Smith, Soares-Santos, Stölzner, Suchyta, Swanson, Tarle,
  Thomas, To, Troxel, Tröster, Tutusaus, van~den Busch, Varga, Walker,
  Weaverdyck, Wechsler, Weller, Wiseman, Wright, Yanny, Yin, Yoon, Zhang, \&
  Zuntz}]{survey2023des}
Abbott, T. M.~C., Aguena, M., Alarcon, A., {et~al.} 2023, DES Y3 + KiDS-1000:
  Consistent cosmology combining cosmic shear surveys

\bibitem[{Aghanim {et~al.}(2020)Aghanim, Akrami, Ashdown, Aumont, Baccigalupi,
  Ballardini, Banday, Barreiro, Bartolo, \& et~al.}]{planck_2018}
Aghanim, N., Akrami, Y., Ashdown, M., {et~al.} 2020, A\&A, 641, A6

\bibitem[{Aihara {et~al.}(2017)Aihara, Arimoto, Armstrong, Arnouts, Bahcall,
  Bickerton, Bosch, Bundy, Capak, Chan, \& et~al.}]{Aihara_2017}
Aihara, H., Arimoto, N., Armstrong, R., {et~al.} 2017, PASJ, 70

\bibitem[{Amon {et~al.}(2022)Amon, Gruen, Troxel, MacCrann, Dodelson, Choi,
  Doux, Secco, Samuroff, Krause, Cordero, Myles, DeRose, Wechsler, Gatti,
  Navarro-Alsina, Bernstein, Jain, Blazek, Alarcon, Fert{\'{e}}, Lemos, Raveri,
  Campos, Prat, S{\'{a}}nchez, Jarvis, Alves, Andrade-Oliveira, Baxter,
  Bechtol, Becker, Bridle, Camacho, Rosell, Kind, Cawthon, Chang, Chen,
  Chintalapati, Crocce, Davis, Diehl, Drlica-Wagner, Eckert, Eifler,
  Elvin-Poole, Everett, Fang, Fosalba, Friedrich, Gaztanaga, Giannini, Gruendl,
  Harrison, Hartley, Herner, Huang, Huff, Huterer, Kuropatkin, Leget, Liddle,
  McCullough, Muir, Pandey, Park, Porredon, Refregier, Rollins, Roodman,
  Rosenfeld, Ross, Rykoff, Sanchez, Sevilla-Noarbe, Sheldon, Shin, Troja,
  Tutusaus, Tutusaus, Varga, Weaverdyck, Yanny, Yin, Zhang, Zuntz, Aguena,
  Allam, Annis, Bacon, Bertin, Bhargava, Brooks, Buckley-Geer, Burke,
  Carretero, Costanzi, da~Costa, Pereira, Vicente, Desai, Dietrich, Doel,
  Ferrero, Flaugher, Frieman, Garc{\'{\i}}a-Bellido, Gaztanaga, Gerdes,
  Giannantonio, Gschwend, Gutierrez, Hinton, Hollowood, Honscheid, Hoyle,
  James, Kron, Kuehn, Lahav, Lima, Lin, Maia, Marshall, Martini, Melchior,
  Menanteau, Miquel, Mohr, Morgan, Ogando, Palmese, Paz-Chinch{\'{o}}n,
  Petravick, Pieres, Romer, Sanchez, Scarpine, Schubnell, Serrano, Smith,
  Soares-Santos, Tarle, Thomas, To, \& and}]{Amon_2022}
Amon, A., Gruen, D., Troxel, M., {et~al.} 2022, PRD, 105

\bibitem[{Anchordoqui {et~al.}(2021)Anchordoqui, Di~Valentino, Pan, \&
  Yang}]{Anchordoqui_2021}
Anchordoqui, L.~A., Di~Valentino, E., Pan, S., \& Yang, W. 2021, JHEAP, 32,
  28–64

\bibitem[{Asgari {et~al.}(2021)Asgari, Lin, Joachimi, Giblin, Heymans,
  Hildebrandt, Kannawadi, Stölzner, Tröster, van~den Busch, \&
  et~al.}]{Asgari_2021}
Asgari, M., Lin, C.-A., Joachimi, B., {et~al.} 2021, A\&A, 645, A104

\bibitem[{Baldauf {et~al.}(2011)Baldauf, Seljak, Senatore, \&
  Zaldarriaga}]{Baldauf_2011}
Baldauf, T., Seljak, U., Senatore, L., \& Zaldarriaga, M. 2011, JCAP, 2011,
  031–031

\bibitem[{Baleato~Lizancos \& White(2023)}]{Baleato_Lizancos_2023}
Baleato~Lizancos, A. \& White, M. 2023, JCAP, 2023, 044

\bibitem[{Bernstein(2010)}]{Bernstein_2010}
Bernstein, G.~M. 2010, MNRAS, 406, 2793–2804

\bibitem[{Blazek {et~al.}(2019)Blazek, MacCrann, Troxel, \& Fang}]{Blazek_2019}
Blazek, J.~A., MacCrann, N., Troxel, M., \& Fang, X. 2019, PRD, 100

\bibitem[{Brewer {et~al.}(2010)Brewer, Pártay, \& Csányi}]{Brewer2010}
Brewer, B.~J., Pártay, L.~B., \& Csányi, G. 2010, Stat. Comput., 21,
  649–656

\bibitem[{Britt {et~al.}(2024)Britt, Gruen, Friedrich, Yuan, \&
  Ried~Guachalla}]{britt2024bounds}
Britt, D., Gruen, D., Friedrich, O., Yuan, S., \& Ried~Guachalla, B. 2024,
  A\&A, 689, A253

\bibitem[{Campos {et~al.}(2023)Campos, Samuroff, \& Mandelbaum}]{Campos_2023}
Campos, A., Samuroff, S., \& Mandelbaum, R. 2023, MNRAS, 525, 1885

\bibitem[{Cawthon {et~al.}(2022)Cawthon, Elvin-Poole, Porredon, Crocce,
  Giannini, Gatti, Ross, Rykoff, Rosell, DeRose, Lee, Rodriguez-Monroy, Amon,
  Bechtol, Vicente, Gruen, Morgan, Sanchez, Sanchez, Sevilla-Noarbe, Abbott,
  Aguena, Allam, Annis, Avila, Bacon, Bertin, Brooks, Burke, Kind, Carretero,
  Castander, Choi, Costanzi, da~Costa, Pereira, Dawson, Desai, Diehl, Eckert,
  Everett, Ferrero, Fosalba, Frieman, Garc{\'{\i} }a-Bellido, Gaztanaga,
  Gruendl, Gschwend, Gutierrez, Hinton, Hollowood, Honscheid, Huterer, James,
  Kim, Kneib, Kuehn, Kuropatkin, Lahav, Lima, Lin, Maia, Melchior, Menanteau,
  Miquel, Mohr, Muir, Myles, Palmese, Pandey, Paz-Chinch{\'{o}}n, Percival,
  Plazas, Roodman, Rossi, Scarpine, Serrano, Smith, Soares-Santos, Suchyta,
  Swanson, Tarle, To, Troxel, \& and}]{Cawthon_2022}
Cawthon, R., Elvin-Poole, J., Porredon, A., {et~al.} 2022, MNRAS, 513, 5517

\bibitem[{Chintalapati {et~al.}(2022)Chintalapati, Gutierrez, \&
  Wang}]{Chintalapati_2022}
Chintalapati, P., Gutierrez, G., \& Wang, M. 2022, PRD, 105

\bibitem[{{Chisari} {et~al.}(2018){Chisari}, {Richardson}, {Devriendt},
  {Dubois}, {Schneider}, {Le Brun}, {Beckmann}, {Peirani}, {Slyz}, \&
  {Pichon}}]{Chisari2018}
{Chisari}, N.~E., {Richardson}, M.~L.~A., {Devriendt}, J., {et~al.} 2018,
  \mnras, 480, 3962

\bibitem[{Chopin \& Robert(2010)}]{Chopin_2010}
Chopin, N. \& Robert, C.~P. 2010, Biometrika, 97, 741–755

\bibitem[{Collaboration {et~al.}(2016)Collaboration, Aghamousa, Aguilar, Ahlen,
  Alam, Allen, Prieto, Annis, Bailey, Balland, Ballester, Baltay, Beaufore,
  Bebek, Beers, Bell, Bernal, Besuner, Beutler, Blake, Bleuler, Blomqvist,
  Blum, Bolton, Briceno, Brooks, Brownstein, Buckley-Geer, Burden, Burtin,
  Busca, Cahn, Cai, Cardiel-Sas, Carlberg, Carton, Casas, Castander,
  Cervantes-Cota, Claybaugh, Close, Coker, Cole, Comparat, Cooper, Cousinou,
  Crocce, Cuby, Cunningham, Davis, Dawson, de~la Macorra, Vicente, Delubac,
  Derwent, Dey, Dhungana, Ding, Doel, Duan, Ealet, Edelstein, Eftekharzadeh,
  Eisenstein, Elliott, Escoffier, Evatt, Fagrelius, Fan, Fanning, Farahi,
  Farihi, Favole, Feng, Fernandez, Findlay, Finkbeiner, Fitzpatrick, Flaugher,
  Flender, Font-Ribera, Forero-Romero, Fosalba, Frenk, Fumagalli, Gaensicke,
  Gallo, Garcia-Bellido, Gaztanaga, Fusillo, Gerard, Gershkovich, Giannantonio,
  Gillet, de~Rivera, Gonzalez-Perez, Gott, Graur, Gutierrez, Guy, Habib,
  Heetderks, Heetderks, Heitmann, Hellwing, Herrera, Ho, Holland, Honscheid,
  Huff, Hutchinson, Huterer, Hwang, Laguna, Ishikawa, Jacobs, Jeffrey,
  Jelinsky, Jennings, Jiang, Jimenez, Johnson, Joyce, Jullo, Juneau, Kama,
  Karcher, Karkar, Kehoe, Kennamer, Kent, Kilbinger, Kim, Kirkby, Kisner,
  Kitanidis, Kneib, Koposov, Kovacs, Koyama, Kremin, Kron, Kronig,
  Kueter-Young, Lacey, Lafever, Lahav, Lambert, Lampton, Landriau, Lang, Lauer,
  Goff, Guillou, Suu, Lee, Lee, Leitner, Lesser, Levi, L'Huillier, Li, Liang,
  Lin, Linder, Loebman, Lukić, Ma, MacCrann, Magneville, Makarem, Manera,
  Manser, Marshall, Martini, Massey, Matheson, McCauley, McDonald, McGreer,
  Meisner, Metcalfe, Miller, Miquel, Moustakas, Myers, Naik, Newman, Nichol,
  Nicola, da~Costa, Nie, Niz, Norberg, Nord, Norman, Nugent, O'Brien, Oh,
  Olsen, Padilla, Padmanabhan, Padmanabhan, Palanque-Delabrouille, Palmese,
  Pappalardo, Pâris, Park, Patej, Peacock, Peiris, Peng, Percival, Perruchot,
  Pieri, Pogge, Pollack, Poppett, Prada, Prakash, Probst, Rabinowitz, Raichoor,
  Ree, Refregier, Regal, Reid, Reil, Rezaie, Rockosi, Roe, Ronayette, Roodman,
  Ross, Ross, Rossi, Rozo, Ruhlmann-Kleider, Rykoff, Sabiu, Samushia, Sanchez,
  Sanchez, Schlegel, Schneider, Schubnell, Secroun, Seljak, Seo, Serrano,
  Shafieloo, Shan, Sharples, Sholl, Shourt, Silber, Silva, Sirk, Slosar, Smith,
  Smoot, Som, Song, Sprayberry, Staten, Stefanik, Tarle, Tie, Tinker, Tojeiro,
  Valdes, Valenzuela, Valluri, Vargas-Magana, Verde, Walker, Wang, Wang,
  Weaver, Weaverdyck, Wechsler, Weinberg, White, Yang, Yeche, Zhang, Zhao,
  Zheng, Zhou, Zhou, Zhu, Zou, \& Zu}]{desicollaboration2016desi}
Collaboration, D., Aghamousa, A., Aguilar, J., {et~al.} 2016, The DESI
  Experiment Part I: Science,Targeting, and Survey Design

\bibitem[{Collaboration: {et~al.}(2016)Collaboration:, Abbott, Abdalla,
  Aleksić, Allam, Amara, Bacon, Balbinot, Banerji, Bechtol, Benoit-Lévy,
  Bernstein, Bertin, Blazek, Bonnett, Bridle, Brooks, Brunner, Buckley-Geer,
  Burke, Caminha, Capozzi, Carlsen, Carnero-Rosell, Carollo, Carrasco-Kind,
  Carretero, Castander, Clerkin, Collett, Conselice, Crocce, Cunha, D'Andrea,
  da~Costa, Davis, Desai, Diehl, Dietrich, Dodelson, Doel, Drlica-Wagner,
  Estrada, Etherington, Evrard, Fabbri, Finley, Flaugher, Foley, Fosalba,
  Frieman, García-Bellido, Gaztanaga, Gerdes, Giannantonio, Goldstein, Gruen,
  Gruendl, Guarnieri, Gutierrez, Hartley, Honscheid, Jain, James, Jeltema,
  Jouvel, Kessler, King, Kirk, Kron, Kuehn, Kuropatkin, Lahav, Li, Lima, Lin,
  Maia, Makler, Manera, Maraston, Marshall, Martini, McMahon, Melchior, Merson,
  Miller, Miquel, Mohr, Morice-Atkinson, Naidoo, Neilsen, Nichol, Nord, Ogando,
  Ostrovski, Palmese, Papadopoulos, Peiris, Peoples, Percival, Plazas, Reed,
  Refregier, Romer, Roodman, Ross, Rozo, Rykoff, Sadeh, Sako, Sánchez,
  Sanchez, Santiago, Scarpine, Schubnell, Sevilla-Noarbe, Sheldon, Smith,
  Smith, Soares-Santos, Sobreira, Soumagnac, Suchyta, Sullivan, Swanson, Tarle,
  Thaler, Thomas, Thomas, Tucker, Vieira, Vikram, Walker, Wechsler, Weller,
  Wester, Whiteway, Wilcox, Yanny, Zhang, \& Zuntz}]{DES2016}
Collaboration:, D. E.~S., Abbott, T., Abdalla, F.~B., {et~al.} 2016, MNRAS,
  460, 1270

\bibitem[{Cordero {et~al.}(2022)Cordero, Harrison, Rollins, Bernstein, Bridle,
  Alarcon, Alves, Amon, Andrade-Oliveira, Camacho, Campos, Choi, DeRose,
  Dodelson, Eckert, Eifler, Everett, Fang, Friedrich, Gruen, Gruendl, Hartley,
  Huff, Krause, Kuropatkin, MacCrann, McCullough, Myles, Pandey, Raveri,
  Rosenfeld, Rykoff, S{\'{a} }nchez, S{\'{a}}nchez, Sevilla-Noarbe, Sheldon,
  Troxel, Wechsler, Yanny, Yin, Zhang, Aguena, Allam, Bertin, Brooks, Burke,
  Rosell, Kind, Carretero, Castander, Cawthon, Costanzi, da~Costa,
  da~Silva~Pereira, Vicente, Diehl, Dietrich, Doel, Elvin-Poole, Ferrero,
  Flaugher, Fosalba, Frieman, Garcia-Bellido, Gerdes, Gschwend, Gutierrez,
  Hinton, Hollowood, Honscheid, Hoyle, James, Kuehn, Lahav, Maia, March,
  Menanteau, Miquel, Morgan, Muir, Palmese, Paz-Chinchon, Pieres,
  Malag{\'{o}}n, S{\'{a}}nchez, Scarpine, Serrano, Smith, Soares-Santos,
  Suchyta, Swanson, Tarle, Thomas, To, \& and}]{Cordero_2022}
Cordero, J.~P., Harrison, I., Rollins, R.~P., {et~al.} 2022, MNRAS, 511, 2170

\bibitem[{Cranmer {et~al.}(2020)Cranmer, Brehmer, \& Louppe}]{Cranmer_2020}
Cranmer, K., Brehmer, J., \& Louppe, G. 2020, PPNAS, 117, 30055–30062

\bibitem[{Cui {et~al.}(2014)Cui, Borgani, \& Murante}]{Cui_2014}
Cui, W., Borgani, S., \& Murante, G. 2014, MNRAS, 441, 1769–1782

\bibitem[{Dalal {et~al.}(2023)Dalal, Li, Nicola, Zuntz, Strauss, Sugiyama,
  Zhang, Rau, Mandelbaum, Takada, More, Miyatake, Kannawadi, Shirasaki,
  Taniguchi, Takahashi, Osato, Hamana, Oguri, Nishizawa, Malag\'on, Sunayama,
  Alonso, Slosar, Luo, Armstrong, Bosch, Hsieh, Komiyama, Lupton, Lust,
  MacArthur, Miyazaki, Murayama, Nishimichi, Okura, Price, Tait, Tanaka, \&
  Wang}]{dalal2023hyper}
Dalal, R., Li, X., Nicola, A., {et~al.} 2023, PRD, 108, 123519

\bibitem[{de~Jong {et~al.}(2012)de~Jong, Bellido-Tirado, Chiappini, Depagne,
  Haynes, Johl, Schnurr, Schwope, Walcher, Dionies, \& et~al.}]{de_Jong_2012}
de~Jong, R.~S., Bellido-Tirado, O., Chiappini, C., {et~al.} 2012, SPIE

\bibitem[{Dekel \& Lahav(1999)}]{Dekel_1999}
Dekel, A. \& Lahav, O. 1999, ApJ, 520, 24–34

\bibitem[{DeRose {et~al.}(2022)DeRose, Wechsler, Becker, Rykoff, Pandey,
  MacCrann, Amon, Myles, Krause, Gruen, Jain, Troxel, Prat, Alarcon, Sánchez,
  Blazek, Crocce, Giannini, Gatti, Bernstein, Zuntz, Dodelson, Fang, Friedrich,
  Secco, Elvin-Poole, Porredon, Everett, Choi, Harrison, Cordero,
  Rodriguez-Monroy, McCullough, Cawthon, Chen, Alves, Andrade-Oliveira,
  Bechtol, Camacho, Campos, Rosell, Kind, Diehl, Drlica-Wagner, Eckert, Eifler,
  Gruendl, Hartley, Huang, Huff, Kuropatkin, Raveri, Rosenfeld, Ross, Sanchez,
  Sevilla-Noarbe, Sheldon, Yanny, Yin, Zhang, Fosalba, Aguena, Allam, Annis,
  Avila, Bacon, Bhargava, Brooks, Buckley-Geer, Burke, Carretero, Castander,
  Chang, Costanzi, da~Costa, Pereira, De~Vicente, Desai, Dietrich, Doel,
  Evrard, Ferrero, Ferté, Flaugher, Frieman, García-Bellido, Gaztanaga,
  Giannantonio, Gschwend, Gutierrez, Hinton, Hollowood, Honscheid, Huterer,
  James, Kuehn, Lahav, Lima, Maia, Marshall, Melchior, Menanteau, Miquel, Mohr,
  Morgan, Palmese, Paz-Chinchón, Petravick, Pieres, Malagón, Sanchez,
  Scarpine, Serrano, Smith, Soares-Santos, Suchyta, Tarle, Thomas, To, \&
  Varga}]{2021arXiv210513547D}
DeRose, J., Wechsler, R., Becker, M., {et~al.} 2022, PRD, 105

\bibitem[{DeRose {et~al.}(2019)DeRose, Wechsler, Becker, Busha, Rykoff,
  MacCrann, Erickson, Evrard, Kravtsov, Gruen, Allam, Avila, Bridle, Brooks,
  Buckley-Geer, Rosell, Kind, Carretero, Castander, Cawthon, Crocce, da~Costa,
  Davis, Vicente, Dietrich, Doel, Drlica-Wagner, Fosalba, Frieman,
  Garcia-Bellido, Gutierrez, Hartley, Hollowood, Hoyle, James, Krause, Kuehn,
  Kuropatkin, Lima, Maia, Menanteau, Miller, Miquel, Ogando, Malagón, Romer,
  Sanchez, Schindler, Serrano, Sevilla-Noarbe, Smith, Suchyta, Swanson, Tarle,
  \& Vikram}]{derose2019buzzard}
DeRose, J., Wechsler, R.~H., Becker, M.~R., {et~al.} 2019, The Buzzard Flock:
  Dark Energy Survey Synthetic Sky Catalogs

\bibitem[{Ding {et~al.}(2018)Ding, Seo, Vlah, Feng, Schmittfull, \&
  Beutler}]{Ding_2018}
Ding, Z., Seo, H.-J., Vlah, Z., {et~al.} 2018, MNRAS

\bibitem[{{Eifler} {et~al.}(2021){Eifler}, {Miyatake}, {Krause}, {Heinrich},
  {Miranda}, {Hirata}, {Xu}, {Hemmati}, {Simet}, {Capak}, {Choi}, {Dor{\'e}},
  {Doux}, {Fang}, {Hounsell}, {Huff}, {Huang}, {Jarvis}, {Kruk}, {Masters},
  {Rozo}, {Scolnic}, {Spergel}, {Troxel}, {von der Linden}, {Wang}, {Weinberg},
  {Wenzl}, \& {Wu}}]{2021MNRAS.507.1746E}
{Eifler}, T., {Miyatake}, H., {Krause}, E., {et~al.} 2021, \mnras, 507, 1746

\bibitem[{Etherington(2019)}]{etherington2019mahalanobis}
Etherington, T.~R. 2019, PeerJ, 7, e6678

\bibitem[{Feroz {et~al.}(2009)Feroz, Hobson, \& Bridges}]{Feroz_2009}
Feroz, F., Hobson, M.~P., \& Bridges, M. 2009, MNRAS, 398, 1601–1614

\bibitem[{Foreman-Mackey {et~al.}(2013)Foreman-Mackey, Hogg, Lang, \&
  Goodman}]{Foreman_Mackey_2013}
Foreman-Mackey, D., Hogg, D.~W., Lang, D., \& Goodman, J. 2013, PASP, 125, 306

\bibitem[{{Friedrich} {et~al.}(2018){Friedrich}, {Gruen}, {DeRose}, {Kirk},
  {Krause}, {McClintock}, {Rykoff}, {Seitz}, {Wechsler}, {Bernstein}, {Blazek},
  {Chang}, {Hilbert}, {Jain}, {Kovacs}, {Lahav}, {Abdalla}, {Allam}, {Annis},
  {Bechtol}, {Benoit-L{\'e}vy}, {Bertin}, {Brooks}, {Carnero Rosell}, {Carrasco
  Kind}, {Carretero}, {Cunha}, {D'Andrea}, {da Costa}, {Davis}, {Desai},
  {Diehl}, {Dietrich}, {Drlica-Wagner}, {Eifler}, {Fosalba}, {Frieman},
  {Garc{\'{\i}}a-Bellido}, {Gaztanaga}, {Gerdes}, {Giannantonio}, {Gruendl},
  {Gschwend}, {Gutierrez}, {Honscheid}, {James}, {Jarvis}, {Kuehn},
  {Kuropatkin}, {Lima}, {March}, {Marshall}, {Melchior}, {Menanteau}, {Miquel},
  {Mohr}, {Nord}, {Plazas}, {Sanchez}, {Scarpine}, {Schindler}, {Schubnell},
  {Sevilla-Noarbe}, {Sheldon}, {Smith}, {Soares-Santos}, {Sobreira}, {Suchyta},
  {Swanson}, {Tarle}, {Thomas}, {Troxel}, {Vikram}, {Weller}, \& {DES
  Collaboration}}]{Friedrich2018}
{Friedrich}, O., {Gruen}, D., {DeRose}, J., {et~al.} 2018, PRD, 98, 023508

\bibitem[{Friedrich {et~al.}(2022)Friedrich, Halder, Boyle, Uhlemann, Britt,
  Codis, Gruen, \& Hahn}]{friedrich2021pdf}
Friedrich, O., Halder, A., Boyle, A., {et~al.} 2022, MNRAS, 510, 5069–5087

\bibitem[{Georgiou {et~al.}(2021)Georgiou, Hoekstra, Kuijken, Bilicki, Dvornik,
  Erben, Giblin, Heymans, Hildebrandt, de~Jong, \& et~al.}]{Georgiou_2021}
Georgiou, C., Hoekstra, H., Kuijken, K., {et~al.} 2021, A\&A, 647, A185

\bibitem[{{Gruen} {et~al.}(2018){Gruen}, {Friedrich}, {Krause}, {DeRose},
  {Cawthon}, {Davis}, {Elvin-Poole}, {Rykoff}, {Wechsler}, {Alarcon},
  {Bernstein}, {Blazek}, {Chang}, {Clampitt}, {Crocce}, {De Vicente}, {Gatti},
  {Gill}, {Hartley}, {Hilbert}, {Hoyle}, {Jain}, {Jarvis}, {Lahav}, {MacCrann},
  {McClintock}, {Prat}, {Rollins}, {Ross}, {Rozo}, {Samuroff}, {S{\'a}nchez},
  {Sheldon}, {Troxel}, {Zuntz}, {Abbott}, {Abdalla}, {Allam}, {Annis},
  {Bechtol}, {Benoit-L{\'e}vy}, {Bertin}, {Bridle}, {Brooks}, {Buckley-Geer},
  {Carnero Rosell}, {Carrasco Kind}, {Carretero}, {Cunha}, {D'Andrea}, {da
  Costa}, {Desai}, {Diehl}, {Dietrich}, {Doel}, {Drlica-Wagner}, {Fernandez},
  {Flaugher}, {Fosalba}, {Frieman}, {Garc{\'{\i}}a-Bellido}, {Gaztanaga},
  {Giannantonio}, {Gruendl}, {Gschwend}, {Gutierrez}, {Honscheid}, {James},
  {Jeltema}, {Kuehn}, {Kuropatkin}, {Lima}, {March}, {Marshall}, {Martini},
  {Melchior}, {Menanteau}, {Miquel}, {Mohr}, {Plazas}, {Roodman}, {Sanchez},
  {Scarpine}, {Schubnell}, {Sevilla-Noarbe}, {Smith}, {Smith}, {Soares-Santos},
  {Sobreira}, {Swanson}, {Tarle}, {Thomas}, {Vikram}, {Walker}, {Weller},
  {Zhang}, \& {DES Collaboration}}]{Gruen2018}
{Gruen}, D., {Friedrich}, O., {Krause}, E., {et~al.} 2018, PRD, 98, 023507

\bibitem[{Habeck(2015)}]{Habeck2015}
Habeck, M. 2015, in AIP Conference Proceedings, Vol. 1641 (AIP Publishing LLC),
  121–129

\bibitem[{Harris {et~al.}(2020)Harris, Millman, van~der Walt, Gommers,
  Virtanen, Cournapeau, Wieser, Taylor, Berg, Smith, Kern, Picus, Hoyer, van
  Kerkwijk, Brett, Haldane, del R{\'{\i}}o, Wiebe, Peterson,
  G{\'{e}}rard-Marchant, Sheppard, Reddy, Weckesser, Abbasi, Gohlke, \&
  Oliphant}]{Harris_2020}
Harris, C.~R., Millman, K.~J., van~der Walt, S.~J., {et~al.} 2020, Nature, 585,
  357

\bibitem[{Hesterberg(2003)}]{importance_sampling1}
Hesterberg, T.~C. 2003, Advances in Importance Sampling

\bibitem[{Heymans {et~al.}(2021)Heymans, Tröster, Asgari, Blake, Hildebrandt,
  Joachimi, Kuijken, Lin, Sánchez, van~den Busch, \& et~al.}]{Heymans_2021}
Heymans, C., Tröster, T., Asgari, M., {et~al.} 2021, A\&A, 646, A140

\bibitem[{Hildebrandt {et~al.}(2021)Hildebrandt, van~den Busch, Wright, Blake,
  Joachimi, Kuijken, Tröster, Asgari, Bilicki, de~Jong, \&
  et~al.}]{Hildebrandt_2021}
Hildebrandt, H., van~den Busch, J.~L., Wright, A.~H., {et~al.} 2021, A\&A, 647,
  A124

\bibitem[{{Hinton}(2016)}]{chainconsumer}
{Hinton}, S.~R. 2016, JOSS, 1, 00045

\bibitem[{Hirata \& Seljak(2004)}]{Hirata2004}
Hirata, C.~M. \& Seljak, U. c.~v. 2004, PRD, 70, 063526

\bibitem[{Ho {et~al.}(2024)Ho, Bartlett, Chartier, Cuesta-Lazaro, Ding, Lapel,
  Lemos, Lovell, Makinen, Modi, Pandya, Pandey, Perez, Wandelt, \&
  Bryan}]{Ho_2024}
Ho, M., Bartlett, D.~J., Chartier, N., {et~al.} 2024, OJAP, 7

\bibitem[{Huang {et~al.}(2019)Huang, Eifler, Mandelbaum, \&
  Dodelson}]{Huang_2019}
Huang, H.-J., Eifler, T., Mandelbaum, R., \& Dodelson, S. 2019, MNRAS, 488,
  1652

\bibitem[{Huang {et~al.}(2017)Huang, Leauthaud, Murata, Bosch, Price, Lupton,
  Mandelbaum, Lackner, Bickerton, Miyazaki, \& et~al.}]{Huang_2017}
Huang, S., Leauthaud, A., Murata, R., {et~al.} 2017, PASJ, 70

\bibitem[{Hunter(2007)}]{matpltlib}
Hunter, J.~D. 2007, CiSE, 9, 90

\bibitem[{Huterer \& Turner(2001)}]{Huterer_2001}
Huterer, D. \& Turner, M.~S. 2001, PRD, 64

\bibitem[{Ishikawa {et~al.}(2021)Ishikawa, Okumura, Oguri, \&
  Lin}]{ishikawa2021halomodel}
Ishikawa, S., Okumura, T., Oguri, M., \& Lin, S.-C. 2021, ApJ, 922, 23

\bibitem[{Ivanov {et~al.}(2024)Ivanov, Cuesta-Lazaro, Mishra-Sharma, Obuljen,
  \& Toomey}]{Ivanov2024}
Ivanov, M.~M., Cuesta-Lazaro, C., Mishra-Sharma, S., Obuljen, A., \& Toomey,
  M.~W. 2024, PRD, 110

\bibitem[{Ivezi\'{c} {et~al.}(2019)Ivezi\'{c}, Kahn, Tyson, Abel, Acosta,
  Allsman, Alonso, AlSayyad, Anderson, Andrew, \& et~al.}]{Ivezi2019}
Ivezi\'{c}, v., Kahn, S.~M., Tyson, J.~A., {et~al.} 2019, ApJ, 873, 111

\bibitem[{Joachimi {et~al.}(2021)Joachimi, Lin, Asgari, Tröster, Heymans,
  Hildebrandt, Köhlinger, Sánchez, Wright, Bilicki, Blake, van~den Busch,
  Crocce, Dvornik, Erben, Getman, Giblin, Hoekstra, Kannawadi, Kuijken,
  Napolitano, Schneider, Scoccimarro, Sellentin, Shan, von Wietersheim-Kramsta,
  \& Zuntz}]{Joachimi_2021}
Joachimi, B., Lin, C.-A., Asgari, M., {et~al.} 2021, A \& A, 646, A129

\bibitem[{Kiessling {et~al.}(2015)Kiessling, Cacciato, Joachimi, Kirk,
  Kitching, Leonard, Mandelbaum, Schaefer, Sifon, Brown, \&
  et~al.}]{Kiessling_2015}
Kiessling, A., Cacciato, M., Joachimi, B., {et~al.} 2015, Space Sci. Rev., 193,
  67–136

\bibitem[{Krause {et~al.}(2021)Krause, Fang, Pandey, Secco, Alves, Huang,
  Blazek, Prat, Zuntz, Eifler, MacCrann, DeRose, Crocce, Porredon, Jain,
  Troxel, Dodelson, Huterer, Liddle, Leonard, Amon, Chen, Elvin-Poole, Ferté,
  Muir, Park, Samuroff, Brandao-Souza, Weaverdyck, Zacharegkas, Rosenfeld,
  Campos, Chintalapati, Choi, Valentino, Doux, Herner, Lemos, Mena-Fernández,
  Omori, Paterno, Rodriguez-Monroy, Rogozenski, Rollins, Troja, Tutusaus,
  Wechsler, Abbott, Aguena, Allam, Andrade-Oliveira, Annis, Bacon, Baxter,
  Bechtol, Bernstein, Brooks, Buckley-Geer, Burke, Rosell, Kind, Carretero,
  Castander, Cawthon, Chang, Costanzi, da~Costa, Pereira, Vicente, Desai,
  Diehl, Doel, Everett, Evrard, Ferrero, Flaugher, Fosalba, Frieman,
  García-Bellido, Gaztanaga, Gerdes, Giannantonio, Gruen, Gruendl, Gschwend,
  Gutierrez, Hartley, Hinton, Hollowood, Honscheid, Hoyle, Huff, James, Kuehn,
  Kuropatkin, Lahav, Lima, Maia, Marshall, Martini, Melchior, Menanteau,
  Miquel, Mohr, Morgan, Myles, Palmese, Paz-Chinchón, Petravick, Pieres,
  Malagón, Sanchez, Scarpine, Schubnell, Serrano, Sevilla-Noarbe, Smith,
  Soares-Santos, Suchyta, Tarle, Thomas, To, Varga, \& Weller}]{krause2021}
Krause, E., Fang, X., Pandey, S., {et~al.} 2021, Dark Energy Survey Year 3
  Results: Multi-Probe Modeling Strategy and Validation

\bibitem[{Kullback \& Leibler(1951)}]{Kullback1951}
Kullback, S. \& Leibler, R.~A. 1951, Ann. Math. Stat., 22, 79–86

\bibitem[{Lamman {et~al.}(2023)Lamman, Eisenstein, Aguilar, Brooks, de~la
  Macorra, Doel, Font-Ribera, Gontcho, Honscheid, Kehoe, Kisner, Kremin,
  Landriau, Levi, Miquel, Moustakas, Palanque-Delabrouille, Poppett, Schubnell,
  \& Tarl{\'{e} }}]{Lamman_2023}
Lamman, C., Eisenstein, D., Aguilar, J.~N., {et~al.} 2023, MNRAS, 522, 117

\bibitem[{Lamman {et~al.}(2024)Lamman, Tsaprazi, Shi, Šarčević, Pyne,
  Legnani, \& Ferreira}]{lamman2023ia}
Lamman, C., Tsaprazi, E., Shi, J., {et~al.} 2024, OJAP, 7

\bibitem[{Lange(2023)}]{Lange2023}
Lange, J.~U. 2023, MNRAS, 525, 3181–3194

\bibitem[{Laureijs {et~al.}(2011)Laureijs, Amiaux, Arduini, Auguères,
  Brinchmann, Cole, Cropper, Dabin, Duvet, Ealet, Garilli, Gondoin, Guzzo,
  Hoar, Hoekstra, Holmes, Kitching, Maciaszek, Mellier, Pasian, Percival,
  Rhodes, Criado, Sauvage, Scaramella, Valenziano, Warren, Bender, Castander,
  Cimatti, Fèvre, Kurki-Suonio, Levi, Lilje, Meylan, Nichol, Pedersen, Popa,
  Lopez, Rix, Rottgering, Zeilinger, Grupp, Hudelot, Massey, Meneghetti,
  Miller, Paltani, Paulin-Henriksson, Pires, Saxton, Schrabback, Seidel, Walsh,
  Aghanim, Amendola, Bartlett, Baccigalupi, Beaulieu, Benabed, Cuby, Elbaz,
  Fosalba, Gavazzi, Helmi, Hook, Irwin, Kneib, Kunz, Mannucci, Moscardini, Tao,
  Teyssier, Weller, Zamorani, Osorio, Boulade, Foumond, Giorgio, Guttridge,
  James, Kemp, Martignac, Spencer, Walton, Blümchen, Bonoli, Bortoletto,
  Cerna, Corcione, Fabron, Jahnke, Ligori, Madrid, Martin, Morgante, Pamplona,
  Prieto, Riva, Toledo, Trifoglio, Zerbi, Abdalla, Douspis, Grenet, Borgani,
  Bouwens, Courbin, Delouis, Dubath, Fontana, Frailis, Grazian, Koppenhöfer,
  Mansutti, Melchior, Mignoli, Mohr, Neissner, Noddle, Poncet, Scodeggio,
  Serrano, Shane, Starck, Surace, Taylor, Verdoes-Kleijn, Vuerli, Williams,
  Zacchei, Altieri, Sanz, Kohley, Oosterbroek, Astier, Bacon, Bardelli, Baugh,
  Bellagamba, Benoist, Bianchi, Biviano, Branchini, Carbone, Cardone, Clements,
  Colombi, Conselice, Cresci, Deacon, Dunlop, Fedeli, Fontanot, Franzetti,
  Giocoli, Garcia-Bellido, Gow, Heavens, Hewett, Heymans, Holland, Huang,
  Ilbert, Joachimi, Jennins, Kerins, Kiessling, Kirk, Kotak, Krause, Lahav, van
  Leeuwen, Lesgourgues, Lombardi, Magliocchetti, Maguire, Majerotto, Maoli,
  Marulli, Maurogordato, McCracken, McLure, Melchiorri, Merson, Moresco,
  Nonino, Norberg, Peacock, Pello, Penny, Pettorino, Porto, Pozzetti,
  Quercellini, Radovich, Rassat, Roche, Ronayette, Rossetti, Sartoris,
  Schneider, Semboloni, Serjeant, Simpson, Skordis, Smadja, Smartt, Spano,
  Spiro, Sullivan, Tilquin, Trotta, Verde, Wang, Williger, Zhao, Zoubian, \&
  Zucca}]{laureijs2011euclid}
Laureijs, R., Amiaux, J., Arduini, S., {et~al.} 2011, Euclid Definition Study
  Report

\bibitem[{Lemos {et~al.}(2021)Lemos, Raveri, Campos, Park, Chang, Weaverdyck,
  Huterer, Liddle, Blazek, Cawthon, \& et~al.}]{Lemos_2021}
Lemos, P., Raveri, M., Campos, A., {et~al.} 2021, MNRAS, 505, 6179–6194

\bibitem[{Li {et~al.}(2022)Li, Miyatake, Luo, More, Oguri, Hamana, Mandelbaum,
  Shirasaki, Takada, Armstrong, Kannawadi, Takita, Miyazaki, Nishizawa,
  Plazas~Malagon, Strauss, Tanaka, \& Yoshida}]{li2021threeyear}
Li, X., Miyatake, H., Luo, W., {et~al.} 2022, PASJ, 74, 421–459

\bibitem[{{LSST Science Collaboration} {et~al.}(2009){LSST Science
  Collaboration}, {Abell}, {Allison}, {Anderson}, {Andrew}, {Angel}, {Armus},
  {Arnett}, {Asztalos}, {Axelrod}, \& et~al.}]{2009arXiv0912.0201L}
{LSST Science Collaboration}, {Abell}, P.~A., {Allison}, J., {et~al.} 2009,
  ArXiv e-prints [\eprint[arXiv]{0912.0201}]

\bibitem[{MacCrann {et~al.}(2021)MacCrann, Becker, McCullough, Amon, Gruen,
  Jarvis, Choi, Troxel, Sheldon, Yanny, Herner, Dodelson, Zuntz, Eckert,
  Rollins, Varga, Bernstein, Gruendl, Harrison, Hartley, Sevilla-Noarbe,
  Pieres, Bridle, Myles, Alarcon, Everett, Sánchez, Huff, Tarsitano, Gatti,
  Secco, Abbott, Aguena, Allam, Annis, Bacon, Bertin, Brooks, Burke,
  Carnero Rosell, Carrasco Kind, Carretero, Costanzi, Crocce, Pereira,
  De Vicente, Desai, Diehl, Dietrich, Doel, Eifler, Ferrero, Ferté, Flaugher,
  Fosalba, Frieman, García-Bellido, Gaztanaga, Gerdes, Giannantonio, Gschwend,
  Gutierrez, Hinton, Hollowood, Honscheid, James, Lahav, Lima, Maia, March,
  Marshall, Martini, Melchior, Menanteau, Miquel, Mohr, Morgan, Muir, Ogando,
  Palmese, Paz-Chinchón, Plazas, Rodriguez-Monroy, Roodman, Samuroff, Sanchez,
  Scarpine, Serrano, Smith, Soares-Santos, Suchyta, Swanson, Tarle, Thomas, To,
  \& Wilkinson}]{maccrann2020des}
MacCrann, N., Becker, M.~R., McCullough, J., {et~al.} 2021, MNRAS, 509,
  3371–3394

\bibitem[{MacCrann {et~al.}(2018)MacCrann, DeRose, Wechsler, Blazek, Gaztanaga,
  Crocce, Rykoff, Becker, Jain, Krause, Eifler, Gruen, Zuntz, Troxel,
  Elvin-Poole, Prat, Wang, Dodelson, Kravtsov, Fosalba, Busha, Evrard, Huterer,
  Abbott, Abdalla, Allam, Annis, Avila, Bernstein, Brooks, Buckley-Geer, Burke,
  Rosell, Kind, Carretero, Castander, Cawthon, Cunha, D’Andrea, da Costa,
  Davis, De Vicente, Diehl, Doel, Frieman, García-Bellido, Gerdes, Gruendl,
  Gutierrez, Hartley, Hollowood, Honscheid, Hoyle, James, Jeltema, Kirk, Kuehn,
  Kuropatkin, Lima, Maia, Marshall, Menanteau, Miquel, Plazas, Roodman,
  Sanchez, Scarpine, Schubnell, Sevilla-Noarbe, Smith, Smith, Soares-Santos,
  Sobreira, Suchyta, Swanson, Tarle, Thomas, Walker, \& Weller}]{MacCrann_2018}
MacCrann, N., DeRose, J., Wechsler, R.~H., {et~al.} 2018, MNRAS, 480,
  4614–4635

\bibitem[{Mahalanobis(1936)}]{mahalanobis1936generalized}
Mahalanobis, P.~C. 1936, Proc. Natl. Acad. Sci. India, 2, 49

\bibitem[{{Mandelbaum} {et~al.}(2019){Mandelbaum}, {Blazek}, {Chisari},
  {Collett}, {Galbany}, {Gawiser}, {Hlo{\v{z}}ek}, {Kim}, {Leonard}, {Lochner},
  {Mandelbaum}, {Newman}, {Perrefort}, {Schmidt}, {Singh}, {Sullivan}, \& {LSST
  Dark Energy Science Collaboration}}]{mandelbaum2019widefield}
{Mandelbaum}, R., {Blazek}, J., {Chisari}, N.~E., {et~al.} 2019, \baas, 51, 363

\bibitem[{Mummery {et~al.}(2017)Mummery, McCarthy, Bird, \&
  Schaye}]{Mummery_2017}
Mummery, B.~O., McCarthy, I.~G., Bird, S., \& Schaye, J. 2017, MNRAS, 471,
  227–242

\bibitem[{Myles {et~al.}(2021)Myles, Alarcon, Amon, S{\'{a} }nchez, Everett,
  DeRose, McCullough, Gruen, Bernstein, Troxel, Dodelson, Campos, MacCrann,
  Yin, Raveri, Amara, Becker, Choi, Cordero, Eckert, Gatti, Giannini, Gschwend,
  Gruendl, Harrison, Hartley, Huff, Kuropatkin, Lin, Masters, Miquel, Prat,
  Roodman, Rykoff, Sevilla-Noarbe, Sheldon, Wechsler, Yanny, Abbott, Aguena,
  Allam, Annis, Bacon, Bertin, Bhargava, Bridle, Brooks, Burke, Rosell, Kind,
  Carretero, Castander, Conselice, Costanzi, Crocce, da~Costa, Pereira, Desai,
  Diehl, Eifler, Elvin-Poole, Evrard, Ferrero, Fert{\'{e}}, Flaugher, Fosalba,
  Frieman, Garc{\'{\i}}a-Bellido, Gaztanaga, Giannantonio, Hinton, Hollowood,
  Honscheid, Hoyle, Huterer, James, Krause, Kuehn, Lahav, Lima, Maia, Marshall,
  Martini, Melchior, Menanteau, Mohr, Morgan, Muir, Ogando, Palmese,
  Paz-Chinch{\'{o}}n, Plazas, Rodriguez-Monroy, Samuroff, Sanchez, Scarpine,
  Secco, Serrano, Smith, Soares-Santos, Suchyta, Swanson, Tarle, Thomas, To,
  Varga, Weller, \& Wester}]{Myles_2021}
Myles, J., Alarcon, A., Amon, A., {et~al.} 2021, MNRAS, 505, 4249

\bibitem[{Newman \& Gruen(2022)}]{Newman_2022}
Newman, J.~A. \& Gruen, D. 2022, ARA\&A, 60, 363

\bibitem[{Nguyen {et~al.}(2024)Nguyen, Schmidt, Tucci, Reinecke, \&
  Kostić}]{nguyen2024informationextractedgalaxyclustering}
Nguyen, N.-M., Schmidt, F., Tucci, B., Reinecke, M., \& Kostić, A. 2024, PRL,
  133

\bibitem[{Novaes {et~al.}(2024)Novaes, Thiele, Armijo, Cheng, Cowell, Marques,
  Ferreira, Shirasaki, Osato, \& Liu}]{novaes2024cosmologyhscy1weak}
Novaes, C.~P., Thiele, L., Armijo, J., {et~al.} 2024, Cosmology from HSC Y1
  Weak Lensing with Combined Higher-Order Statistics and Simulation-based
  Inference

\bibitem[{pandas~development team(2020)}]{reback2020pandas}
pandas~development team, T. 2020, pandas-dev/pandas: Pandas

\bibitem[{Pandey {et~al.}(2022)Pandey, Krause, DeRose, MacCrann, Jain, Crocce,
  Blazek, Choi, Huang, To, Fang, Elvin-Poole, Prat, Porredon, Secco,
  Rodriguez-Monroy, Weaverdyck, Park, Raveri, Rozo, Rykoff, Bernstein,
  S{\'{a}}nchez, Jarvis, Troxel, Zacharegkas, Chang, Alarcon, Alves, Amon,
  Andrade-Oliveira, Baxter, Bechtol, Becker, Camacho, Campos, Rosell, Kind,
  Cawthon, Chen, Chintalapati, Davis, Valentino, Diehl, Dodelson, Doux,
  Drlica-Wagner, Eckert, Eifler, Elsner, Everett, Farahi, Fert{\'{e}}, Fosalba,
  Friedrich, Gatti, Giannini, Gruen, Gruendl, Harrison, Hartley, Huff, Huterer,
  Kovacs, Leget, McCullough, Muir, Myles, Navarro-Alsina, Omori, Rollins,
  Roodman, Rosenfeld, Sevilla-Noarbe, Sheldon, Shin, Troja, Tutusaus, Varga,
  Wechsler, Yanny, Yin, Zhang, Zuntz, Abbott, Aguena, Allam, Annis, Bacon,
  Bertin, Brooks, Burke, Carretero, Conselice, Costanzi, da~Costa, Pereira,
  Vicente, Dietrich, Doel, Evrard, Ferrero, Flaugher, Frieman,
  Garc{\'{\i}}a-Bellido, Gaztanaga, Gerdes, Giannantonio, Gschwend, Gutierrez,
  Hinton, Hollowood, Honscheid, James, Jeltema, Kuehn, Kuropatkin, Lahav, Lima,
  Lin, Maia, Marshall, Melchior, Menanteau, Miller, Miquel, Mohr, Morgan,
  Palmese, Paz-Chinch{\'{o}}n, Petravick, Pieres, Malag{\'{o}}n, Sanchez,
  Scarpine, Serrano, Smith, Soares-Santos, Suchyta, Tarle, Thomas, \&
  and}]{Pandey_2022}
Pandey, S., Krause, E., DeRose, J., {et~al.} 2022, PRD, 106

\bibitem[{Percival {et~al.}(2021)Percival, Friedrich, Sellentin, \&
  Heavens}]{percival2021matching}
Percival, W.~J., Friedrich, O., Sellentin, E., \& Heavens, A. 2021, MNRAS, 510,
  3207–3221

\bibitem[{Raveri \& Hu(2019)}]{Raveri_2019}
Raveri, M. \& Hu, W. 2019, PRD, 99

\bibitem[{Rosell {et~al.}(2021)Rosell, Rodriguez-Monroy, Crocce, Elvin-Poole,
  Porredon, Ferrero, Mena-Fernández, Cawthon, De Vicente, Gaztanaga, Ross,
  Sanchez, Sevilla-Noarbe, Alves, Andrade-Oliveira, Asorey, Avila,
  Brandao-Souza, Camacho, Chan, Ferté, Muir, Riquelme, Rosenfeld, Cid,
  Hartley, Weaverdyck, Abbott, Aguena, Allam, Annis, Bertin, Brooks,
  Buckley-Geer, Burke, Calcino, Carollo, Kind, Carretero, Castander, Choi,
  Conselice, Costanzi, da Costa, da Silva Pereira, Davis, Desai, Diehl,
  Doel, Drlica-Wagner, Eckert, Everett, Evrard, Flaugher, Fosalba, Frieman,
  Garcia-Bellido, Gerdes, Giannantonio, Glazebrook, Gruen, Gruendl, Gschwend,
  Gutierrez, Hinton, Hollowood, Honscheid, Hoyle, Huterer, James, Kim, Krause,
  Kuehn, Lahav, Lewis, Lidman, Lima, Maia, Malik, Marshall, Menanteau, Miquel,
  Mohr, Moller, Morgan, Ogando, Palmese, Paz-Chinchon, Percival, Pieres,
  Malagón, Roodman, Scarpine, Schubnell, Serrano, Sharp, Sheldon, Smith,
  Soares-Santos, Suchyta, Swanson, Tarle, Thomas, To, Tucker, Tucker, Uddin, \&
  Varga}]{rosell2021dark}
Rosell, A.~C., Rodriguez-Monroy, M., Crocce, M., {et~al.} 2021, MNRAS, 509,
  778–799

\bibitem[{Rozo {et~al.}(2016)Rozo, Rykoff, Abate, Bonnett, Crocce, Davis,
  Hoyle, Leistedt, Peiris, Wechsler, \& et~al.}]{Rozo_2016}
Rozo, E., Rykoff, E.~S., Abate, A., {et~al.} 2016, MNRAS, 461, 1431–1450

\bibitem[{Samuroff {et~al.}(2023)Samuroff, Mandelbaum, Blazek, Campos,
  MacCrann, Zacharegkas, Amon, Prat, Singh, Elvin-Poole, Ross, Alarcon, Baxter,
  Bechtol, Becker, Bernstein, Rosell, Kind, Cawthon, Chang, Chen, Choi, Crocce,
  Davis, DeRose, Dodelson, Doux, Drlica-Wagner, Eckert, Everett, Fert{\'{e} },
  Gatti, Giannini, Gruen, Gruendl, Harrison, Herner, Huff, Jarvis, Kuropatkin,
  Leget, Lemos, McCullough, Myles, Navarro-Alsina, Pandey, Porredon, Raveri,
  Rodriguez-Monroy, Rollins, Roodman, Rossi, Rykoff, S{\'{a}}nchez, Secco,
  Sevilla-Noarbe, Sheldon, Shin, Troxel, Tutusaus, Weaverdyck, Yanny, Yin,
  Zhang, Zuntz, Aguena, Alves, Annis, Bacon, Bertin, Bocquet, Brooks, Burke,
  Carretero, Costanzi, da~Costa, Pereira, Vicente, Desai, Diehl, Dietrich,
  Doel, Ferrero, Flaugher, Frieman, Garc{\'{\i}}a-Bellido, Hinton, Hollowood,
  Honscheid, James, Kuehn, Lahav, Marshall, Melchior, Mena-Fern{\'{a}}ndez,
  Menanteau, Miquel, Newman, Palmese, Pieres, Malag{\'{o}}n, Sanchez, Scarpine,
  Smith, Suchyta, Swanson, Tarle, \& and}]{Samuroff_2023}
Samuroff, S., Mandelbaum, R., Blazek, J., {et~al.} 2023, MNRAS, 524, 2195

\bibitem[{Sartoris {et~al.}(2020)Sartoris, Biviano, Rosati, Mercurio, Grillo,
  Ettori, Nonino, Umetsu, Bergamini, Caminha, \& et~al.}]{Sartoris2020}
Sartoris, B., Biviano, A., Rosati, P., {et~al.} 2020, A\&A, 637, A34

\bibitem[{Scherrer \& Weinberg(1998)}]{Scherrer_1998}
Scherrer, R.~J. \& Weinberg, D.~H. 1998, ApJ, 504, 607–611

\bibitem[{Schmidt(2021)}]{Schmidt_2021}
Schmidt, F. 2021, JCAP, 2021, 033

\bibitem[{Secco {et~al.}(2022)Secco, Samuroff, Krause, Jain, Blazek, Raveri,
  Campos, Amon, Chen, Doux, Choi, Gruen, Bernstein, Chang, DeRose, Myles,
  Ferté, Lemos, Huterer, Prat, Troxel, MacCrann, Liddle, Kacprzak, Fang,
  Sánchez, Pandey, Dodelson, Chintalapati, Hoffmann, Alarcon, Alves,
  Andrade-Oliveira, Baxter, Bechtol, Becker, Brandao-Souza, Camacho,
  Carnero~Rosell, Carrasco~Kind, Cawthon, Cordero, Crocce, Davis, Di~Valentino,
  Drlica-Wagner, Eckert, Eifler, Elidaiana, Elsner, Elvin-Poole, Everett,
  Fosalba, Friedrich, Gatti, Giannini, Gruendl, Harrison, Hartley, Herner,
  Huang, Huff, Jarvis, Jeffrey, Kuropatkin, Leget, Muir, Mccullough,
  Navarro~Alsina, Omori, Park, Porredon, Rollins, Roodman, Rosenfeld, Ross,
  Rykoff, Sanchez, Sevilla-Noarbe, Sheldon, Shin, Troja, Tutusaus, Varga,
  Weaverdyck, Wechsler, Yanny, Yin, Zhang, Zuntz, Abbott, Aguena, Allam, Annis,
  Bacon, Bertin, Bhargava, Bridle, Brooks, Buckley-Geer, Burke, Carretero,
  Costanzi, da~Costa, De~Vicente, Diehl, Dietrich, Doel, Ferrero, Flaugher,
  Frieman, García-Bellido, Gaztanaga, Gerdes, Giannantonio, Gschwend,
  Gutierrez, Hinton, Hollowood, Honscheid, Hoyle, James, Jeltema, Kuehn, Lahav,
  Lima, Lin, Maia, Marshall, Martini, Melchior, Menanteau, Miquel, Mohr,
  Morgan, Ogando, Palmese, Paz-Chinchón, Petravick, Pieres, Plazas~Malagón,
  Rodriguez-Monroy, Romer, Sanchez, Scarpine, Schubnell, Scolnic, Serrano,
  Smith, Soares-Santos, Suchyta, Swanson, Tarle, Thomas, \& To}]{secco2021dark}
Secco, L., Samuroff, S., Krause, E., {et~al.} 2022, PRD, 105

\bibitem[{Seo \& Eisenstein(2007)}]{Seo_2007}
Seo, H. \& Eisenstein, D.~J. 2007, ApJ, 665, 14–24

\bibitem[{Sevilla-Noarbe {et~al.}(2021)Sevilla-Noarbe, Bechtol, Carrasco~Kind,
  Carnero~Rosell, Becker, Drlica-Wagner, Gruendl, Rykoff, Sheldon, Yanny, \&
  et~al.}]{Sevilla_Noarbe_2021}
Sevilla-Noarbe, I., Bechtol, K., Carrasco~Kind, M., {et~al.} 2021, ApJ
  Supplement Series, 254, 24

\bibitem[{Siegmund(1976)}]{Siegmund_1976}
Siegmund, D. 1976, Ann. Stat., 4, 673–684

\bibitem[{Skilling(2004)}]{Skilling2004}
Skilling, J. 2004, in AIP Conference Proceedings, Vol. 735 (AIP), 395–405

\bibitem[{Speagle(2020)}]{Speagle_2020}
Speagle, J.~S. 2020, MNRAS, 493, 3132–3158

\bibitem[{Sugiyama {et~al.}(2020)Sugiyama, Takada, Kobayashi, Miyatake,
  Shirasaki, Nishimichi, \& Park}]{Sugiyama_2020}
Sugiyama, S., Takada, M., Kobayashi, Y., {et~al.} 2020, PRD, 102

\bibitem[{Sugiyama {et~al.}(2022)Sugiyama, Takada, Miyatake, Nishimichi,
  Shirasaki, Kobayashi, Mandelbaum, More, Takahashi, Osato, Oguri, Coupon,
  Hikage, Hsieh, Komiyama, Leauthaud, Li, Luo, Lupton, Murayama, Nishizawa,
  Park, Price, Simet, Speagle, Strauss, \& Tanaka}]{Sugiyama_2022}
Sugiyama, S., Takada, M., Miyatake, H., {et~al.} 2022, PRD, 105

\bibitem[{Szewciw {et~al.}(2022)Szewciw, Beltz-Mohrmann, Berlind, \&
  Sinha}]{szewciw2021accurate}
Szewciw, A.~O., Beltz-Mohrmann, G.~D., Berlind, A.~A., \& Sinha, M. 2022, ApJ,
  926, 15

\bibitem[{Talts {et~al.}(2018)Talts, Betancourt, Simpson, Vehtari, \&
  Gelman}]{2018Talts}
Talts, S., Betancourt, M., Simpson, D., Vehtari, A., \& Gelman, A. 2018,
  Validating Bayesian Inference Algorithms with Simulation-Based Calibration

\bibitem[{Tanaka {et~al.}(2017)Tanaka, Coupon, Hsieh, Mineo, Nishizawa,
  Speagle, Furusawa, Miyazaki, \& Murayama}]{Tanaka_2017}
Tanaka, M., Coupon, J., Hsieh, B.-C., {et~al.} 2017, PASJ, 70

\bibitem[{To {et~al.}(2023)To, Rozo, Krause, Wu, Wechsler, \& Salcedo}]{To2023}
To, C.-H., Rozo, E., Krause, E., {et~al.} 2023, JCAP, 2023, 016

\bibitem[{{Tr{\"o}ster} {et~al.}(2022){Tr{\"o}ster}, {Mead}, {Heymans}, {Yan},
  {Alonso}, {Asgari}, {Bilicki}, {Dvornik}, {Hildebrandt}, {Joachimi},
  {Kannawadi}, {Kuijken}, {Schneider}, {Shan}, {van Waerbeke}, \&
  {Wright}}]{2022A&A...660A..27T}
{Tr{\"o}ster}, T., {Mead}, A.~J., {Heymans}, C., {et~al.} 2022, \aap, 660, A27

\bibitem[{Troxel \& Ishak(2015)}]{Troxel_2015}
Troxel, M. \& Ishak, M. 2015, Physics Reports, 558, 1–59

\bibitem[{van Daalen {et~al.}(2019)van Daalen, McCarthy, \&
  Schaye}]{van_Daalen_2019}
van Daalen, M.~P., McCarthy, I.~G., \& Schaye, J. 2019, MNRAS, 491, 2424–2446

\bibitem[{Velliscig {et~al.}(2014)Velliscig, van Daalen, Schaye, McCarthy,
  Cacciato, Le~Brun, \& Vecchia}]{Velliscig_2014}
Velliscig, M., van Daalen, M.~P., Schaye, J., {et~al.} 2014, MNRAS, 442,
  2641–2658

\bibitem[{Virtanen {et~al.}(2020)Virtanen, Gommers, Oliphant, Haberland, Reddy,
  Cournapeau, Burovski, Peterson, Weckesser, Bright, {van der Walt}, Brett,
  Wilson, Millman, Mayorov, Nelson, Jones, Kern, Larson, Carey, Polat, Feng,
  Moore, {VanderPlas}, Laxalde, Perktold, Cimrman, Henriksen, Quintero, Harris,
  Archibald, Ribeiro, Pedregosa, {van Mulbregt}, \& {SciPy 1.0
  Contributors}}]{2020SciPy}
Virtanen, P., Gommers, R., Oliphant, T.~E., {et~al.} 2020, Nature Methods, 17,
  261

\bibitem[{Voivodic \& Barreira(2021)}]{Voivodic_2021}
Voivodic, R. \& Barreira, A. 2021, JCAP, 2021, 069

\bibitem[{Walther {et~al.}(2021)Walther, Armengaud, Ravoux,
  Palanque-Delabrouille, Yèche, \& Lukić}]{Walther_2021}
Walther, M., Armengaud, E., Ravoux, C., {et~al.} 2021, JCAP, 2021, 059

\bibitem[{Wechsler {et~al.}(2022)Wechsler, DeRose, Busha, Becker, Rykoff, \&
  Evrard}]{wechsler2021addgals}
Wechsler, R.~H., DeRose, J., Busha, M.~T., {et~al.} 2022, ApJ, 931, 145

\bibitem[{Yao {et~al.}(2024)Yao, Blancard, \&
  Domke}]{yao2024simulationbasedstacking}
Yao, Y., Blancard, B. R.-S., \& Domke, J. 2024, Simulation-based stacking

\bibitem[{Yao \& Domke(2023)}]{yao2023discriminative}
Yao, Y. \& Domke, J. 2023, in Thirty-seventh Conference on Neural Information
  Processing Systems

\end{thebibliography}


\begin{appendix}

\section{Volume of re-scaled biased confidence interval}
\label{app:volume}

Here we give a derivation for Eq. \ref{eqn:volume_rescaling}, the re-scaling factor of the confidence interval volume when accounting for a bias by applying the change in confidence level according to the prescription given in Sect. \ref{sec:crit_2}.

The volume of the $n$-dimensional ellipsoid is given by
\begin{equation}
    V_n =\frac{2}{n} \frac{\pi^{n/2}}{\Gamma(n/2)} a_1 \cdot a_2 \cdot ... \cdot a_n \; ,
\end{equation}
where $a_i$ is the $i$-th semi-axis.

We assumed the ellipsoid under consideration is the one defined by the target parameter covariance matrix. What the re-scaling does it to change the length of the semi-axes by a constant factor, namely by the ratio of the inverse CDF of the non-central $\chi^2$ distribution with non-centrality parameter equal to the bias $\lambda=x^2$ defined in Eq. \ref{eq:bias}, to the inverse CDF of the central $\chi^2$ distribution.

\begin{align}
    V_n^{\rm extended} &=\frac{2}{n} \frac{\pi^{n/2}}{\Gamma(n/2)} a_1 \cdot a_2 \cdot ... \cdot a_n \cdot \nonumber \\
    & \left(\frac{{\rm CDF}^{-1}(0.68; k = n, \lambda)}{{\rm CDF}^{-1}(0.68; k = n, 0)}\right)^{n/2} \\
    &=V_n \left(\frac{{\rm CDF}^{-1}(0.68; k = n, \lambda)}{{\rm CDF}^{-1}(0.68; k = n, 0)}\right)^{n/2} \nonumber.
\end{align}
Since the denominator does not depend on the bias $\lambda$ of the respective prior, it is left out from Eq. \ref{eqn:volume_rescaling}.


\section{Summary of the DSS and its stochasticity models}
\label{app:APP_2}

The spatial distribution of galaxies and matter can be described by their respective density contrast fields
\begin{equation}
    \delta(x) = \frac{\rho(x) - \overline{\rho}}{\overline{\rho}} ,
\end{equation}
where $\rho(x)$ is the density at position $x$ and $\overline{\rho}$ is the mean density, of either matter ($m$) or galaxies ($g$). 

The distribution of matter $\delta_m$ cannot be directly observed, but it is traced by galaxies $\delta_g$ distributed in the sky. 
On large scales, the relation between galaxies and matter density contrasts can be approximated as linear expansion
\begin{equation}
    \delta_g \approx b \cdot \delta_m ,
    \label{eq:bias}
\end{equation}
with $b$ being the linear galaxy bias.  

We started by considering a distribution of galaxies in a fixed range around a redshift $z_{f}$, that we call foreground galaxy population. 
The 2D density field of the position of galaxies is obtained by applying a circular top-hat filter: 
at redshift $z_{f}$, we would count how many of the galaxies are inside of it. 
With this information, the sky can be divided into quintiles by ranked spatial number density. 
Next, we add another set of galaxies located at a higher redshift $z_b$ ($z_f < z_b$) that we call the background galaxy population. 

The light emitted by the background galaxies will be subject to gravitational shear when passing through the tidal gravitational field of the foreground. 
The effect will be stronger in zones where the density of foreground galaxies is higher. 
To evaluate this, we measure the tangential shear that the background sources will suffer in each quintile of foreground density. 
The observed lensed galaxies will trace the distribution and number density of the source galaxies.
This information is synthesized in a data vector of both the distribution of foreground galaxies and the lensing signals.  

If galaxy counts and the matter density were perfectly correlated, then a split of the sky by galaxy density would be identical to a split by the matter density. 
In a more realistic scenario, however, there is scatter of galaxy count in a volume at fixed matter density. 
Reasons for this include the discrete sampling of the galaxy density field by individual galaxies, higher order galaxy bias terms that are averaged over in projection, intrinsic stochasticity in three dimensions, or observational systematics.  

It is often assumed that the distribution of galaxy counts $N_g$ traces a smooth field $\delta_g=(1+b\delta_m) \overline{N}_g$, which is related to the matter overdensity $\delta_m$ in that region by a linear bias $b$, and it is given by Poisson shot-noise:
\begin{equation}
    P(N_g|\delta_m) = \text{Poisson}[ N_g,  (1+b\delta_m) \overline{N}_g] ,
\end{equation}
where $\overline{N}_g$ is the mean count of galaxies within the volume of such a region\footnote{Here, $\mathrm{Poisson}[k,\lambda]$ denotes the probability of drawing a count $k$ from a Poisson distribution with mean $\lambda$.}. 
This means that the variance of $N_g$ for fixed $\delta_g$ satisfies
\begin{equation}
    \frac{\text{Var}[N_g|\delta_g]}{\langle N_g| \delta_g \rangle} = 1 .
\end{equation}
When assuming a deterministic relationship between $\delta_m$ and $\delta_g$, a consequence is that the variance of $N_g$ for fixed $\delta_m$ is also 
\begin{equation}
    \frac{\text{Var}[N_g|\delta_m]}{\langle N_g| \delta_m \rangle} = 1 .
    \label{eq:Poissonian}
\end{equation}
If the strong assumption of linear, deterministic bias is not fulfilled, DSS analysis needs to consider more degrees of freedom in Eq. \ref{eq:Poissonian} to account for the effect of stochasticity.


\subsection{$\alpha$ model: Parametric model for non-Poissonianity} 

As a way of generalizing the galaxy-matter connection, it is possible to add two stochasticity parameters, $\alpha_0$ and $\alpha_1$, to obtain an equation for a generalized Poisson distribution:
\begin{equation}
    \frac{\text{Var}[N_g|\delta_m]}{\langle N_g| \delta_m \rangle} = \alpha_0 + \alpha_1 \delta_m ,
    \label{eq:Poissonian_Extension}
\end{equation}
in which, for $\alpha_0 = 1.0$ and $\alpha_1 = 0.0$, we recover Eq. \ref{eq:Poissonian}. 

The original prior distributions for the $\alpha$ parameters assigned in the original work from \citet{Friedrich2018, Gruen2018} are described in Table \ref{tab:table_1}. 
The only actual constraint on stochasticity is that $\alpha_0>0$, but the boundary $0.1 < \alpha_0$ was originally motivated with numerical arguments. 
See Appendix E from \citet{Friedrich2018} for more details on the derivation for the other boundaries.


\subsection{$r$ model: Correlation $r \neq 1$ between galaxy density and matter density}
\label{app:APP_r_model}

The second approach used in DSS is the introduction of a free parameter to the baseline linear bias model: 
a correlation coefficient $r$ between the random fields $\delta_g$ and $\delta_m$ (described in Sect. IV of \citealt{Friedrich2018}). 
This new degree of freedom captures a possible stochasticity. 
For $r=1$ the random fields $\delta_g$ and $\delta_m$ would be perfectly correlated. 
The covariance of $\delta_g$ and $\delta_m$ can be parametrized by the correlation coefficient
\begin{equation}
    r = \dfrac{\langle \delta_g \delta_m \rangle}{\sqrt{ \langle \delta^2_g \rangle \langle \delta^2_m \rangle}}.
\end{equation}

This can lead to a $\delta_m$ dependence of the ratio in Eq. \ref{eq:Poissonian} and the general model for the galaxy count:
\begin{equation}
    \text{Var}[N_g] = \overline{N}_g + \overline{N}_g^2 b^2 \text{Var}[\delta_m],
\end{equation}
\begin{equation}
    \text{Cov}[N_g, \delta_m] = \overline{N}_g b r \text{Var}[\delta_m].
\end{equation}

Mathematically, the possible values that $r$ could take are $-1 \leq r \leq 1$. 
For galaxies one expects a positive correlation to matter density, so the original prior for $r$ followed a uniform distribution $ \Pi(r) \sim \mathcal{U}$[0.0,1.0]. 


\section{ITEM priors with the MAP and mean estimators}
\label{app:APP_3}

\begin{figure}
  \includegraphics[width=0.95\columnwidth]{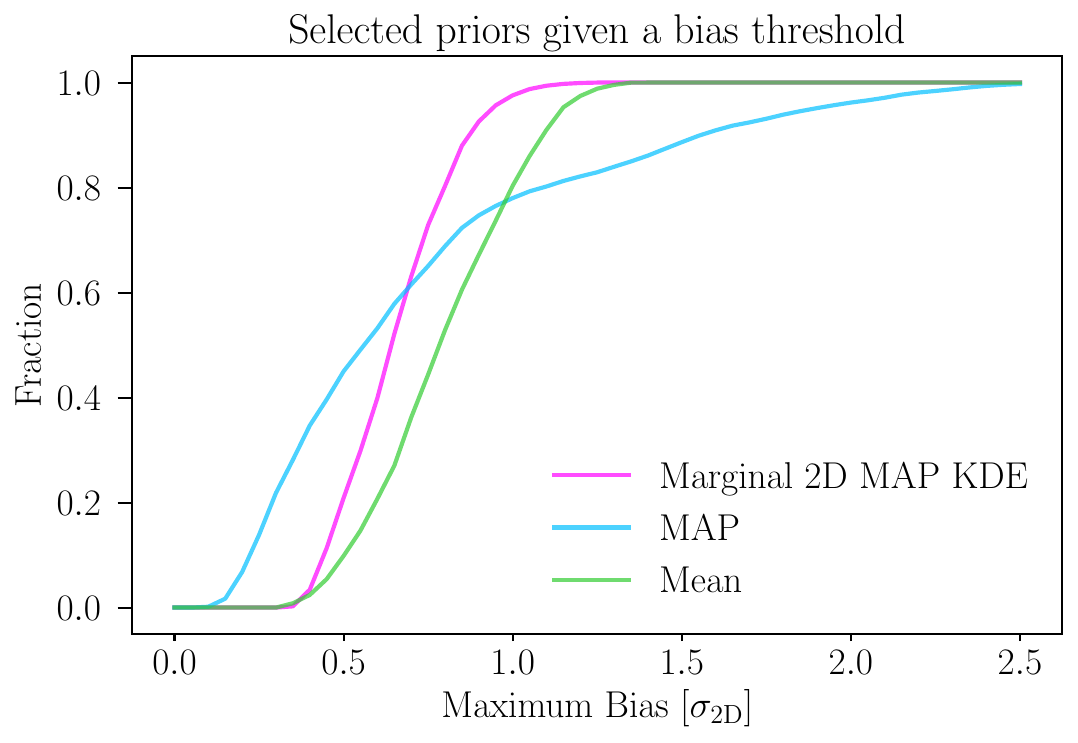}
  \centering
    \caption{Fraction of the total number of priors from Sect. \ref{sec:ITEM_alpha} that have a maximum bias smaller than $\sigma_{2D}$. 
    In magenta, light blue, and light green we show the results for the 2D marginal MAP, the MAP, and the mean estimators respectively.
    }
    \label{fi:biases_alpha_all}
\end{figure}

In this appendix we show a complementary ITEM prior analysis using the maximum a posteriori (MAP) and mean as point estimators on the $\alpha$ model and contrast their performance with respect to the marginal MAP estimator used in the main text of this paper.
We use specific colors for each point estimator, as shown in Fig. \ref{fi:biases_alpha_all}. 

\begin{figure}
\centering
    \includegraphics[width=0.8\columnwidth]{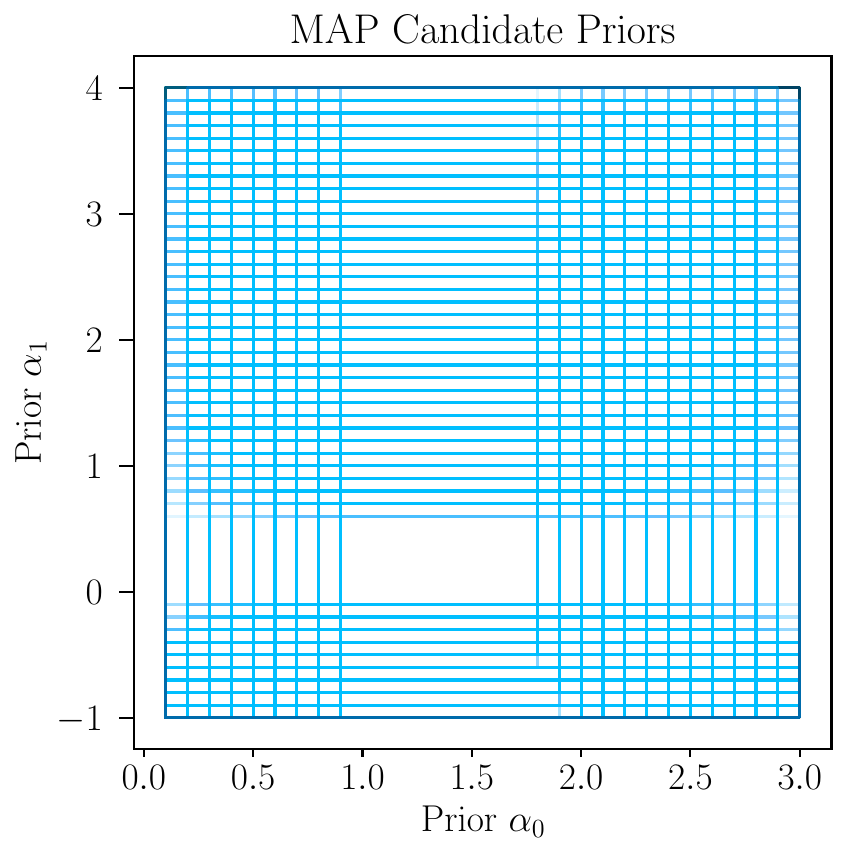}
    \includegraphics[width=0.8\columnwidth]{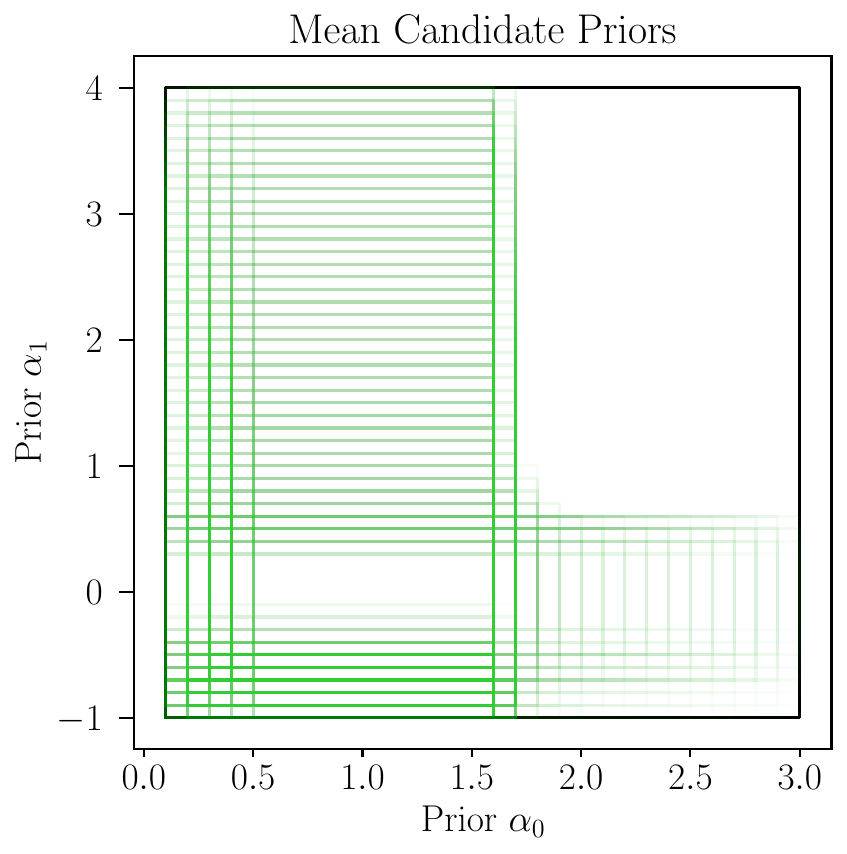}
   \caption{Priors that passed the first filter for each point estimator. In other words, these priors had a bias less than 0.4$\sigma_{2D}$.
   }
  \label{fi:priors_filter_1}
\end{figure}

The question of which point estimator to use for cosmological analysis is still in debate. 
DES used both the marginal MAP and the mean for presenting their summary cosmological statistics (\citealt{krause2021,Amon_2022, secco2021dark}).
Both the MAP and the 2D marginal MAP are non-trivial to estimate and sensitive to the algorithm used, as studied in \citet{survey2023des}.

In our study, the MAP was obtained using the maximum value output from \textsc{chainconsumer} \citep{chainconsumer}, which \citet{survey2023des} found to have a Gaussian kernel density estimation that leads to $\sim$10$\%$ larger errors.
On the other hand, the 2D marginal MAP was estimated using  \textsc{chainconsumer}'s Gaussian kernel density estimate (KDE) to smooth the marginalized posteriors and therefore estimate the marginal MAP. 
The estimate of this point is affected by the scale; therefore, after studying the impact of the bandpass, it was decided to keep the default values presented by \textsc{chainconsumer} in their API (kde = 1.0 in \textsc{v0.33.0}). 
We use the same candidate priors as in Sect. \ref{sec:ITEM_alpha}, but we first studied the $\sigma_{2D}$ biases for each model realization, but now using additional point estimators.

\begin{figure}
    \includegraphics[width=0.95\columnwidth]{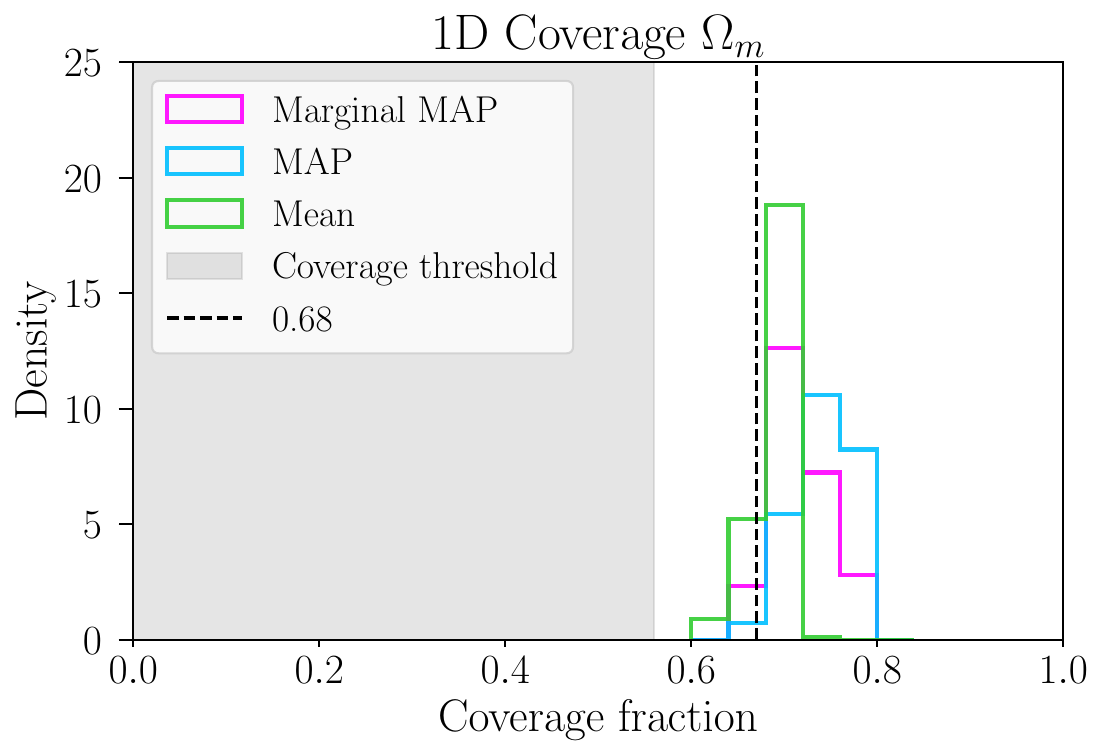}
    \includegraphics[width=0.95\columnwidth]{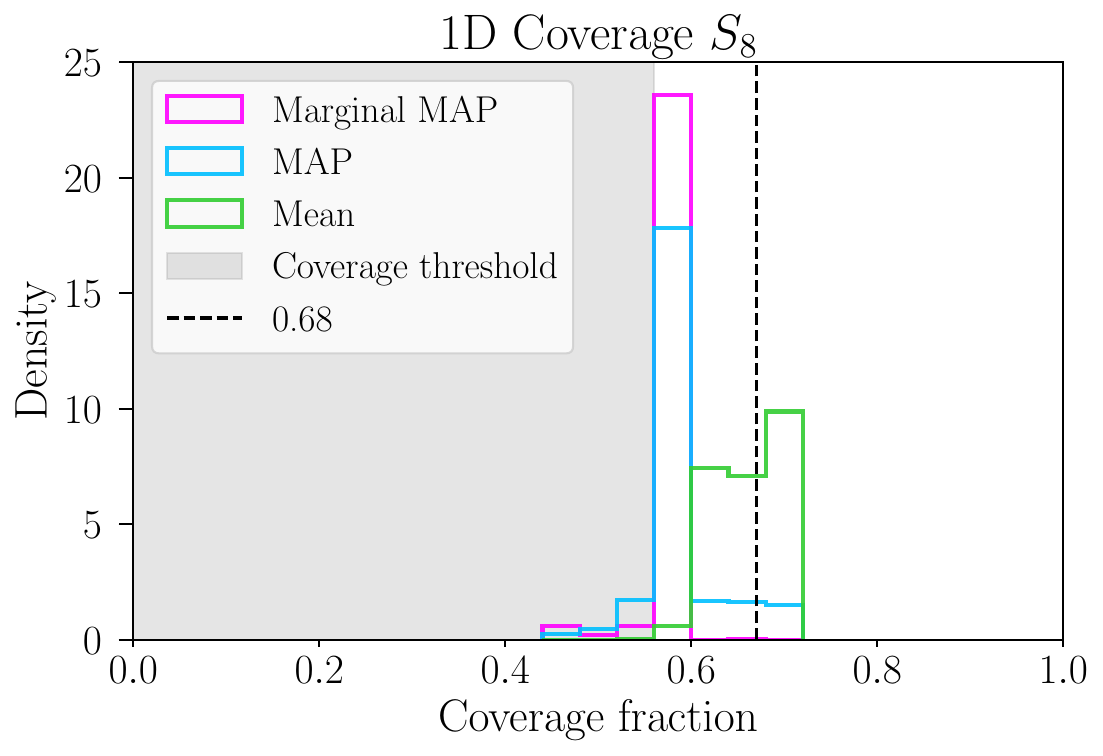}
   \caption{
   One-dimensional coverages for the candidate priors that passed the filter 1 for the different point estimators. 
   In the upper panel we show the 1D coverage on $\Omega_m$, while in the lower panel, we display the 1D coverage on $S_8$. 
   The histograms have been normalized to contrast their distribution along the vertical axis. 
   }
  \label{fi:coverages_alpha}
\end{figure}
In Fig. \ref{fi:biases_alpha_all}, we show the fraction of priors as a function of the maximum bias of the two stochasticity realizations.
The MAP curve, in blue, resulted in both the lowest and largest maximum biases possible due to the prior volume effect present on the whole dimensional space.
In magenta we have the 2D marginal MAP curve.
This is the point estimator that returned the most constrained values of biases (all are covered with $\sim$1.0 $\sigma_{\rm 2D}$).
The advantage of using the 2D marginal MAP rather than the MAP as the main statistics in this context is that the 2D marginal MAP will not return values as biased as the ones calculated using the MAP in the presence of projection effects. 
Finally, in green, the mean estimator had the worst bias due to volume effects happening when marginalizing the 1D posteriors. 

\begin{table}
\centering
\caption{\label{t7}
Number of priors passing each filter given a point estimator.
}
\label{tab:table_4}
\begin{tabular}{c|c|c}
\textbf{Point estimator} & \textbf{Filter 1} & \textbf{Filter 2} \\ \hline
2D marginal MAP  &  1,980 &  1,868 \\
MAP  &  20,160 &  18,232 \\
mean  &  1,401 &  1,400       
\end{tabular}
\end{table}

\begin{figure}
  \includegraphics[width=0.8\columnwidth]{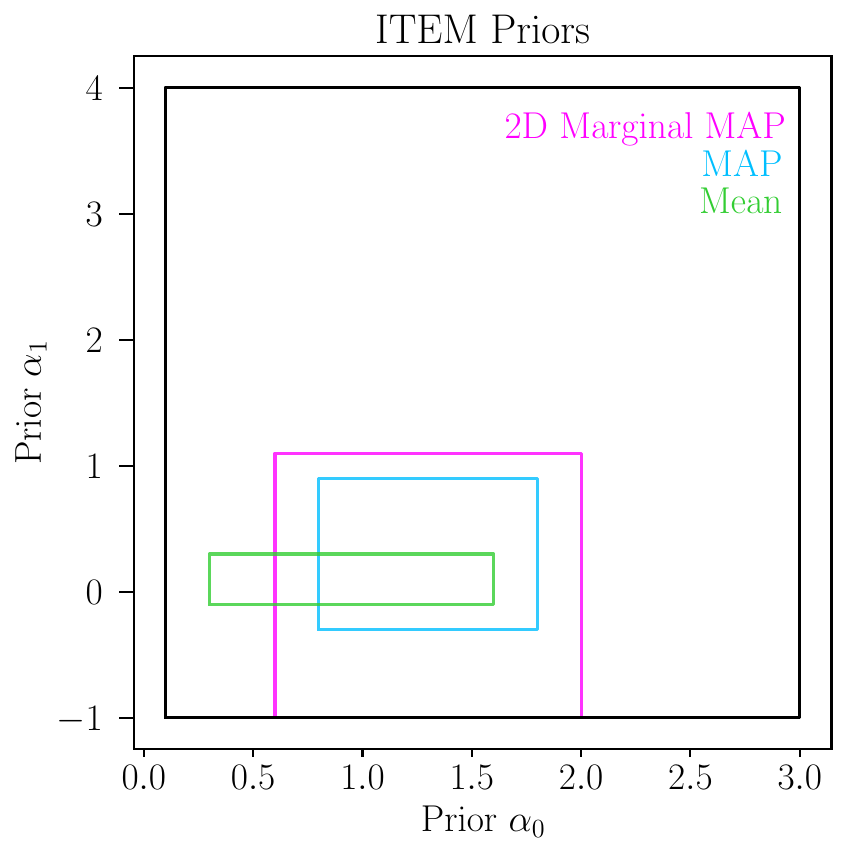}
  \centering
    \caption{
    Distribution of the ITEM priors for the different tested point estimators.}
    \label{fi:item_priors_all}
\end{figure}

Following Sect. \ref{sec:ITEM_alpha}, we used a bias of 0.4$\sigma_{\rm 2D}$ for each estimator.
The number of candidate priors that passed those bias thresholds are listed in Table \ref{tab:table_4}, and are shown in Fig. \ref{fi:A_2} and \ref{fi:priors_filter_1}.
The next step in determining the ITEM priors is to study the fraction of priors that have a minimum frequentist coverage. 
The 1D coverage distributions are visible in Fig. \ref{fi:coverages_alpha}. 
Following Sect. \ref{sec:ITEM_alpha}, we required the candidate priors to have the same coverage fractions for the noisy realizations, in other words at least a 56\% (14 out of 25 realizations) coverage for $\Omega_m$ and $S_8$.
This lower limit on coverages decreased the number of priors as shown in the last column of Table \ref{tab:table_4}.
The last step of the ITEM prior implementation is to minimize the total uncertainty of the posterior.
In Fig. \ref{fi:item_priors_all}, we show the resulting priors on both stochasticity parameters $\alpha_0$ and $\alpha_1$, which differ depending on the point estimator.

\begin{figure*}
  \begin{subfigure}[b]{0.5\textwidth}
    \centering\includegraphics[width=0.96\columnwidth]{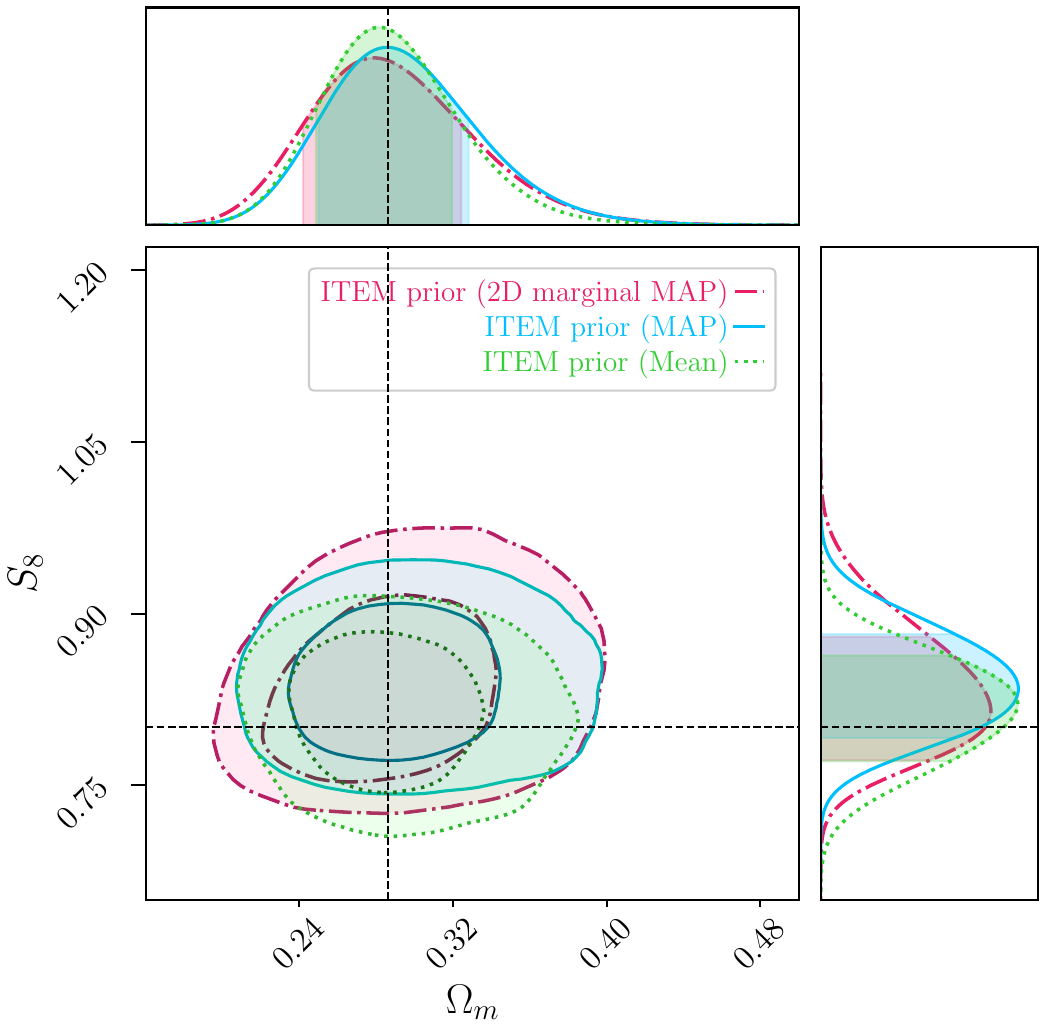}
  \end{subfigure}
  \begin{subfigure}[b]{0.5\textwidth}
    \includegraphics[width=\columnwidth]{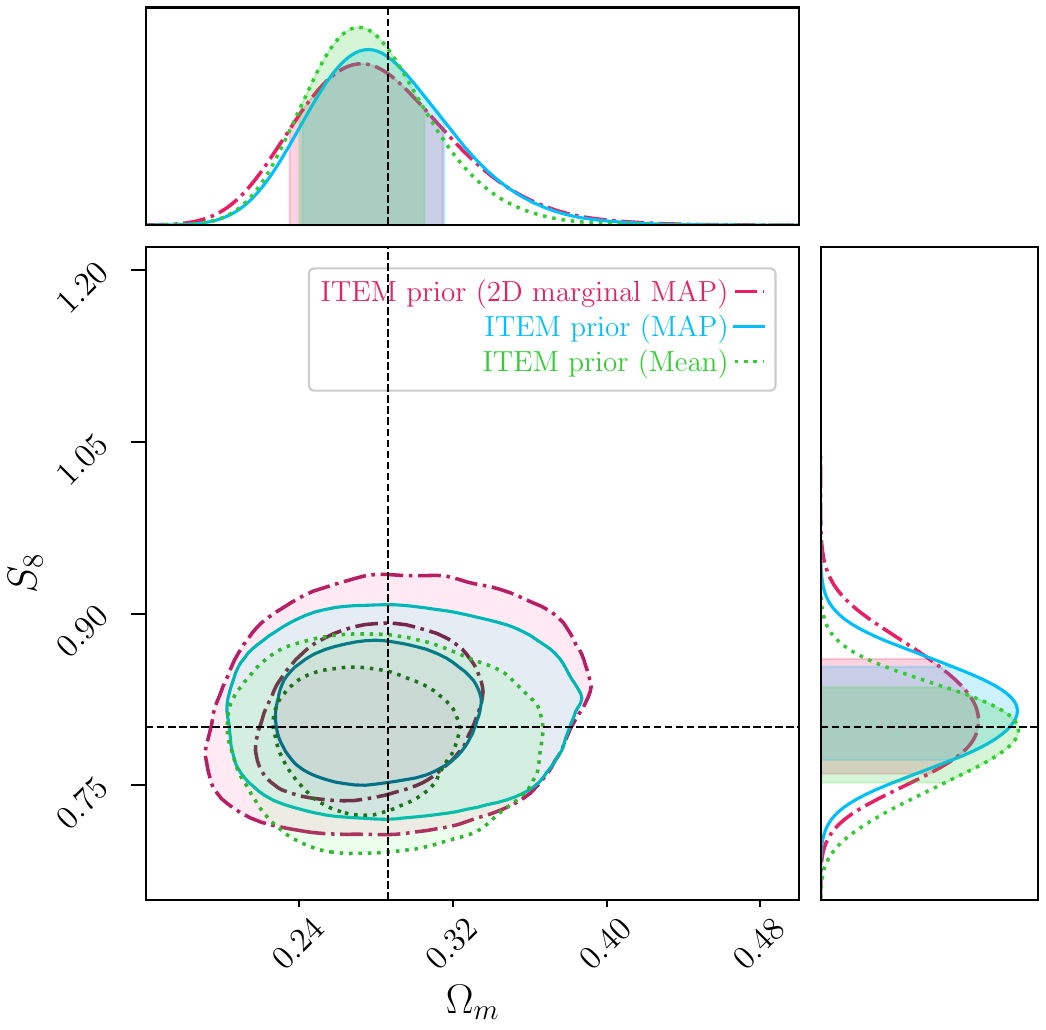}
  \end{subfigure}
  \caption{
  Resulting marginalized posteriors for the ITEM priors obtained with different point estimators.
  Similar to Fig. \ref{fi:ITEM_marginalized_posteriors}, the $1$$\sigma$ and $2$$\sigma$ confidence levels had been widened to account for the bias of each case. 
  }
  \label{fi:item_porteriors}
\end{figure*}

The ITEM priors for both $\alpha_0$ and $\alpha_1$ have substantially smaller widths with respect to the one from the original prior (see Table \ref{tab:table_2}), especially the case of the mean.
In particular, the upper limit on the parameter $\alpha_1$ is reduced to less than half of the original prior width in all cases.
In Table \ref{tab:table_5}, we display the main statistics for the ITEM priors obtained using different point estimators as a reference, and, in Fig. \ref{fi:item_porteriors} we show their marginalized posteriors. 

\begin{table*}
    \centering
    \caption{\label{t7} ITEM priors for the $\alpha$ model for different point estimators.
    }
        \label{tab:table_5}
    \begin{tabular}{lcccccccc}
        \hline
        Parameter & Prior & Prior width & Bias [$\sigma_{2D}$] & $\Omega_m$ coverage & $S_8$ coverage & $\overline{\sigma}_{\Omega_m}$  & $\overline{\sigma}_{S_8}$  & $\tilde{V}/\tilde{V}_{\rm Original}$ \\ 
        \hline
        & & & & & & & \\
        \textbf{2D MAP} & & & & & & & & \\
        $\alpha_0$ & $\mathcal{U}$[0.6, 2.0] & 1.4 & 
        \Large{$\underset{0.33/0.39}{}$}
        & 
        \Large{$\underset{80 \% / 64 \%}{}$}
        &  
        \Large{$\underset{68 \% / 56 \%}{}$}
        & 
        \Large{$\underset{0.040}{}$}
        & 
        \Large{$\underset{0.049}{}$}
        & 
        \Large{$\underset{0.58}{}$} \\
        $\alpha_1$ & $\mathcal{U}$[-1.0, 1.1] & 2.1 & & & & \\  
        & & & & & & & \\
        \textbf{MAP} & & & & & & & & \\
        $\alpha_0$ & $\mathcal{U}$[0.8, 1.8] & 1.0 
        & 
        \Large{$\underset{0.30 / 0.40}{}$}
        & 
        \Large{$\underset{80 \% / 80 \%}{}$}
        &  
        \Large{$\underset{80 \% / 56 \%}{}$}
        & 
        \Large{$\underset{0.037}{}$}
        & 
        \Large{$\underset{0.040}{}$}
        & 
        \Large{$\underset{0.52}{}$} \\
        $\alpha_1$ & $\mathcal{U}$[-0.3, 0.9] & 1.2 & & & & & & \\
        & & & & & & & \\
        \textbf{mean} & & & & & & & \\
        $\alpha_0$ & $\mathcal{U}$[0.3, 1.6] & 1.3 
        & 
        \Large{$\underset{0.35 / 0.40}{}$}
        & 
        \Large{$\underset{68 \% / 72 \%}{}$}
        &  
        \Large{$\underset{80 \% / 64 \%}{}$}
        & 
        \Large{$\underset{0.033}{}$}
        & 
        \Large{$\underset{0.040}{}$}
        & 
        \Large{$\underset{0.34}{}$} \\
        $\alpha_1$ & $\mathcal{U}$[-0.1, 0.3] & 0.4 & & & & & & \\ 
        & & & & & & & \\
        \hline
    \end{tabular}
\end{table*}


All the biases for both nuisance result in similar values and we obtained consistent coverages despite the use of varying different priors.
The $\overline{\sigma}$ uncertainties for the ITEM priors and the total uncertainty $\tilde{V}/\tilde{V}_{\rm Original}$ decreased consistently with the prior volume reduction. 
The marginal posteriors from Fig. \ref{fi:item_porteriors} are also in agreement with what we would expect from divergent priors: the three of them are different.


\end{appendix}

\end{document}